%% file: main.tex
\RequirePackage{tikz}

\documentclass{article}

\usepackage[margin=1in]{geometry}
\usepackage[toc]{appendix}
\usepackage[backend=biber, style=numeric-comp, sortcites=false,sorting=none,uniquelist=false,natbib=true,doi=true,isbn=false,url=false,eprint=false,pagetracker, ibidtracker=constrict, giveninits=true,maxbibnames=10,maxcitenames=1]{biblatex}
\DeclareSourcemap{
  \maps{
    \map{
      \step[fieldset=entrysubtype, null]
      \step[fieldset=address, null]
      \step[fieldset=location, null]
    }
  }
}

\usepackage[colorlinks=true,citecolor=blue]{hyperref}
\usepackage{textcomp}
\usepackage{textgreek}
\usepackage{latexsym}
\usepackage[T1]{fontenc}
\usepackage{bbm}
\usepackage{float}
\usepackage{wrapfig}

\usepackage{tikz}
\usepackage{pgfplots}
\usetikzlibrary{shapes,positioning,calc,arrows.meta,arrows,patterns,patterns.meta,fit}
\usepackage{tikzscale}
\usepackage{subcaption}

\usepackage{longtable}
\usepackage{makecell}
\usepackage{multirow}
\usepackage{multicol}

\usepackage{amsmath,bm}
\usepackage{amssymb}
\usepackage{dsfont}
\usepackage[english]{babel}
\usepackage{url}  
\usepackage{graphicx}  
\usepackage{url}
\usepackage{tabularx}
\usepackage{booktabs}
\usepackage{array}

\usepackage[acronym,nomain,nonumberlist]{glossaries}
\makeglossaries

\input{acronyms_en}

\DeclareMathOperator*{\argmax}{arg\,max}

\title{Balancing Specialization and Adaptation in a Transforming Scientific Landscape}

\newcommand*{\affaddr}[1]{#1} 
\newcommand*{\affmark}[1][*]{\textsuperscript{#1}}
\newcommand*{\email}[1][*]{\texttt{#1}}

\author{Lucas Gautheron\protect\affmark[1,2]}

\date{}

\begin{document}
\maketitle


\begin{abstract}
How do scientists navigate between the need to capitalize on their prior knowledge through specialization, and the urge to adapt to evolving research opportunities? Drawing from diverse perspectives on adaptation, this paper proposes an unsupervised Bayesian approach motivated by Optimal Transport of the evolution of scientists' research portfolios in response to transformations in their field. The model relies on $186,162$ scientific abstracts and authorship data to evaluate the influence of intellectual, social, and institutional resources on scientists' trajectories within a cohort of $2\,094$ high-energy physicists between 2000 and 2019. Using Inverse Optimal Transport, the reallocation of research efforts is shown to be shaped by learning costs, thus enhancing the utility of the scientific capital disseminated among scientists. Two dimensions of social capital, namely ``diversity'' and ``power'', have opposite associations with the magnitude of change in scientists' research interests: while ``diversity'' is associated with change and expanded research portfolios, ``power'' is associated with more stable research agendas. Social capital plays a more crucial role in shifts between cognitively distant research areas. More generally, this work suggests new approaches for understanding, measuring and modeling collective adaptation using Optimal Transport.
\end{abstract}

\ \\

  \email{lucas.gautheron@gmail.com}\\ORCID: 0000-0002-3776-3373\\
 \affaddr{\affmark[1]Interdisciplinary Centre for Science and Technology Studies (IZWT), University of Wuppertal, Germany}\\
 \affaddr{\affmark[2]Département d'Études Cognitives, École Normale Supérieure, Paris, France}



\section{Introduction}

Scientists are subject to conflicting incentives. On the one hand, they must work within the realm of their expertise, where they can most effectively exploit their prior knowledge and compete with peers; this conservative preference for familiar research topics is at the root of specialization. On the other hand, scientists are simultaneously compelled to revise their research interests to engage with more promising research areas in order to benefit from more exposure or to secure funding. Thus, in some instances, specialization is at odds with the need to adapt to the decline of certain research opportunities and the growth of new ones. How do scientists navigate the trade-off between specialization (i.e. the concentration of their intellectual resources within a narrow cognitive range) and adaptation (i.e. the need to adjust these resources to new realities)? This conflict differs from the ``essential tension'' between ``tradition'' and ``innovation'' proposed by \citet{Kuhn1997}, or that between ``exploration'' and ``exploitation'' \citep{March1991}, which have both been explored quantitatively in previous works \citep{Foster2015,Jia2017,Aleta2019,Zeng2019,Tripodi2020,Singh2024,liu2024science}
. First, ``adaptation'' is not tantamount to innovation or disruption, for it can be a conformist move (e.g. as a result of a bandwagon effect \citep{Fujimura1988}). Moreover,  unlike ``exploration'', adaptation is not identical to a search strategy in a static landscape \citep{Galesic2023}, but rather the convergence towards a new state more congruent with current realities. Disruptions due to breakthroughs in Machine Learning or challenges due to climate change urge to understand how scientists adapt to changing circumstances. Therefore, the present paper investigates scientists' responses to changes in their field (whether driven by epistemic or institutional factors), and the effect of their capital (intellectual, social, or institutional) on their ability to adapt. Drawing insights on adaptation from cultural evolution and institutional change, we develop an unsupervised Bayesian approach to analyze changes in scientists' research agenda while measuring the effect of ``capital'' (intellectual or social) on their individual trajectories. The model is applied to a cohort of high-energy physicists between the years 2000 and 2019, a time during which the historical driver of progress in the field -- particle accelerators -- have been contested by emerging astrophysical experiments, thus transforming the landscape of opportunities.

Our approach reveals trends in the field: the boom of dark matter research -- fueled by shifts away from the physics of neutrinos and the electroweak sector -- and the partial disintegration of string theory into the study of black holes and holography/dualities. More importantly, this analysis also shows that changes in scientists' research portfolios are shaped by learning costs, as scientific communities adapting to new circumstances address an ``Optimal Transport'' problem by reallocating research efforts efficiently. Optimal Transport is a mathematical framework initially concerned with the optimal displacement and allocation of resources \citep{monge1781memoire,kantorovich2006translocation,Peyr2019}, and has since then found wide-ranging applications. We show that it also provides a characterization of scientists' behavior, as driven by the need to maximize the utility of their scientific capital under changing circumstances. Moreover, the comparative analysis shows that two dimensions of social capital, namely ``diversity'' and ``power'' \citep{Abbasi2014}, have opposite associations with change. While ``diversity'' of social capital  -- the extent to which scientists have access to diverse cognitive resources via their collaborators -- is correlated with greater change and further diversification of scientists' research interests, ``power'' -- roughly speaking, the size of their network -- is associated with more stability in their research interests. Social capital has a stronger association with transfers between research areas that are more cognitively ``distant''. There is no discernible effect of institutional stability after controlling for academic age (although affiliation data is a bit noisy in the dataset). Overall, we contribute: i) a conceptual account of the features of change in scientists' research interests; ii) a novel methodological approach that introduces a model of scientists' trajectories connected to Optimal Transport and measures of intellectual capital, social capital, diversity, and power; and finally, iii) some empirical evidence from high-energy physics. More generally, this paper addresses the relative lack of empirical works within the body of literature that investigates science as a cultural evolutionary system \citep{Wu2023}. It demonstrates that Optimal Transport provides an insightful description of certain aspects of collective adaptation, but also computational tools (such as Probabilistic Inverse Optimal Transport \citep{pmlr-v162-chiu22b} and OT based measures of change) for measuring adaptive behavior, and more generally, mobility in physical and abstract spaces.

In what follows, Sections \ref{sec:review} and \ref{sec:conceptual} summarize previous research and lay out the conceptual background on which the analysis rests, and Section \ref{sec:hep} introduces the context of high-energy physics to which the model is applied. 
Section \ref{sec:methods} elaborates the methodology: the data (\ref{sec:data}), the topic model approach for measuring authors' research portfolios (\ref{sec:topics}), the proposed measures of intellectual and social capital (\ref{sec:capital}), and the model of scientists' trajectories (\ref{sec:model}).
Section \ref{sec:results} presents the results: i) the transfers of attention from one research area to another due to changes in the scientific landscape; ii) the structuring role of learning costs in the observed patterns of adaptation (\ref{sec:optimal-transport}), and iii) the effect of capital (intellectual and social) and institutional stability on physicists' strategies (\ref{sec:comparative}). 

\subsection{\label{sec:review}Empirical background}

Several works have investigated the evolution of scientists' research interests. For instance, by mapping the trajectories of 103,246 physicists over 26 years using the \gls{aps} dataset and its topic classification (the \gls{pacs}), \citet{Aleta2019} demonstrated that a majority of physicists gradually migrate to entirely different topics by the end of their careers while often staying within the same general area. They reveal differences between subfields of physics, such that ``exploitation'' (i.e. specialization, as opposed to the ``exploration'' of new topics) is especially prevalent in particle physics.  Using the same data, \citet{Jia2017} instead find an exponentially decaying distribution of changes in scientists' interests. 
Previous works generally agree, however, on the graduality of change in research topics \citep{Jia2017,Aleta2019,Zeng2019}, as previously observed by  \citet{Gieryn1978}. Recognizing that scientists typically investigate several research questions in parallel, \citeauthor{Gieryn1978} proposed four mechanisms of gradual change, including ``accretion'' (a problem is added to their ``problem set''),  ``selective substitution'' (one problem is replaced by another), and ``selective disengagement'' (one problem is neglected).

While \citep{Jia2017,Aleta2019,Zeng2019} document the structure of changes in scientists' interests, they do not relate these transformations to changes in epistemic and institutional context, or to the scientists' incentives and resources. 
\citet{Tripodi2020} have taken a step in this direction. Using the \gls{aps} dataset, they show that physicists are more likely to explore areas to which they are connected via their collaborators, and highlight the crucial importance of collaborations in the expansion of research portfolios, especially for the exploration of research areas distant from one's core specialization. 
However, their work does not primarily address the \textit{transformations} of scientists research portfolios throughout time -- they do not quantify ``change'' --, and they recognize the need for further longitudinal analyses. Finally, previous works have explored the connection between spatial mobility patterns and scientific mobility using gravity or radiation models \citep{Singh2024,liu2024science}. In particular, \citealt{Singh2024} used such methods to compare the characteristics of two types of scientists, ``explorers'' as opposed to ``exploiters''.

The present paper complements previous works on changes in scientists' research in several ways. First, since the focus is on adaptation strategies, the core of our approach is both comparative (as in \citealt{Tripodi2020}, and unlike other previous works) \textit{and} longitudinal (unlike \citealt{Tripodi2020}, although their paper includes longitudinal robustness checks). Second, this contribution evaluates previously unexplored aspects, such as the choice between expansion or consolidation of research portfolios and the effect of affiliation stability. 
Third, this work relates the findings to the epistemic context of the field and its transformations by performing the analysis at a circumscribed scale (high-energy physics). Fourth, this work does \textit{not} rely primarily on the \gls{aps} dataset and \gls{pacs} categories, on which most previous works depend \citep{Jia2017,Aleta2019,Battiston2019,Tripodi2020}, or any other pre-existing classification of the literature. Research areas are clustered using an unsupervised topic model, such that this approach measures linguistic change, which is arguably a more direct proxy of cognitive change. Fifth, this paper is the first application of the Probabilistic Inverse Optimal Transport approach from \citealt{pmlr-v162-chiu22b}, which provides an alternative to other approaches to mobility (e.g. gravity models). Finally, the proposed approach is grounded in theory, by operationalizing concepts such as capital \citep{Bourdieu1980,Bourdieu1986}, and by exploiting theoretical insights from diverse approaches to ``adaptation''.

\subsection{\label{sec:conceptual}Conceptual framework}

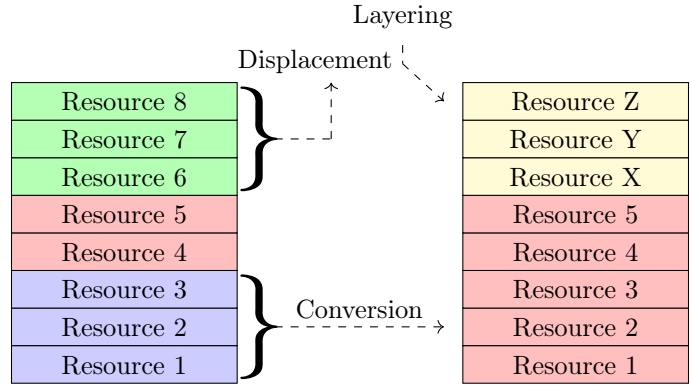
\begin{figure}[h]
    \centering
    \begin{tikzpicture}
    \foreach \i in {1,2,3}
        \draw [fill=blue!20] (0,{(\i-1)*0.5}) rectangle (3,{\i*0.5}) node[pos=.5] (res\i) {Resource \i};

    \foreach \i in {4,5}
        \draw [fill=red!25] (0,{(\i-1)*0.5}) rectangle (3,{\i*0.5}) node[pos=.5] (res\i) {Resource \i};



    \foreach \i in {6,7,8}
        \draw [fill=green!30] (0,{(\i-1)*0.5}) rectangle (3,{\i*0.5}) node[pos=.5] (res\i) {Resource \i};

        
    \draw [->, dashed] (3.5,0.75) -- (5.75,0.75) node[midway, above] {Conversion};
     \draw [->, dashed] (4.25,3.25) -- (4.25,{8.5*0.5-0.25}) node[above,yshift=0,xshift=-6] {Displacement};
        
    \draw [-, dashed] (3.5,3.25) -- (4.25,3.25);

    \node [fit=(res1) (res2) (res3)] (fitblue) {};
    \path let \p1=(fitblue.north west), \p2 = (fitblue.south west) in
       node [right of=fitblue, xshift=2.25em] {%
       \pgfmathsetmacro\heightoffit{.6*(\y1-\y2)}%
       \resizebox{!}{\heightoffit pt}{\}}%
     };

    \node [fit=(res6) (res7) (res8)] (fitgreen) {};
    \path let \p1=(fitgreen.north west), \p2 = (fitgreen.south west) in
       node [right of=fitgreen, xshift=2.25em] {%
       \pgfmathsetmacro\heightoffit{.6*(\y1-\y2)}%
       \resizebox{!}{\heightoffit pt}{\}}%
     };


    \draw [-, dashed] (5.2,{8.5*0.5+0.25}) -- (5.2,{8.5*0.5}) node[above,yshift=10] {Layering};
    
    
    \draw [->, dashed] (5.2,{8.5*0.5}) -- (5.75,{7.5*0.5});
    
    

    \foreach \i in {1,2,3,4,5}
        \draw [fill=red!25] (6,{(\i-1)*0.5}) rectangle (9,{\i*0.5}) node[pos=.5] {Resource \i};

    \draw [fill=yellow!20] (6,{(6-1)*0.5}) rectangle (9,{6*0.5}) node[pos=.5] {Resource X};

    \draw [fill=yellow!20] (6,{(7-1)*0.5}) rectangle (9,{7*0.5}) node[pos=.5] {Resource Y};

    \draw [fill=yellow!20] (6,{(8-1)*0.5}) rectangle (9,{8*0.5}) node[pos=.5] {Resource Z};

    

    
    \end{tikzpicture} 

    \caption{\textbf{Changes in a scientist's ``research portfolio'' over time.} Colors designate research areas. Resources entail any intellectual or methodological assets that a scientist uses to investigate problems in each area. ``Conversion'' repurposes knowledge to new goals. ``Displacement'' is the replacement of certain research interests with little or no transfer of prior knowledge, as illustrated by the green area. Layering is the introduction of new research interests via the addition of new knowledge.}
    \label{fig:research-agenda}
\end{figure}

A central dilemma of adaptation consists in choosing which  resources to leverage among those already available (although those may be suboptimal or irrelevant under new circumstances) and which resources to abandon and replace with others (which may be inefficiently costly). 
By adapting gradually, scientists can strategically retain  the benefits of ``problem retention'' (e.g., the exploitation of ``accumulated skills and resources'' in one area, or ``of an established research network'', \citealt[p.~106]{Gieryn1978}) while progressively investing resources in new research directions.
This is illustrated in Figure \ref{fig:research-agenda}, 
which represents the research portfolios of one scientist during two distinct time periods. Cells indicate the resources exploited by the scientist (e.g., concepts, models, methods, etc.) and colors indicate to what problem areas this knowledge is applied.
Figure \ref{fig:research-agenda} shows how scientists can enter new research areas by repurposing certain resources to new ends \citep{Mulkay1974,schon1963displacement}. We call this strategy ``\textit{conversion}'', in reference to the typology of incremental institutional change proposed by \citet{mahoney_thelen_2009}\footnote{Indeed, as shown in previous works on the transformations of high-energy physics facilities to photon science instruments \citep{Hallonsten2013,hallonsten2015formation,Heinze2017a}, historical institutionalism can account for gradual adaptations with large cumulative effects taking place in response to scientific and technological change \citep{Heinze2012}. In this paper, we apply the typology of change to \textit{individuals} rather than organizations.}

Not all knowledge can be successfully applied to new research areas: 
as illustrated in Figure \ref{fig:research-agenda}, entering new research areas typically requires ``\textit{layering}''\footnote{Again, borrowing the terminology from historical institutionalism.}, that is, the introduction of new concepts, models, or methods, on top of prior knowledge. The acquisition of knowledge entails learning costs, which can be partially avoided by collaborating with experts in the target domain \citep{Tripodi2020}. Another mode of change is \textit{displacement}, when the replacement of one research area for another involves significant neglect of prior knowledge. 
This may not be the preferred strategy, since it fails to take advantage of accumulated resources. However, certain knowledge may not apply to a new context, or sometimes there might be reasons to suspend a line of research in order to focus on more promising topics. Overall, we expect that these transformations will manifest themselves as changes in scientists' linguistic behavior, i.e., as changes in the vocabulary of their publications. Generally, we expect an important amount of continuity in linguistic behavior, given the need to minimize cognitive learning costs by capitalizing on prior knowledge.

Scientists manage two kinds of assets when navigating the trade-off between specialization and adaptation: their own prior expertise, and the expertise to which they have access through their social network. Both constitute ``capital'' \citep{Bourdieu1986}, i.e. assets that individuals accumulate and leverage in the competitive context of their field. ``Capital''  (whether ``economic'', ``cultural'', ``social'' or even ``symbolic'', cf. \citealt{Bourdieu1986}) defines the scope of scientists' opportunities and therefore their ability to adapt. This paper considers the \textit{intellectual capital} possessed by scientists in the form of scientific knowledge, and \textit{social capital}. 
Measures that represent these concepts will be proposed, and their effect on the magnitude of transfers of attention across research areas will be evaluated. Emphasis will be put on the divide underlined in \citealt{Abbasi2014} between two dimensions of social capital, namely ``power'' (roughly speaking, network size in the present paper) and ``diversity'' (of cognitive resources). Group diversity is generally recognized as a factor of adaptation in an evolving environment or in the context of collective problem-solving \citep{Smaldino2023,Schimmelpfennig2021,Muthukrishna2016,page2008difference}. ``Power'' is also plausibly associated with higher abilities.

While capital defines scientists' opportunities, it is not sufficient to explain \textit{why} scientists \textit{do} turn to new research areas or not, which also requires understanding actors' incentives and why they must respond to these incentives. Consequently, the present paper also considers the effect of institutional stability and academic age on migrations between research areas, since the need to respond to changes in the epistemic and institutional environment is presumably different for, say, tenured physicists versus postdocs. Moreover, we may assume that younger and older generations play different roles in cultural change and collective adaptation in general \citep{Acerbi2012}. Finally, the effect of productivity will also be considered. 

\subsection{\label{sec:hep}The case of high-energy physics: navigating a changing epistemic landscape}

High-energy physics is a prime example for investigating adaptation in a transforming scientific landscape. As this field relied on the input of increasingly large particle colliders to achieve progress, it has accumulated considerable capital directed towards collider physics in the form of large infrastructure and complex knowledge. These efforts culminated in 2010 with the start of the \gls{lhc}, the largest accelerator ever. However, the LHC has found no evidence for anything that was not already predicted by the Standard Model of particle physics, and it is increasingly plausible that no future accelerator could ever find any evidence for new particles, leading to a situation of ``crisis'' \citep{susy_crisis}. Although the \gls{lhc} will continue to take data for years and plans for successors are being discussed \citep{Roser2023}, some have speculated that particle physics as we know it has come to an end \citep{Harlander2023,Kosyakov2023}
, ``the proscenium
[being] captured by astrophysics and cosmology'' instead \citep{Kosyakov2023}. However, according to physicist Mikhail Shifman the ``pause in accelerator programs we are witnessing now is not necessarily [\dots] the end of explorations at [high energies]''; instead, such explorations ``will continue, perhaps in a new form, with novel devices'' \citep{Shifman2020}. Indeed, new experimental opportunities have emerged in parallel, including gravitational waves astronomy (since 2015 \citep{Abbott2016}), searches for dark matter of astrophysical origin in underground facilities, and more precise observations of the cosmic microwave background (see Figure \ref{fig:experiments}, Appendix \ref{appendix:landscape}). 
Morever, astrophysics seems to be increasingly replacing particle colliders in citations across experimental and theoretical high-energy physics \citep{Gautheron2023}. 
The present paper investigates how high-energy physicists have adapted in reaction to these transformations. 

\section{\label{sec:methods}Methods}

\subsection{\label{sec:data}Data}

Our source is the Inspire HEP database \citep{InspireAPI}. It aggregates \gls{hep} literature from various sources, including the main scientific publishers and arXiv, and has been used in a few works \citep{Gautheron2023,Perovi2016,Chall2019a,Strumia2021,Sikimi2022}. For the literature on \gls{hep}, it is more comprehensive than the often used \gls{aps} dataset which is limited to a few journals\footnote{\url{https://journals.aps.org/datasets}}. 
Moreover, it implements both automatized and manual measures for the disambiguation of author names\footnote{Besides the use of ``advance algorithms'' of author-disambiguation, Inspire invites scientists to correct their own publication record on the website (\url{https://twiki.cern.ch/twiki/pub/Inspire/WebHome/INSPIRE_background.pdf}, June 2014)} 
, thus allowing careers' analyses \citep{Strumia2021} (nevertheless, occasional misidentifications remain possible). The database also contains data on experiments; consequently, the evolution of the landscape of experimental opportunities can be retrieved (see Figure \ref{fig:experiments}, Appendix \ref{appendix:landscape}). 

The analysis includes all papers from the categories ``Theory-HEP'' and ``Phenomenology-HEP'' (inspired from arXiv's categories ``hep-th'' and ``hep-ph''), to which most \gls{hep} publications belong, which amounts to $D=186,162$ articles between 2000 and 2019. The minority of purely experimental high-energy physics publications are excluded: such papers are typically authored by thousands of collaborators, and authorship data provide no information about individual experimentalists' specialization. Therefore, this paper documents how theorists and phenomenologists have adapted to the changes outline above.

For the longitudinal analysis, two time periods are considered. An initial phase (2000-2009) is used to infer a reference ``research agenda'' for each physicist in the cohort, as well as their intellectual and social capital. A late phase (2015-2019) is used to measure how each physicist's research agenda has shifted in comparison to the initial time period, in the context of the changes outlined above. The five-year gap between these two periods allows to measure the cumulative effect of the transformations in the scientific landscape that have unfolded gradually between 2010 and 2015 (had they been sudden, we would not have introduced such a wide gap -- see Figure \ref{fig:experiments}, Appendix \ref{appendix:landscape}), together with the effect of the capital accumulated \textit{prior} to these transformations. Only physicists with $\geq 5$ publications during each time period (2000 to 2009, and 2015 to 2019) are included, resulting in a cohort of $N=2\,094$ physicists. This study therefore considers physicists that have remained dedicated to high-energy physics, thus revealing adaptation and ``survival'' strategies \textit{within} \gls{hep}, excluding authors that exited the field. This author inclusion rule excludes scientists who publish very irregularly; however, although scientists who continuously publish are a minority, they make up most of the publications in their field\footnote{Less than 1\% of scientists active in the years 1996 to 2011 have published every year during this period, and yet they are responsible for 47\% of the publications \citep{Ioannidis2014}; the 13\% of physicists with $\geq 16$ publications between 1985 and 2009 account for 82\% of publications \citep{liu2024science}.}. We do not seek ``representativeness'', but rather to achieve enough variance to uncover patterns among this particular cohort of productive high-energy physicists. The median academic age of the cohort was 23 years in 2015. 49\% of these physicsts have had an affiliation that spanned the entire time period (see Appendix \ref{appendix:sample_characteristics}, Figure \ref{fig:sample_characteristics}). 

\subsection{\label{sec:topics}Measuring research portfolios}

Research portfolios are evaluated in terms of the distribution of keywords (n-grams) that belong to each research area within the scientists' publications' abstracts. Instead of reyling on \gls{pacs} categories and citation data as in most previous works, research areas are extracted with a topic model that recovers latent ``topics'' within the corpus and their vocabulary distributions, while directly classifying keywords into separate research areas. Arguably, texts provide a more direct access to the kinds of knowledge leveraged by a scientist in their publications: linguistic flexibility implies cognitive flexibility and a low commitment to a specific body of knowledge. Moreover, this approach is applicable to a wider range of situations for which only textual data (even short texts) are available. For instance, \gls{pacs} codes are only sparsely available in the recent \gls{hep} literature. A coarse-grained classification of the literature into $K_0=20$ broad ``topics'' is performed. The number of topics is always somewhat arbitrary, but that topic models give some control over the cognitive scale is a feature, rather than a bug\footnote{Previous works based on the \gls{pacs} categories have leveraged the different levels of this hierarchical classification system to investigate different scales.}: as \citet{Gieryn1978} puts it, ``in such analyses [of problem change], empirical findings will in part reflect the defined scope of problem areas'', which is itself arbitrary. In our case, we would ideally like our clustering to be just fine-grained enough to measure the impact of the shifts in the landscape of experimental opportunities that we are interested in (the start of the \gls{lhc}, the rise of new probes of dark matter and black holes, etc.). In this respect, $K_0=20$ turns out to be just sufficient to discern the effects of the transformations in \gls{hep} discussed in Section \ref{sec:hep} as well as the observed evolution in the popularity of various kinds of experiments shown in Appendix \ref{appendix:landscape}, Figure \ref{fig:experiments}. Additional models were trained for robustness assessment, setting different values for $K_0$ (15, 20, 25). More coarse-grained models (using lower values of $K_0$) are typically less able to observe fine-grained patterns of adaptation to changing experimental opportunities, and the initial $K_0=20$ model is better at distinguishing black hole phenomenology from cosmological phenomenology than the most fine-grained model (see Appendix \ref{appendix:topic_comparison}, Figure \ref{fig:topic_experiments}).

We use an embedding model \citealt{Dieng2020}, a recent and straightforward approach that relies on pretrained embeddings representations of the n-grams and provides more reliable classifications for heavy-tailed vocabulary distributions than previous models such as Latent Dirichlet Allocation \citep{Blei2003}. Given the coarseness of the clustering, Language Models were not deemed necessary.
The model is trained on $D=186,162$ abstracts published between 2000 and 2019. Tokens are extracted from the papers' titles and abstracts by filtering n-grams between one and three words matching syntactic expressions susceptible of carrying scientific information (by designating concepts, models, methods, etc.), following the procedure from \citealt{Gautheron2023,omodei_tel-01097702}. Embeddings are learned using a skip-grap model in $L=50$ dimensions (few are needed given the small size of the vocabulary, $V=4,751$; nevertheless, some analyses are re-iterated with $L=150$; see Appendix \ref{appendix:topics}, Figure \ref{fig:word2vec}) \citep{word2vec}). We obtain the topics listed in Appendix \ref{appendix:topics}, Table \ref{table:research_areas}. Four of the 20 resulting topics regroup keywords that do not clearly refer to any specific research area (e.g. ``paper'', ``approach'') and correlate poorly with the \gls{pacs} categories, which suggests their lack of scientific dimension. Consequently, the present analysis only considers the $K=15$ remaining topics that designate actual research areas (as confirmed by their strong tendency to preferably cite themselves, cf. Appendix \ref{appendix:citation_validation}, and their correlation patterns with the \gls{pacs} classification, cf. Appendix \ref{appendix:pacs_validation}). In order to enhance the robustness of the topic removal process, we made sure all retained topics had a maximal loading on the \gls{pacs} classification higher than that of all removed topics.

\begin{figure}[h]
\begin{subfigure}{.45\textwidth}
    \includegraphics[width=1.15\textwidth]{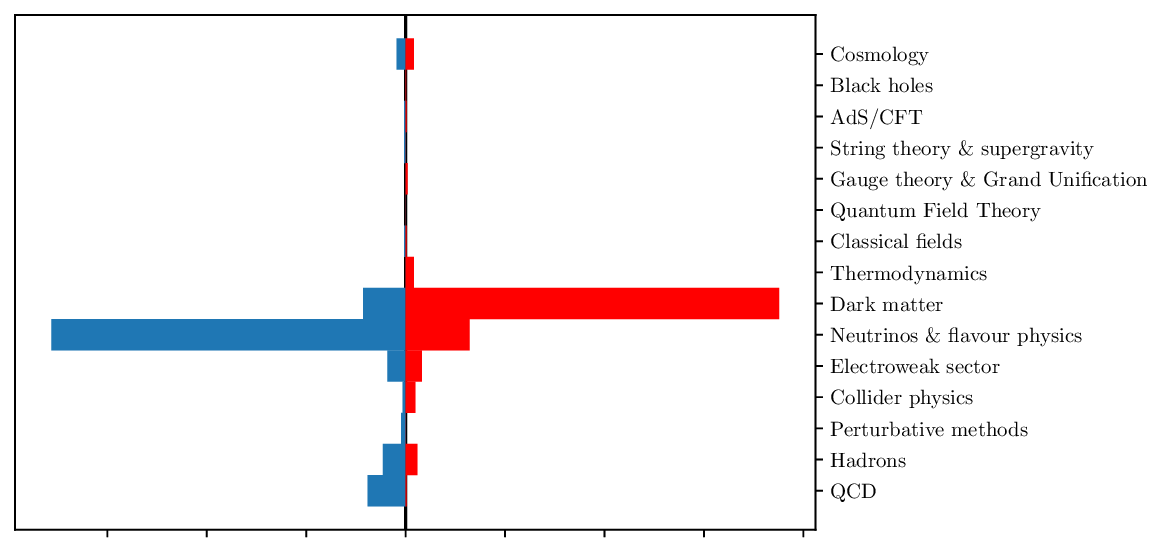}
    \caption{Physicist A. 
    }
    \label{fig:S.Ando.1}
\end{subfigure}\hfill%
\begin{subfigure}{0.45\textwidth}
    \includegraphics[width=1.15\textwidth]{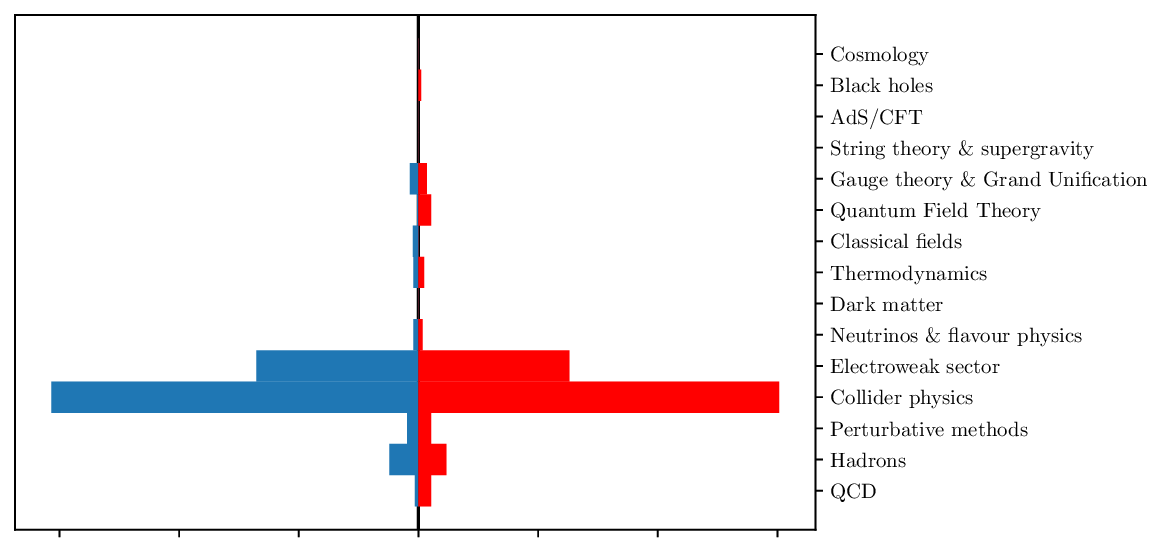}
    \caption{Physicist B. 
    }
    \label{fig:J.F.Beacom.1}
\end{subfigure}
\caption{\textbf{Initial (in blue) and late (in red) research portfolios of two physicists}. Physicist A shifted their focus from neutrinos physics to dark matter\protect\footnotemark. Physicist B, on the other hand, pursued a remarkably stable research agenda.}
\label{fig:turns_neutrinos_dm}
\end{figure}

Then, we derive $n_{dk}$, the amount of keywords in the abstract of $d$ that refer to a research area $k$ (using the method described in Appendix \ref{appendix:keywords}), and consequently $X_{a,k}$, the amount of times keywords (``resources'', i.e. concepts, models, methods, etc.) in relation to research area $k$ have occurred in papers (co-)authored by $a$ in the initial time-period (2000 to 2009). Mathematically, $X_{a,k}=\sum_{d\in [2000,2009],a\in A_d} n_{d,k}$, where $A_d$ is the set of authors of a publication $d$. The matrix $Y_{a,k}$ is derived similarly, using publications from the later time period (2015 to 2019). Research portfolios are then normalized into distributions $x_{ak} \equiv X_{ak}/\sum_{k'} X_{ak'}$ and $y_{ak} \equiv Y_{ak}/\sum_{k'} Y_{ak'}$, thus encoding how scientists divided their attention during each period. This approach ensures that research portfolios are evaluated based on the frequency of keywords that belong to each ``topic'', according to the idea illustrated in Figure \ref{fig:research-agenda}. Therefore, this approach captures variations in the prevalence of different kinds of vocabulary (and thus bodies of knowledge) exploited in scientists' publications. For purposes of illustration, Figure \ref{fig:turns_neutrinos_dm} shows the research portfolios of one physicst who migrated from neutrinos to dark matter physics, and of one physicist who maintained their research agenda over the time periods considered.

\footnotetext{Physicist A's personal website reads: ``I am working on particle astrophysics and cosmology. In particular, \textit{I am interested in dark matter problem in the Universe, and how to probe it using annihilation products such as energetic gamma rays and neutrinos}. [\dots]  I \textit{started my research career by studying supernova neutrinos from various aspects} [our emphases].'' (\url{https://staff.fnwi.uva.nl/s.ando/eng/Research.html}). This is therefore an instance of ``conversion'' of prior knowledge to new purposes, one of the forms of change drawn from historical institutionalism represented in Figure \ref{fig:research-agenda}.}

\subsection{\label{sec:capital}Measuring capital}

As shown by \citet{Schirone2023} in an extensive review of references to Bourdieu in bibliometrics, most mentions of capital focus on symbolic and social capital. Only a dozen works considered cultural/intellectual capital, and none of those proposed a measure that adequately captured the distribution of capital across epistemic domains. Therefore, an alternative unified approach for measuring the distribution of intellectual and social capital is proposed below.

\subsubsection{Intellectual capital}

Intellectual capital is represented by a vector $\bm{I_a}=(I_{ak})$ that measures the concentration of $a$'s intellectual resources in each domain $k \in \{1,\dots,K\}$. It is constructed in a way similar to $\bm{x_{a}}$, summing the contribution of keywords dedicated to each research area in the publications of each author between 2000 and 2009 (thus excluding publications that belong to the outcome research portfolio $\bm{y_a}$), except that publications are now weighed differently depending on the amount of authors. Indeed, publications with fewer co-authors convey more information about each author's own expertise. The weight is $1 \over |A_d|$, where $|A_d|$ is the amount of authors of publication $d$\footnote{A justification for this weight is that the probability that a given author has been responsible for introducing any particular concept or method present in the paper is $\mathcal{O}(1/|A_d|)$. }:

\begin{equation}
    I_{ak} \propto  \sum_{d\in [2000,2009],a\in A_d} \dfrac{n_{d,k}}{|A_d|}
\end{equation}

$\bm{I_a}$ is normalized, such that $\sum_k I_{ak}=1$; therefore, $\bm{I_a}$ only captures the ways scientists divide their cognitive resources between research areas, rather than the ``absolute magnitude'' of their knowledge of each area (by contrast, the measure of  semantic capital proposed in \citealt{Roth2010} measures total knowledge but cannot capture diversity). 

\subsubsection{Social capital}

Many measures of scientists' social capital have been proposed \citep{Abbasi2014,Schirone2023}, the simplest being the amount of collaborators of a scientist (i.e. degree centrality in the co-authorship network \citealt{Roth2010}). Other measures revolve around \textit{betweeness} centrality, which captures the extent to which an actor ``bridges'' a network (e.g. ``brokerage'', i.e., the ability of an individual to overcome ``structural holes'' in a social network \citep{BurtBrokerage2007}). \citet{Abbasi2014} distinguish two general approaches to social capital, depending on whether the emphasis is placed on ``power'' versus ``diversity''. Measures of social capital (as those discussed in \citealt{Abbasi2014}) typically represent social capital by single scalars; however, social capital has multiple dimensions. In fact, according to \citet{Bourdieu1980}, ``the volume of social capital possessed by a particular agent [\dots] depends on the extent of the network of links that he can effectively mobilize, and on the volume of capital (economic, cultural or symbolic) possessed by each of those to whom he is linked''. In that respect, social capital can come in different forms depending on the resources being leveraged via one's network. 
In the following, we focus on the intellectual dimension of social capital, which we represent by a vector $\bf{S}_{a}$  
defined as the sum of the intellectual capital of $a$'s collaborators, weighted by the strength of their relationship:

\begin{equation}
    \bm{S}_{a} \equiv \sum_{c \in \text{co-authors}(a)} w_{ac} \bm{I}_{c\backslash a}
\end{equation}

$\bm{I}_{c\backslash a}$ is the intellectual capital of $c$, evaluated by excluding papers co-authored with $a$ (in order to disentangle the effect of an author's own knowledge and that available to them via their collaborators). Collaborators outside the cohort are taken into account. The weight $w_{ac}$, which represents the strength of the relationship between $a$ and $c$, is defined as:

\begin{equation}
    w_{ac} \equiv \max_{d|\{a,c\} \subset A_d} \frac{1}{|A_d|-1}
    \label{eq:weighing_scheme}
\end{equation}

Where $A_d$ is the set of the co-authors of a paper $d$. This weighing scheme -- inspired from \citealt{Newman2004} -- captures the fact that a paper with, say, two co-authors, signals a stronger relationship between the authors than a publication with a dozen authors.\footnote{Assuming that in a collaboration, each author interacts with a constant amount of co-authors in practice (regardless of the total amount of co-authors), then the probability that they had interactions with one specific co-author in particular is $\propto \frac{1}{|A_d|-1}$. 
} However, it does not take into account the recency and frequency of collaborations.

\subsubsection{Diversity and power}

Measures of ``diversity'' (and ``power'') can be readily derived from $\bm{I_a}$ (and $\bm{S_a}$). A common measure of diversity is the Shannon entropy $H$ \citep{Jost2006}. Let 
 $D(\bm{I_a})=\exp{H(\bm{I_a})}$ be the diversity of intellectual capital (and $D(\bm{S_a})=\exp{H(\bm{S_a})}$ that of social capital). Roughly speaking, these are measures of how many research areas scientists have
divided their cognitive or social resources among. In the cohort, individuals typically have cognitive resources in several research areas ($\mu_{D(\bm{I_a})}=5.6,\sigma=2.0$), and social capital is even more diverse ($\mu_{D(\bm{S_a})}=7.8,\sigma=2.1$). In fact, intuitively,  scientific collaborations enable individuals to take advantage of their group's diversity. Furthermore, $D(\bm{I_a})$ and $D(\bm{S_a})$ are highly correlated ($R=0.75$); indeed, individuals with more diverse expertise are more able to engage with diverse collaborators. Since the diversity of social capital is mostly expected to enhance individuals' abilities when it exceeds that of their own knowledge, from now on, only \textit{excess} social capital diversity $D^{\ast}(\bm{S_a})$ (defined as the residuals of the linear regression of $D(\bm{S_a})$ against $D(\bm{I_a})$, by ordinary least squares) is considered\footnote{This approach aims to address the difficulty of determining the direction of the causal relationship between social resources and research interests raised by \citet{Tripodi2020}.}.

The ``power'' dimension of social capital is evaluated as the \textit{magnitude} of social capital:

\begin{equation}
    P(\bm{S_a})\equiv \sum_k S_{ak} = \sum_{c \in \text{co-authors}(a)} w_{ac}
\end{equation}

``Power'' is therefore the amount of collaborators weighed by the strength of each relationship. Our measures of diversity and power depart from \citealt{Abbasi2014}, which conflates diversity with network size and power with performance. By combining semantic and authorship data, our approach assesses diversity more directly. 

Alternative measures of diversity and power are considered for robustness assessment. The alternative measure of diversity is based on Stirling's index, and the alternative measure of power uses the notion of brokerage. All these measures are defined and compared in Appendix \ref{appendix:capital_validation}. 

\subsection{\label{sec:model}Modelling trajectories}

The model for the late research portfolio $\bm{Y_a}$ is schematically illustrated in Figures \ref{fig:model_schema_a},\ref{fig:model_schema_b},\ref{fig:model_schema_c}, and a more formal representation is given in Figure \ref{fig:model_structure}. It captures the idea expressed in Figure \ref{fig:research-agenda} that research portfolios are transformed via strategic transfers of knowledge and attention from one research area to another. Occurrences of keywords that belong to each research area $k$ in papers by $a$ in the late time period, $\bm{Y_{a}} \in \mathbb{N}^K$, are assumed to be drawn from a hierarchical multinomial logistic model. $\bm{Y_{a}}$ results from a linear combination of the initial research portfolio, $\bm{x_{a}}$, and a mixing matrix $\theta_a$ that measures the fraction of attention redistributed from each research area to another. $\theta_a$ is drawn from a hierarchical process, thus capturing the ``average'' cohort behavior. 
Formally speaking, $\bm{Y_{a}}$ is assumed to derive from a multinomial process involving linear combinations of $(x_{ak})$:

\begin{equation}
    \bm{Y_a} \sim \text{multinomial}(\sum_{k=1}^{K} x_{ak}\theta_{ak1} ,\dots,\sum_{k=1}^{K}x_{ak}\theta_{akK})
\end{equation}

Where $\theta_{akk'}$ is the fraction of attention to a topic $k$ by $a$ that has been redirected to a topic $k'$. 
$\theta$ is a function of intellectual capital $\bm{I_a}$ and social capital $\bm{S_a}$ according to the following generalized linear model:

\begin{equation}
    \bm{\theta_{ak}} = \text{softmax}\left(\beta_{ak1} + \gamma_{k1} I_{a1} + \delta_{k1} S_{a1}, \dots,\beta_{akK} + \gamma_{kK} I_{aK} + \delta_{kK} S_{aK}\right)
    \label{eq:glm}
\end{equation}

$\delta_{kk'}$ is the effect of the scientists' social capital in a research area $k$ on the magnitude of transfers from $k$ to $k'$. Similarly, $\gamma_{kk'}$ is the effect of having more expertise in $k'$ (intellectual capital) on shifts from $k$ to $k'$. High values of the diagonal elements of $\gamma$ would imply that physicists are more conservative towards research areas in which they concentrate more expertise. The coefficients $\beta_{akk'}$ encode the average behavior of the cohort plus individual deviations to the average behavior that are unexplained by the covariates\footnote{The priors for this hierarchical model are:

\begin{align}
    \beta_{akk'} &\sim \mathcal{N}(\mu_{k k'},\sigma_{k k'}) \text{ for } 1\leq k' \leq K-1 \text{ and } \beta_{ak K} = \mu_{kK}\label{eq:beta}\\
    \mu_{k k'} &\sim \mathcal{N}(\lambda \times  \nu_{kk'},1) \text{ (average behavior)}\\
    \delta_{kk'} &\sim \mathcal{N}(\delta_0+\lambda ' \times \nu_{kk'},1) \text { (effect of social capital)}\label{eq:social_capital_nu}\\
    \gamma_{kk'} &\sim \mathcal{N}(0,1)  \text { (effect of intellectual capital)}\\
    \sigma &\sim \mathrm{Exponential}(1)
\end{align}

Where $\nu_{kk'}$ is the fraction of physicists with expertise in $k$ (that is, with more intellectual capital than average in $k$) who also have expertise in $k'$. 
Priors must be thought thoroughly, as certain invariances can lead to identification issues -- for instance, shifting $\mu$ by a constant does not change the likelihood.
}\textsuperscript{,}\footnote{The fit is performed with Stan's Hamiltonian Monte-Carlo sampler (HMC is better behaved than Gibbs for such problems).
}. 
The ability of the model to predict individual trajectories is assessed in Appendix \ref{appendix:model-performance}, for various temporal segmentations of the initial and late research portfolios and topic models. The model is better at predicting individual trajectories for larger adaptive responses and longer time-scales.  

Our model is strongly connected to Optimal Transport \citep{muzellec2017tsallis,li2019learning}, which seeks optimal ways to  ``transport'' an input distribution (say, $\bm{x}=(x_k)$) to a target distribution (e.g., $\bm{y}=(y_{k'}$)) through ``transfers'' $(\theta_{kk'})$ across their components while minimizing a cost function $\sum_{k,k'} x_{k} \theta_{kk'}C_{kk'}$ (where $C_{kk'}$ is a cost matrix) \citep{Peyr2019}. The difference is that the proposed Bayesian approach estimates transfers $\theta_{kk'}$ by minimizing a likelihood rather than a cost function. However, the connection with Optimal Transport suggests an economic interpretation of the reallocation of research efforts, which will be leveraged in Section \ref{sec:optimal-transport} to show that patterns of change in research interests are shaped by learning costs. 

    \tikzstyle{block} = [draw,minimum size=2em, minimum width=3em]
    \tikzstyle{block1} = [block,fill=blue!20]
    \tikzstyle{block2} = [block,fill=red!20]
    \tikzstyle{block3} = [block,fill=green!20]
    \tikzstyle{block4} = [block,fill=cyan!20]
    \tikzstyle{block5} = [block,fill=gray!20]
    \tikzstyle{block6} = [block,fill=yellow!20]
    
    \tikzstyle{pre} = []
    \tikzstyle{post} = []
    \tikzstyle{traj} = []
    \tikzstyle{pre2} = []
    \tikzstyle{post2} = []
    \tikzstyle{traj2} = []
    \tikzstyle{pre3} = []
    \tikzstyle{post3} = []
    \tikzstyle{traj3} = []
 \begin{figure}[h]
 \centering
     \begin{subfigure}[b]{0.25\textwidth}
          \centering
          \resizebox{0.8\linewidth}{!}{
          
\begin{tikzpicture}[>=latex',baseline={(0,0)}]

    \foreach \y in {1,2,3} {
        \node[block\y,pre] at (2,-\y) (input\y) {$X_{a,\y}$};
        \node[block\y,post] at (6,-\y) (output\y) {$Y_{a,\y}$};;
    }
    \node[block5,pre] at (2,-4) (input5) {\dots};
    \node[block5,post] at (6,-4) (output5) {\dots};
    \node[block6,pre] at (2,-5) (input6) {$X_{a,K}$};
    \node[block6,post] at (6,-5) (output6) {$Y_{a,K}$};
        \foreach \x in {1,2,3} {
        \draw[->,traj] (input1.east) -- (output\x.west) node[midway,above=-0.2em,sloped] {$\theta_{a,1,\x}$};
    }
    \draw[->,traj] (input1.east) -- (output5.west) node[midway,above=-0.2em,sloped] {$\theta_{a,1,\dots}$};
    \draw[->,traj] (input1.east) -- (output6.west) node[midway,above=-0.2em,sloped] {$\theta_{a,1,K}$};



    \tikzstyle{s}=[shift={(0mm,\radius)}]
\end{tikzpicture}

          }  
          \caption{\label{fig:model_schema_a}}
     \end{subfigure}
     \begin{subfigure}[b]{0.25\textwidth}
          \centering
          \resizebox{0.8\linewidth}{!}{
          
\begin{tikzpicture}[>=latex',baseline={(0,0)}]

    \foreach \y in {1,2,3} {
        \node[block\y,pre2] at (2,-\y) (input\y) {$X_{a,\y}$};
        \node[block\y,post2] at (6,-\y) (output\y) {$Y_{a,\y}$};;
    }
    \node[block5,pre2] at (2,-4) (input5) {\dots};
    \node[block5,post2] at (6,-4) (output5) {\dots};
    \node[block6,pre2] at (2,-5) (input6) {$X_{a,K}$};
    \node[block6,post2] at (6,-5) (output6) {$Y_{a,K}$};

    \foreach \x in {1,2,3} {
        \draw[->,traj2] (input2.east) -- (output\x.west) node[midway,above=-0.2em,sloped] {$\theta_{a,2,\x}$};
    }
    \draw[->,traj2] (input2.east) -- (output5.west) node[midway,above=-0.2em,sloped] {$\theta_{a,2,\dots}$};
    \draw[->,traj2] (input2.east) -- (output6.west) node[midway,above=-0.2em,sloped] {$\theta_{a,2,K}$};

    \tikzstyle{s}=[shift={(0mm,\radius)}]
\end{tikzpicture}
          }  
          \caption{\label{fig:model_schema_b}}
     \end{subfigure}
     \begin{subfigure}[b]{0.25\textwidth}
          \centering
          \resizebox{0.8\linewidth}{!}{
          
\begin{tikzpicture}[>=latex',baseline={(0,0)}]

    \foreach \y in {1,2,3} {
        \node[block\y,pre3] at (2,-\y) (input\y) {$X_{a,\y}$};
        \node[block\y,post3] at (6,-\y) (output\y) {$Y_{a,\y}$};;
    }
    \node[block5,pre3] at (2,-4) (input5) {\dots};
    \node[block5,post3] at (6,-4) (output5) {\dots};
    \node[block6,pre3] at (2,-5) (input6) {$X_{a,K}$};
    \node[block6,post3] at (6,-5) (output6) {$Y_{a,K}$};
    
    \foreach \x in {1,2,3} {
        \draw[->,traj3] (input3.east) -- (output\x.west) node[midway,above=-0.2em,sloped] {$\theta_{a,3,\x}$};
    }
    \draw[->,traj3] (input3.east) -- (output5.west) node[midway,above=-0.2em,sloped] {$\theta_{a,3,\dots}$};
    \draw[->,traj3] (input3.east) -- (output6.west) node[midway,above=-0.2em,sloped] {$\theta_{a,3,K}$};

    \tikzstyle{s}=[shift={(0mm,\radius)}]
\end{tikzpicture}
          }  
          \caption{\label{fig:model_schema_c}}
     \end{subfigure}\hfill%
\begin{subfigure}[b]{0.22\textwidth}
    \centering
    \begin{tikzpicture}[
    node distance=1.45cm,
    obs/.style={circle,draw,fill=gray!20,inner sep=1pt,minimum size=24pt},
    latent/.style={circle,draw,color=black,inner sep=1pt, minimum size=24pt}
    ]
        \node [obs](X)  {$\bm{x_a}$};
        \node [obs](Y) [right of=X] {$\bm{Y_a}$};
        \node [latent](theta) [above of=Y] {$\theta_a$};
        \node [latent](beta) [above of=theta] {$\beta_a$};
        \node [latent](mu) [above of=beta] {$\mu,\Sigma$};
        \node [obs](I) [left of=theta,yshift=-0.5cm] {$\bm{I_a}$};
        \node [obs](S) [left of=theta,yshift=+0.5cm] {$\bm{S_a}$};
        \draw [->] (X) edge (Y);
        \draw [->] (theta) edge (Y);
        \draw [->] (beta) edge (theta);
        \draw [->] (mu) edge (beta);
        \draw [->] (I) edge (theta);
        \draw [->] (S) edge (theta);

\end{tikzpicture}
    \caption{\label{fig:model_structure}}
\end{subfigure}

     \caption{\textbf{\ref{fig:model_schema_a},\ref{fig:model_schema_b},\ref{fig:model_schema_c}: Transfers of attention across research areas.} $\bm{x_a}$ and $\bm{y_a}$ are the distributions scientist $a$'s attention across research areas in two consecutive time periods. $\bm{\theta_{ak}}=(\theta_{akk'})$ represents the fraction of the attention devoted by an author $a$ to a topic $k$ redirected to topics $k' \in \{1,\dots,K\}$, as scientists repurpose, expand or concentrate their knowledge. By definition, $\sum_{k'} \theta_{akk'}=1$. 
     \\
     \textbf{Figure \ref{fig:model_structure}: Hierarchical model}. $\theta_a$ is drawn from a hierarchical process, with intellectual and social capital ($\bm{I_a}$ and $\bm{S_a}$) as covariates. Observed variables are represented in gray, latent variables in white.}
     \label{fig:ei}
 \end{figure}
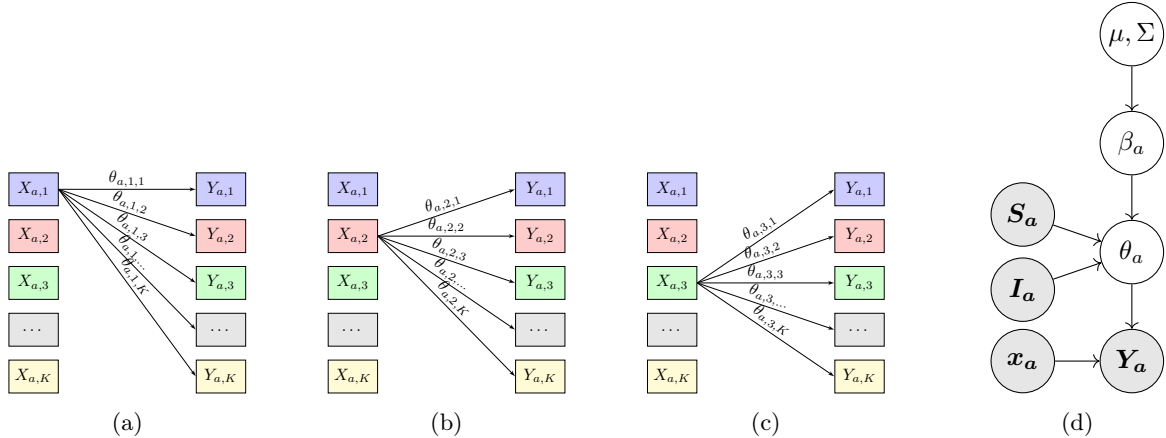

 \section{\label{sec:results}Results}

\begin{figure}[h]
     \centering
\begin{subfigure}[t]{.47\textwidth}
\includegraphics[width=1.05\textwidth,trim={2.1cm 1.5cm 2.1cm 1.5cm},clip]{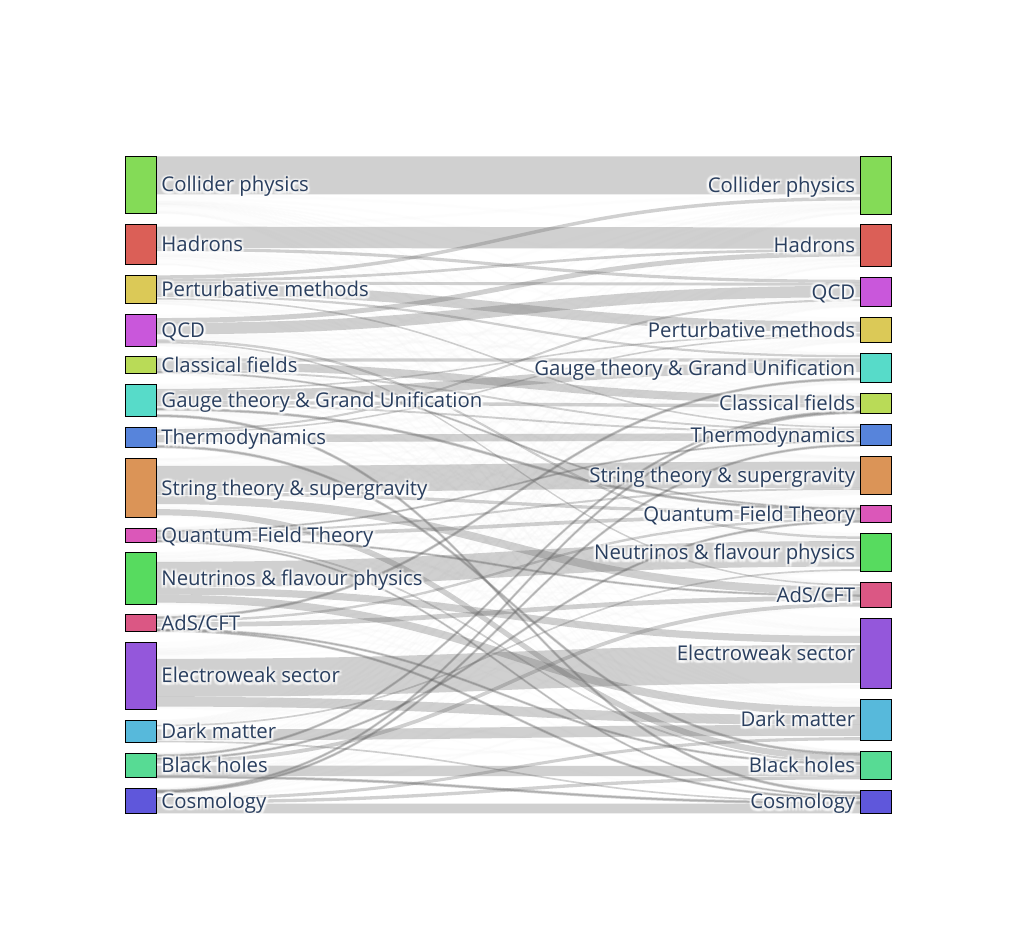}
    \caption{\label{fig:sankey_a}Transfers across all research areas.}
\end{subfigure}\hfill%
\begin{subfigure}[t]{0.47\textwidth}
\includegraphics[width=1.05\textwidth,trim={2.1cm 1.5cm 2.1cm 1.5cm},clip]{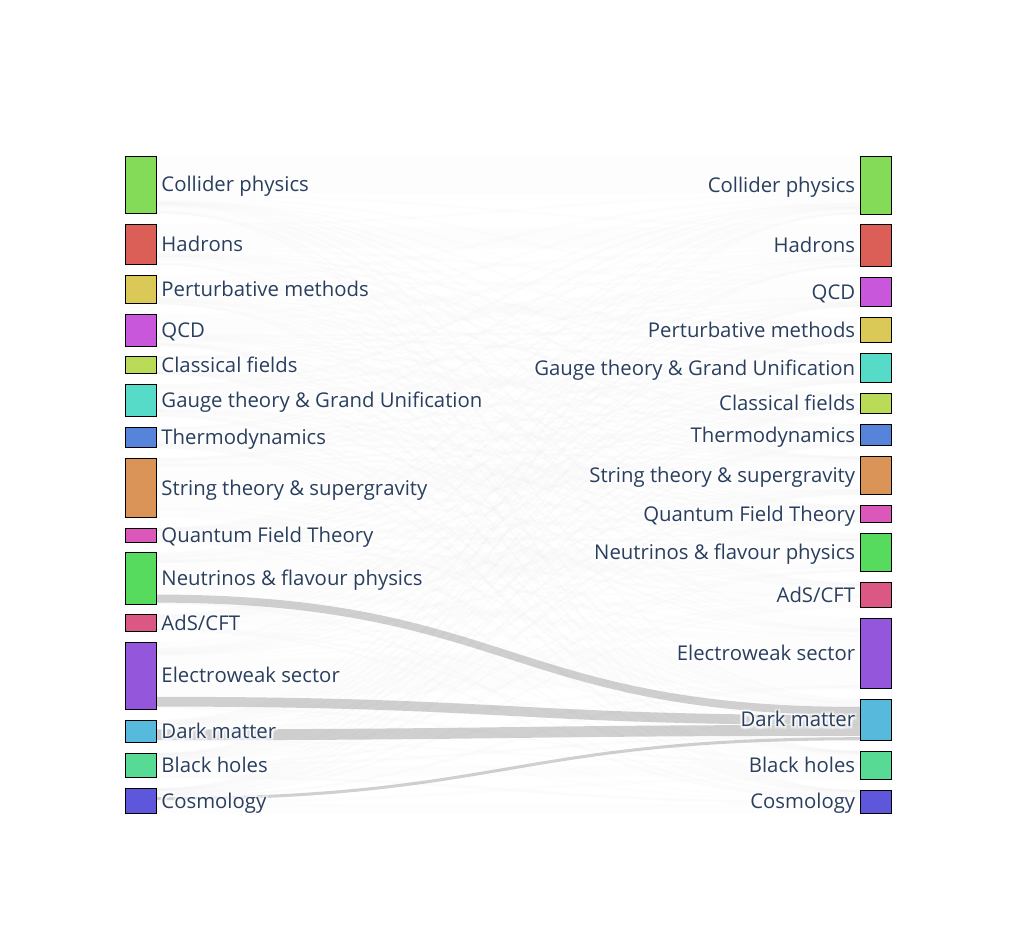}
    \caption{\label{fig:sankey_b}Flows of attention directed towards ``dark matter''.}
\end{subfigure}
     \caption{\textbf{(a) Aggregate transfers of attention across research areas, between 2000-2009 (to the left) and 2015-2019 (to the right)}. Widths of flows are proportional to $\sum_a X_{ak}\theta_{akk'}$. Insignificant transfers (that happen less than expected by chance alone assuming uniform mixing) are transparent. (b) For purposes of clarity, the same figure is repeated to the right, highlighting only the flows directed towards ``Dark matter''.}
     \label{fig:sankey}
 \end{figure}

Figure \ref{fig:sankey_a} shows the aggregate transfers of attention at the level of the cohort revealed by the model. 
The most obvious feature is the remarkable stability of most research areas: indeed, physicists' conservatism toward their research area due to specialization is known to be particularly high in \gls{hep} \citep{Aleta2019}; late research portfolios are largely constrained by prior research interests: they exhibit \textit{path dependence} \citep{Galesic2023} (using Inverse Optimal Transport, Section \ref{sec:optimal-transport} shows that these patterns are structured by learning costs). Conservatism seems especially prevalent in the case of ``collider physics'', a research area dedicated to knowledge specific to particle accelerators. Nevertheless, ``dark matter'' has doubled, fueled by a shift away from ``neutrinos and flavor physics'', and ``electroweak sector'', a phenomenal domain studied at the \gls{lhc} (Figure \ref{fig:sankey_b})\footnote{The electroweak notably includes Higgs physics, which are very prominent at the \gls{lhc}, where the Higgs boson was discovered.}\footnote{The migration of many particle physicists towards dark matter provides an explanation for the persisting schism between two research programs in fundamental physics, namely dark matter particle research and modified gravity. Both research programs seek to explain 
a shared set of anomalies in astronomical observations, and yet their communities communicate very little \citep{Martens2023,martens2022integrating}. Our approach suggests that particle physicists' interest in dark matter is in great part motivated by the fact this is a natural extension of their previous research; particle physicists would therefore not consider the alternative to dark matter (modified gravity), given this topic that would make little use of their expertise.}. This confirms that the cohort has responded to changes in the landscape of experimental opportunities. Moreover, ``string theory and supergravity'' has declined in favor of ``AdS/CFT'' (a research program that explores dualities between theories of quantum gravity and certain types of theories of quantum field theory) and ``black holes'' \footnote{This converges with physicist Peter Woit's controversial assessment that ``string theorists'' are no longer doing string theory per se, though they keep identifying themselves as string theorists. As Peter Woit puts it, citing the 2022 ``Strings'' conference: ``one thing that stands out is that the string theory community has almost completely stopped doing string theory.''; and, ``[presentations' titles] make very clear what the string theory community has found to replace string theory: black holes'' (Woit, 2022, \url{https://www.math.columbia.edu/~woit/wordpress/?p=12981});}. Of course, this rough description must be considered with caution, as interpreting clusters from topic models is notoriously hard, and these topics in particular can regroup quite heterogeneous research programs.

\begin{figure}[!h]
    \centering
    \includegraphics[width=0.8\textwidth,trim={2.1cm 1.5cm 2.1cm 1.5cm},clip]{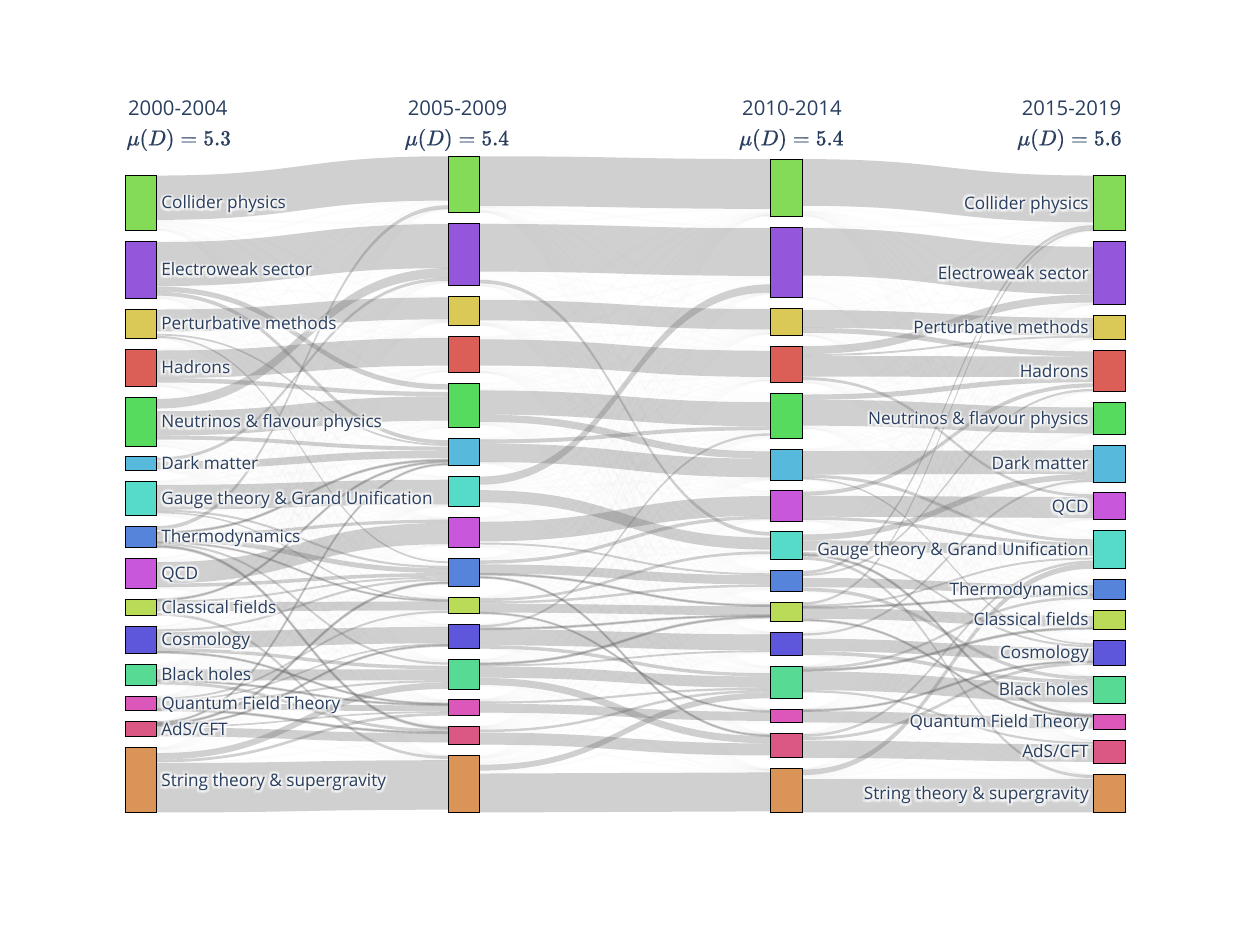}
    \caption{\textbf{Aggregate transfers of attention throughout the years 2000 to 2019, considering four time-intervals of five years each}. $\mu(D)$ is the average portfolio diversity of the cohort during each time-period.}
    \label{fig:sankey_five_years}
\end{figure}

While the paper focuses on two time periods (2000-2009 and 2015-2019), multiple alternative temporal segmentations can be considered. Figure \ref{fig:sankey_five_years} shows the transfers of research attention of a cohort of physicists across four time-periods of five years each. It reveals that the changes outlined above have unfolded rather gradually. The average diversity of physicists' research portfolio ($\mu(D)$, the average of the exponentiated entropy of $\bm{x}$) during each five-year time-bin is also shown. It has gradually increased over the years ($P<10^{-4}$): on average, physicists have \textit{expanded} their portfolio. Interestingly, the average linguistic diversity of each individual paper increased as well (with a confidence level $P<10^{-4}$) from 2.81 topics per paper (2000-2004) to 2.96 (2015-2019). This means physicists diversified their research portfolios in part by diversifying the knowledge leveraged \textit{within} each of their individual papers (rather than solely by writing multiple papers on separate issues).

\subsection{\label{sec:optimal-transport}The structuring effect of learning costs on scientists' behavior}

As research priorities change, research efforts must be efficiently reallocated among scientists, given their prior knowledge. Here, we leverage the similarity between our model of scientists' trajectories and \gls{ot} \citep{muzellec2017tsallis,li2019learning} to formulate the observed behavior in economic terms and demonstrate that the shifts of attention are structured by the minimization of ``learning costs'', thus providing a first-order explanation of the aggregate patterns of change in high energy physicists' research interests. Optimal Transport is a mathematical framework first introduced by Gaspard Monge for finding optimal ways of displacing piles of sand in a military context \citep{monge1781memoire}, and later refined by Leonid Kantorovitch in the context of economic planning \citep{kantorovich2006translocation}.

Let $\bm{x}=\sum_a \bm{x_a}$ be the cohort's initial distribution of attention across research areas (summing over each author $a$), and $\bm{y}=\sum_a \bm{y_a}$ the late distribution. Let us further assume that $\bm{x}$ and $\bm{y}$ can be considered ``fixed'' by the institutions (scientific leadership, laboratories, funding agencies, etc.) that define scientific priorities throughout time. In order to achieve the distribution of research efforts $\bm{y}$ given the previous distribution $\bm{x}$, some scientists must redirect their attention away from certain research areas (for which $y_k<x_k$) and towards more pressing ones (for which $y_k>x_k$). What is the most efficient way to reallocate research efforts and achieve the transition from $\bm{x}$ to $\bm{y}$? Intuitively, research areas should be assigned to scientists in a way that requires as few of them as possible to acquire new knowledge -- in other words, in a way that minimizes learning costs, given the way knowledge is distributed among individuals. This, we show, can be framed as an Optimal Transport problem. Let $T_{kk'}$ be the coupling matrix that encodes how much attention has been shifted from one research area $k$ to a research area $k'$. There are two constraints on $T_{kk'}$: $\sum_{k'} T_{kk'} = x_k$, and $\sum_{k} T_{kk'} = y_{k'}$. These constraints encode the need to adapt (since $\bm{y}\neq \bm{x}$). But $T$ must also minimize learning costs ($C$), which we assume to be linear in $T$ for simplicity\footnote{Roughly speaking, this linear assumption entails that scientists from a given research area are equally able to shift attention to another research area. In practice, some scientists have more abilities to switch to a given research area, and the cost will increase non-linearly as more and more scientists are required to make the transition, including those less able to make the switch.}: $C=\sum_{k,k'}T_{kk'}C_{kk'}$, where $C_{kk'}$ is a cost matrix. The problem of finding the couplings $T_{kk'}$ that minimize the ``transportation'' costs (given a cost matrix $C_{kk'}$ and the constraints on $T_{kk'}$) is an instance of Optimal Transport problem \citep{Peyr2019}. Typical instances of \gls{ot} problems include how to efficiently transport (say, ore from mines to factories) or the optimal assignment of workers to firms \citep{galichon2018optimal}. In our case, the couplings are known (they were recovered by the model from Section \ref{sec:model}), and we want to infer the underlying cost matrix that these couplings minimize. The transfers from a research area $k$ to $k'$ for each individual are simply $x_{ak}\theta_{akk'}$. Summing over individuals yields the coupling matrix $T_{kk'}=\sum_a x_{ak} \theta_{akk'}$, which measures how much attention was shifted away from $k$ and toward $k'$ at the cohort level.

The problem of recovering the cost matrix $C_{kk'}$ given the couplings is an Inverse Optimal Transport problem. Below, this problem is solved using the probabilistic method proposed in \citealt{pmlr-v162-chiu22b}\footnote{It should be noted that the approach by \citet{pmlr-v162-chiu22b} does not entail the assumption that scientists' behavior is perfectly minimizing the cost matrix. Indeed, the optimization problem they consider includes an entropic regularization term; while this term is often introduced for numerical reasons, in the case of human behavior, it can be taken to represent inefficiencies and random deviations from the optimum behavior \citep{dupuy2014personality}.}. 
This method requires to put a prior on $P(C_{kk'})$ -- indeed, infinitely many cost matrices yield the same optimum, and priors are needed to lift this degeneracy. Following \citealt{pmlr-v162-chiu22b}, we consider a prior such that $\sum_{k,k'}C_{kk'}=\mathrm{cst}$. We assume that $\mathbb{E}(C_{kk'})\propto \mathrm{softmax}(\beta \times (1-\nu_{kk'}))$\footnote{This prior has the merit of simplicity -- it is a simple generalized linear model with the desired support.}, where $\nu_{kk'}$ is the fraction of physicists who already held expertise in $k'$ among those who already held expertise in $k$, and $\beta\sim\mathcal{N}(0,1)$ is the effect of learning costs on $C_{kk
}$. If $\nu_{kk'}\sim 1$ (i.e. $1-\nu_{kk'}\sim 0$), then shifting attention from $k$ to $k'$ does not entail the acquisition of additional knowledge, and $C_{kk'}$ should be almost zero. If $\nu_{kk'}=0$ (i.e. $1-\nu_{kk'}= 1$), any scientist shifting from $k$ toward $k'$ must acquire new knowledge, and the cost should be maximal. If actual behaviors \textit{do} involve the minimization of learning costs, we should observe a strong correlation between $C_{kk'}$ and the ``knowledge gap'' ($1-\nu_{kk'}$).

\begin{figure}[!h]
\begin{subfigure}{.475\textwidth}
    \centering
    \includegraphics[height=1\textwidth]{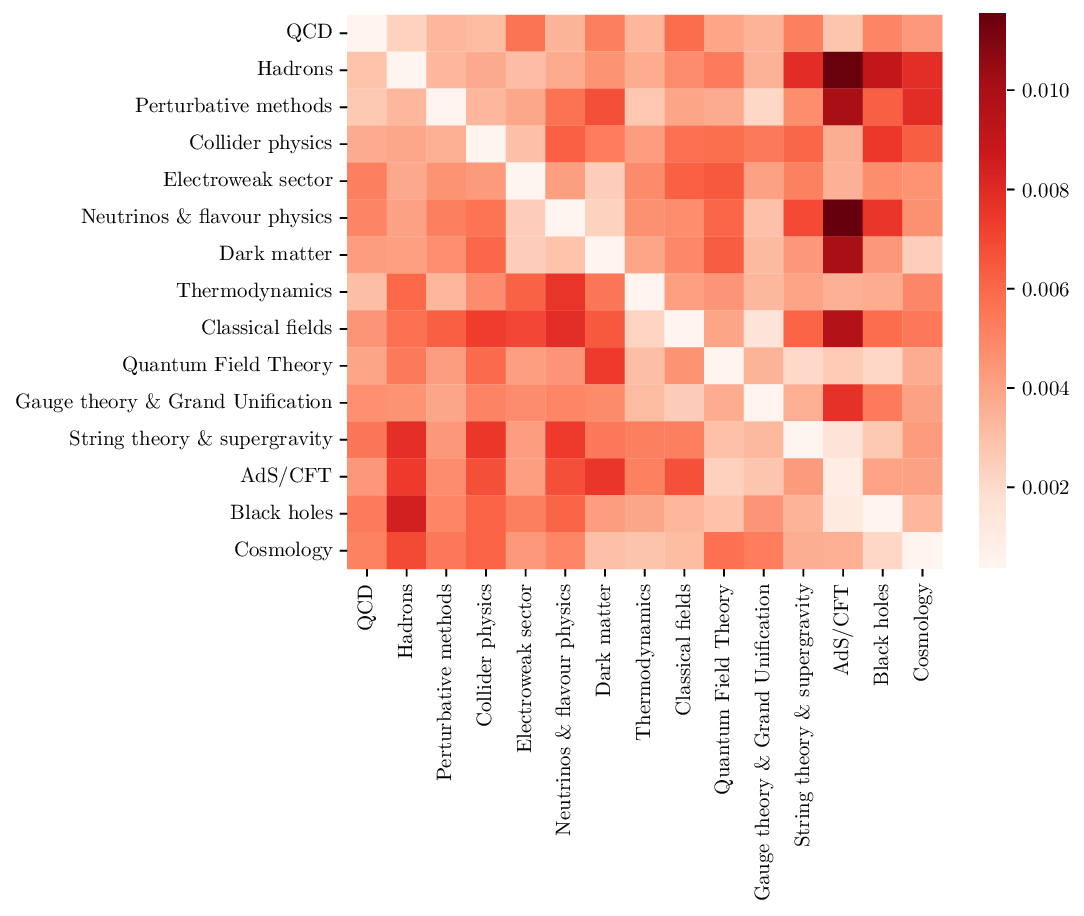}
    \caption{Cost matrix $C_{kk'}$. The cost matrix represents the cost of shifting a unit of cohort's research attention from $k$ to $k'$.}
    \label{fig:cost_matrix}
\end{subfigure}\hfill%
\begin{subfigure}{0.475\textwidth}
    \centering
    \includegraphics[height=1\textwidth]{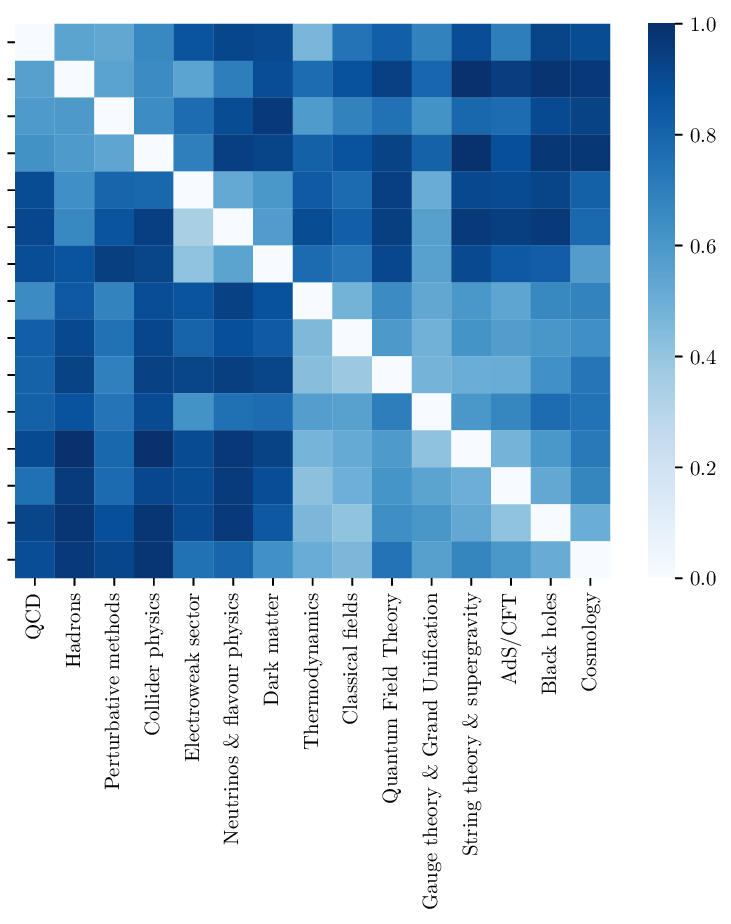}
    \caption{Knowledge gap matrix $1-\nu_{kk'}$, where $\nu_{kk'}$ is the fraction of physicists with expertise in the horizontal research area ($k$) who also hold expertise in the vertical research area ($k'$).}
    \label{fig:nu}
\end{subfigure}
\caption{\textbf{Learning costs and knowledge gap.} The cost of shifting attention from one area to another (left plot) is in great part determined by how frequently individuals hold knowledge in both areas (right plot).}
\label{fig:knowledge}
\end{figure}

Using the ``MetroMC'' algorithm proposed in \citealt{pmlr-v162-chiu22b}, we empirically recover $C_{kk'}$, the underlying cost matrix, as shown in Figure \ref{fig:cost_matrix}. We find a strong correlation with the ``knowledge gap'' (shown in Figure \ref{fig:knowledge}): $R(C_{kk'},1-\nu_{kk'})=0.76$, such that $R^2=0.58$ (Figure \ref{fig:cost_knowledge}, see also Appendix \ref{appendix:optimal_transport}). Using a finer-grained temporal segmentation (2000-2004$\to$2005-2009), the resulting correlation is similar ($R^2=0.62$). The empirical cost of shifting research efforts from one research area to another is therefore shaped by learning costs. The derivation of $C_{kk'}$ is potentially useful: one could in principle predict aggregate transfers of attention given a counterfactual target distribution of research efforts $\bm{y}$ using optimal transport and plugging-in $C_{kk'}$ as the cost matrix \citep{li2019learning}. 

Under changing circumstances, research efforts must be reallocated efficiently. Scientific norms and institutions must address an Optimal Transport problem by providing incentives for scientists to conform to new research imperatives, in a way that factors ``learning costs''. Consequently, shifts between research areas which entail the acquisition of new knowledge (layering) must be less likely that those which can take advantage of prior knowledge (conversion). In the case of \gls{hep}, it does seem that adaptative patterns are structured by this \gls{ot} problem. Interestingly, the matrices in Figure \ref{fig:cost_matrix} and \ref{fig:nu} feature blocks (indicative of an underlying hierarchical structure), such that it is easier for scientists to migrate within than across these blocks. While these observation characterizes the cohort's behavior, drivers of differences among individuals are considered next.

\subsection{\label{sec:comparative}Individual behavior and the effect of capital}

\subsubsection{Magnitude of change and capital}

We propose a change score $c_a$ that measures how much the research agenda of a scientist has changed between the two time periods under consideration, defined as the total variation distance between their initial and late research portfolios:

\begin{equation}
    c_a \equiv \mathrm{d}_{\text{TV}}(\bm{x_a},\bm{y_a}) = \frac{1}{2} \sum_k |y_{ak}-x_{ak}|
\end{equation}

This measure of change is naturally motivated by Optimal Transport: it is the minimal cost of transporting $\bm{x_a}$ to $\bm{y_a}$ if the cost matrix has zeros on the diagonal and ones everywhere else ($C_{kk}=0$ and $C_{kk'}=1$ for $k\neq k'$.). This measure, however, weighs migrations across different research areas identically, regardless of their cognitive proximity. We will therefore also consider a second measure of change, the ``cognitive distance'' ($d_a$), defined as the minimum cost of transporting $\bm{x_a}$ to $\bm{y_a}$ \citep{Peyr2019} induced by the cost matrix empirically recovered (see Figure \ref{fig:cost_matrix}) using the Inverse Optimal Transport approach described in the previous section. Another interesting aspect of Optimal Transport is indeed its ability to provide ``distances'' between distributions that emphasize certain costs in particular \footnote{Given a cost matrix $C_{kk'}$, we can define a measure of the gap from one distribution $\bm{x}$ to another distribution $\bm{y}$, as the minimum cost of displacing $\bm{x}$ to $\bm{y}$:

\begin{equation}
    d(\bm{x},\bm{y}) = \min_{\substack{\substack{\theta_{kk'}\\\theta \bm{1}=\bm{1}, \theta^T \bm{x} = \bm{y}}}} \sum_{kk'} x_k \theta_{kk'} C_{kk'}
\end{equation}

If the cost matrix meets certain properties (such as symmetry), then $d$ is a distance. See \citealt{Peyr2019} for more on the metric properties of \gls{ot}.}.

$c_a$ is comprised between 0 (if research attention has remained identically distributed) and 1 (if the research agenda has been entirely redistributed).
Large values of $c_a$ are rare, with 50\% of authors lying between 0.21 and 0.40  (Appendix \ref{appendix:change-model}, Figure \ref{fig:change_scores}). Although the absolute value of $c_a$ (and $d_a$) has limited interpretability (it depends on the choice of ``cognitive scope''   and the duration of the time periods considered, and $c_a$ is never expected to be exactly zero due to random fluctuations and measurement noise), it allows for comparisons between physicists. We evaluate the effect of several factors on $c_a$: i) the diversity of intellectual capital $D(\bm{I_a})$; ii) the excess diversity of social capital $D^{\ast}(\bm{S_a})$; and iii) the magnitude of social capital (``power''). We also consider the effect of iv) affiliation stability, represented by a binary variable $s_a$ ($s_a=1$ if scientist $a$ has at least one affiliation that spans the whole time range considered, and $s_a=0$ otherwise), the effect of v) academic age, and productivity, estimated from vi) all papers and vii) solo-authored papers. We perform a linear regression of $c_a$ as a function of these variables, adjusting for $Z_a=\argmax_{k} x_{ak}$, i.e. physicists' primary research area over the years 2000 to 2009 (see the model specification in Appendix \ref{appendix:change-model}).

\begin{figure}[h]
    \centering
    \includegraphics[width=0.8\textwidth]{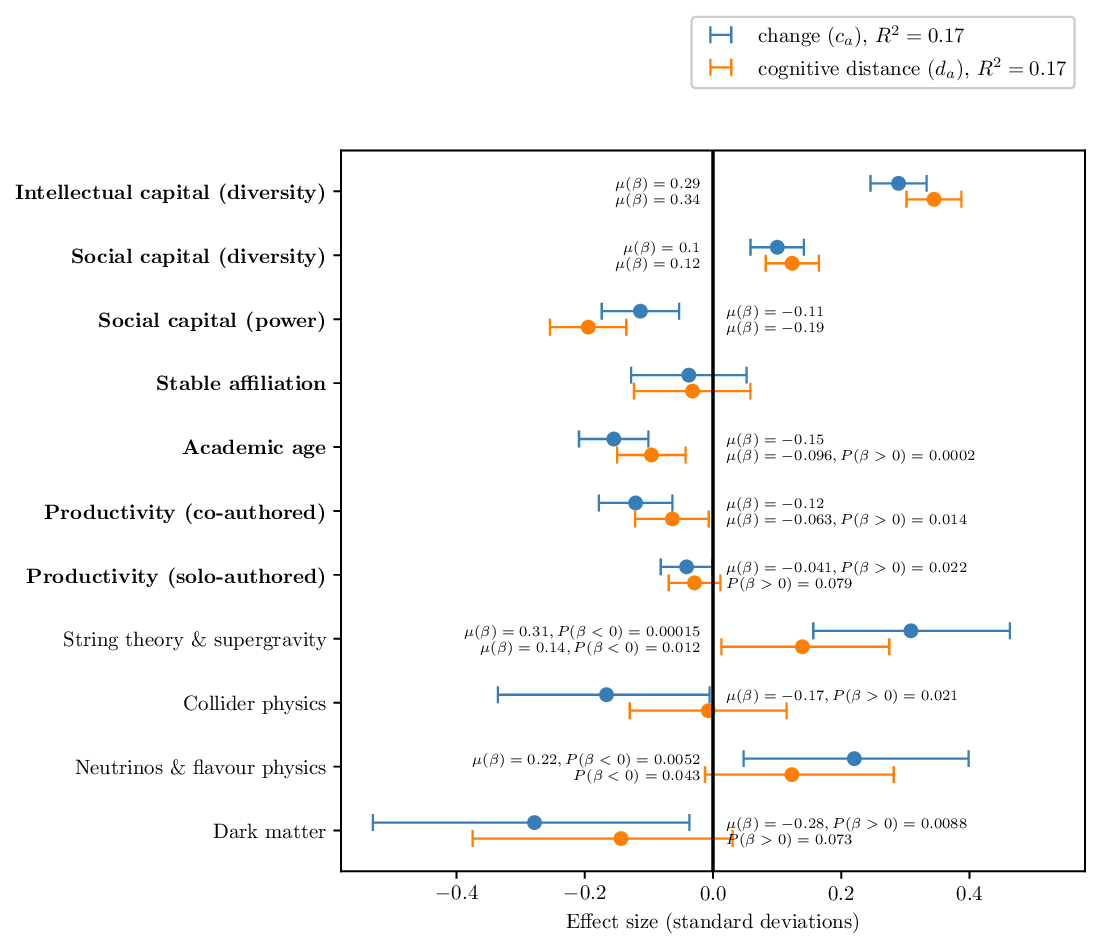}
    \caption{\textbf{Effect of intellectual capital, social capital, and institutional stability the magnitude of changes in high-energy physicists' interests.}. Change ($c_a$) is the total variation distance and weighs migrations across research areas equally. The cognitive distance ($d_a$) gives more weight to migrations across cognitively distant research area. Diversity is evaluated as the exponential of the Shannon entropy of the distribution of intellectual and social capital across research areas. The effect of the primary research area during the initial period (2000-2009) was also included as a control. Three research areas have a significant (95\% CL) effect. Error bars indicate 95\% credible intervals; $\mu(\beta)$ denotes the mean posterior effect size. $R^2=0.17$. Continuous variables (diversity, power and $c_a$) are standardized.}
    \label{fig:change_score_effect}
\end{figure}

The results are shown in Figure \ref{fig:change_score_effect}. The diversity of intellectual capital has a significant positive effect: physicists with resources in many areas tend to revise their research agenda more. There is also evidence of a small but positive effect of diversity of social capital on the magnitude of changes in scientists' research focus (interpreting these results in terms of Optimal Transport, we might say that social capital helps overcome cognitive learning costs). However, the \textit{magnitude} of social capital, ``power'', has a negative direct effect on change. In other words, ``power'' is associated with stability, and ``diversity'' with change. It is noteworthy that these dimensions of social capital have opposite effects. More senior physicists are more conservative toward their research agenda, possibly because they experience less incentive to ``adapt''. This comes in contrast with \cite{Zeng2019}. The difference could stem in the specificity of high-energy physics, and in the fact that change is measured in linguistics terms (instead of relying on citation patterns) at a rather coarse-grained scale in the present paper. There is no discernible effect of affiliation stability after adjusting for academic age. However, affiliation data is a bit noisy, and this could have the consequence of underestimating the effect of institutional stability relative to that of academic age. Finally, productivity (in co-authored papers) is associated with stability. Overall, entrenchment (age, power, productivity, specialization) all drive stability and conservatism. 

Unsurprisingly, both research areas that have shrunk considerably have a significant positive effect on migration scores. ``Collider physics'' and ``dark matter'', on the other hand, have a negative effect on the magnitude of change. All effects combined, physicists whose primary category is ``Collider physics'' are the most conservative, with an average change score 24\% lower than the rest of the cohort; the long time-scales of collider experiments provide stable opportunities to physicists in that area \citep[p.~138]{galison1987how}, thus promoting conservatism. The variance explained remains low ($R^2=0.17$): these factors only partially explain differences between individuals.

$c_a$ neglects the cognitive gap between research areas. Using our alternative measure of change that takes into account cognitive distance ($d_a$),  the diversity of intellectual and social capital has slightly larger effects. The robustness of these results is assessed by using different operationalizations of diversity and power in Appendix \ref{appendix:robustness}), Table \ref{table:summary_change_disruption}. Most findings are stable, except that ``brokerage'' (unlike the magnitude of social capital, i.e. degree centrality) has no \textit{direct} effect on change (beyond the effect of productivity resulting from co-authored publications). 

In addition, in order to exclude the influence of direct collaborations on change, we conducted a second comparative analysis including only scientists' first-authored and last-authored publications in their research portfolios\footnote{It is important to emphasize that alphabetical ordering is still very prevalent in this field \citep{waltman2012empirical}, and therefore this strategy does not address the issue entirely.}. The effect of the diversity of intellectual and social capital is stable; however, the direct effect of power is reduced (Appendix \ref{appendix:robustness}, Tables \ref{table:full_summary_change} and \ref{table:full_summary_disruption}). Moreover, the analysis was reproduced across different choices of temporal segmentation and using different topic models. The previous findings remain stable.

\subsubsection{Diversification versus concentration}

\begin{figure}[h]
    \centering
    \includegraphics[width=0.8\textwidth]{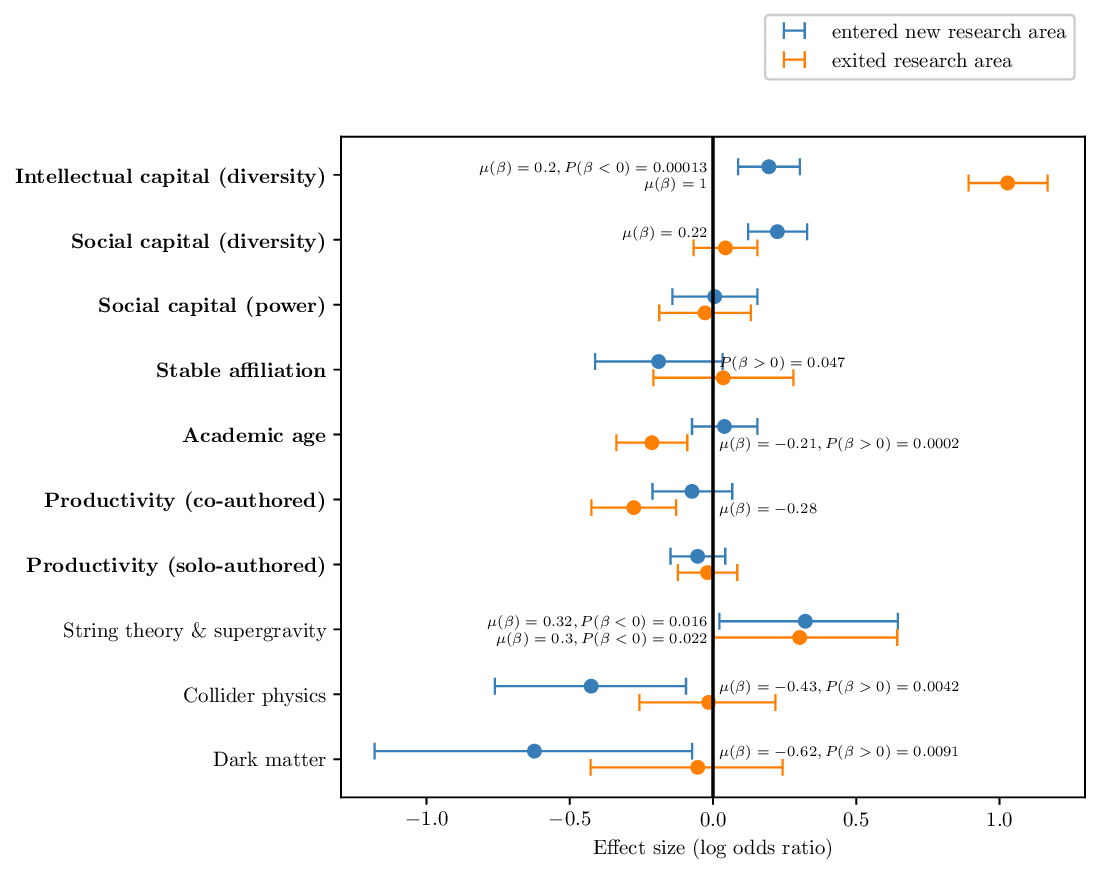}
    \caption{\textbf{Effect of capital and institutional stability on the probability of entering new research areas or exiting previously explored ones.} Bars indicate 95\% credible intervals. $\mu(\beta)$ denotes the mean posterior effect size. Continuous variables (diversity and power) are standardized.}
    \label{fig:diversification_score_effect}
\end{figure}

Research portfolios can be altered by two opposite strategies. One is ``diversification'', i.e. the addition of new research areas. Another is ``concentration'', i.e. the desertion of research areas to focus increasingly on others (Figure \ref{fig:research-agenda} illustrates how this can happen, whether via ``conversion'' or ``displacement''). Figure \ref{fig:diversification_score_effect} shows the effect of the same factors as above on the probability that physicists have i) entered at least one new research area in between the two periods or ii) exited one research area (model description provided in Appendix \ref{appendix:enter-exit-model})\footnote{A research area $k$ is considered ``entered'' by $a$ when $x_{ak}<\frac{1}{N}\sum_{a'}x_{a'k}$ and $y_{ak}>\frac{1}{N}\sum_{a'}y_{a'k}$; conversely, a research area is considered exited when $x_{ak}>\frac{1}{N}\sum_{a'}x_{a'k}$ and $y_{ak}<\frac{1}{N}\sum_{a'}y_{a'k}$.}.

The diversity of intellectual capital has a strong positive effect on the probability of exiting a research area; intuitively, scientists with diverse interests can afford to disengage from certain research areas even if this implies to abandon maladaptive prior knowledge (``displacement''). Excess diversity of social capital increases the probability of entering new research areas, but has no discernible effect on the probability of exiting research areas. In contrast, there is some evidence that ``power'' decreases the probability of leaving a research area. Figure \ref{fig:diversification_score_effect} shows the direct effect of power (controlling for productivity due to co-authored papers), which does not pass the 95\% significance test; however, the \textit{total} effect of power\footnote{That is, assuming the following causal structure:

\centering
\begin{tikzpicture}[>=Stealth, node distance=1cm, thick]

  \node (Power) {Power};
  \node (Productivity) [right=of Power] {Productivity};
  \node (Leaving) [below=1cm of $(Power)!0.5!(Productivity)$,align=center] {Leaving\\ research area};

  \draw [->] (Power) -- (Productivity);
  \draw [->] (Power) -- (Leaving);
  \draw [->] (Productivity) -- (Leaving);

\end{tikzpicture}} on the probability of leaving a research area is significantly negative ($\mu(\beta)=-0.24, P(\beta>0)<10^{-4}$). Presumably, having many collaborators allows scientists to remain committed to many research areas with minimal personal investment, thus stabilizing their research agendas. Moreover, ``power'' also decreases the probability of entering new research areas; this suggests that powerful scientists have less incentives to invest resources in new topics. String theory \& supergravity has a significant positive effect of both entering and exiting research areas (suggesting physicists frequently ``replaced'' it with another topic). 
``Dark matter'' has a negative effect on the probability of entering a new research area, possibly because scientists with prior commitment to this research area had less incentive to diversify their research portfolio and more incentive to focus increasingly on this topic given its success over the following years. These conclusions hold as alternative measures of diversity and power are considered (\ref{appendix:robustness}, Table \ref{table:summary_entered_exited}).

Again, we ran a second analysis considering only scientists' first-authored and last-authored publications. The effects of intellectual and social diversity remain stable. However, the effect of productivity resulting from co-authored papers no longer has an effect on the probability of exiting a research area when only first- and last-authored papers are included in physicists' research portfolios (\ref{table:full_summary_exited}, Appendix \ref{appendix:robustness}. ). The results are generally stable across different topic models and temporal segmentations, except for i) the effect of academic age on the probability of exiting a research area, which is not consistently 95\% CL significant; and ii) the effect of the diversity of social capital on the probability of entering a new research area, which is zero in one particular configuration.

\subsubsection{Why diversity promotes change}

\begin{figure}[h]
\hspace{-2em}
\begin{subfigure}{.5\textwidth}
    \includegraphics[width=1\textwidth]{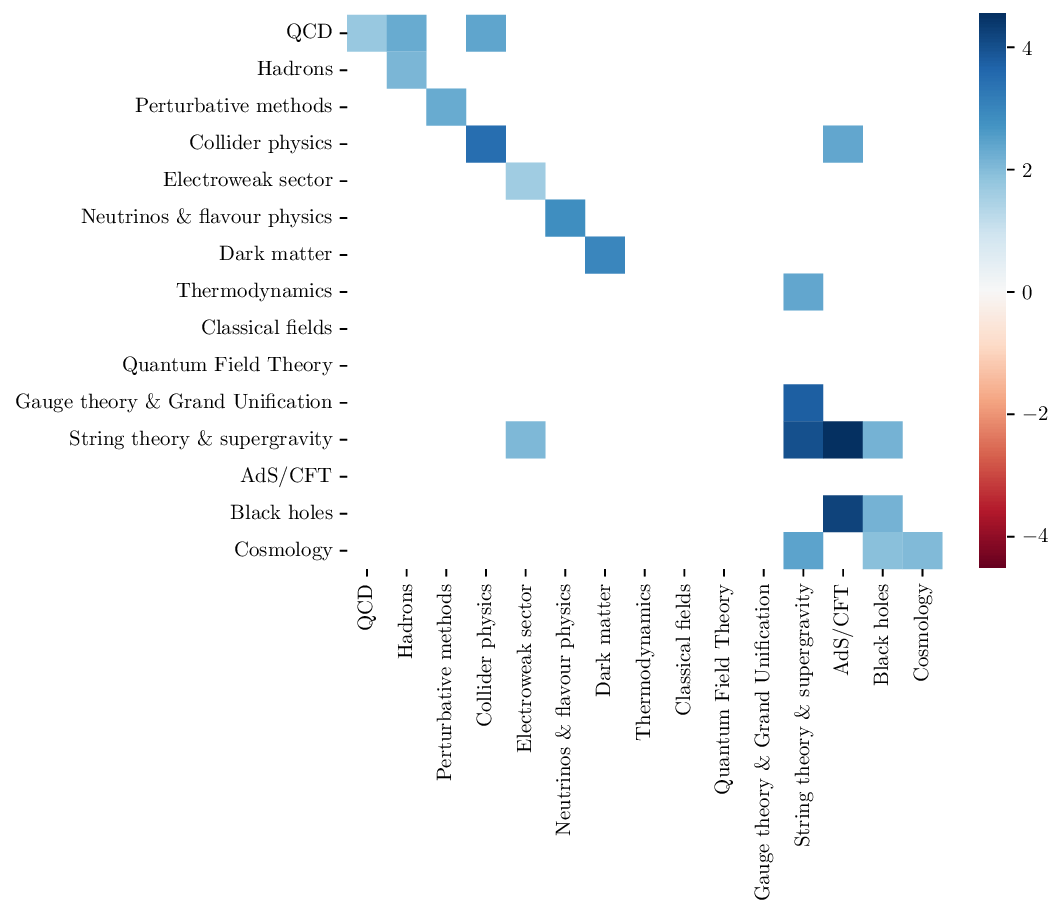}
    \caption{Effect of intellectual capital ($\gamma_{kk'}$).}
    \label{fig:intellectual-capital-effect}
\end{subfigure}%
\begin{subfigure}{.5\textwidth}
    \includegraphics[width=1\textwidth]{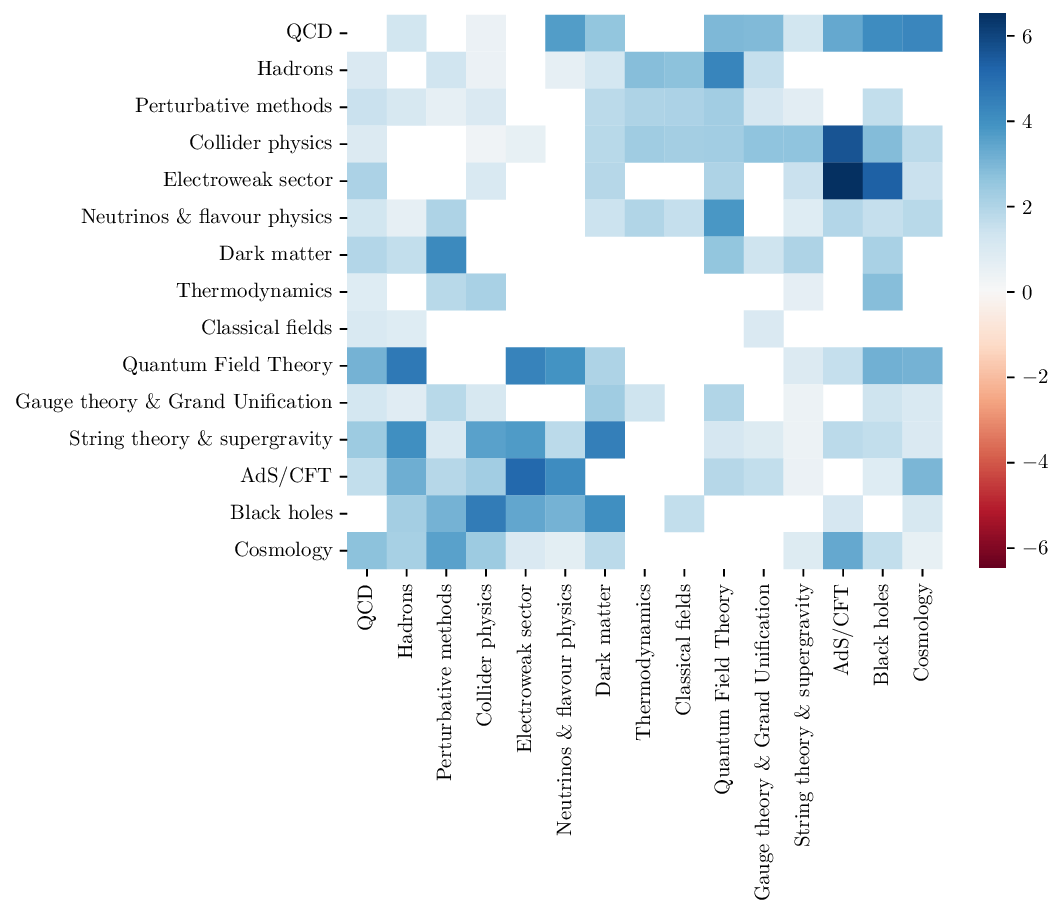}
    \caption{Effect of social capital ($\delta_{kk'}$).}
    \label{fig:social-capital-effect}
\end{subfigure}
\caption{\textbf{Effect of intellectual and social capital on transfers across research areas}. Rows represent research areas of origin and columns represent target areas. Effect sizes are expressed in log-odds ($\log {[\theta_{akk'}/(1-\theta_{akk'})]}$, where $\theta_{akk'}$ is the fraction of attention to $k$ redirected to $k'$) per unit of intellectual or social capital in the target research area $k'$. Effects that are not significant (at the 95\% credible level) are displayed in white for purposes of clarity.}
\label{fig:test}
\end{figure}

Access to diverse cognitive resources is associated with change. To see why, it is insightful to look into how the concentration of scientists' intellectual capital in each research area $k$ ($\bm{I_a}=(I_{ak})$) affects their trajectories. In the model introduced in Section \ref{sec:model}, the diagonal coefficients of the $\gamma$ matrix measure the effect of having intellectual resources in a certain area on the commitment to this area. As shown in Figure \ref{fig:intellectual-capital-effect}, most coefficients on the diagonal are significantly positive: physicists with a strong specialization in a research area tend to preserve their specialization in this area.

The effect of social capital on transfers across research areas ($\delta_{kk'}$, Eq. \ref{eq:glm}), is shown in Figure \ref{fig:social-capital-effect}. Statistically significant effects are always positive: scientists tend to redirect attention to research areas in which they have more collaborators involved, in line with a very recent observation by \citeauthor{Venturini2024} \citealt{Venturini2024}.

Moreover, we find a strong correlation between the effect of social capital ($\delta_{kk'}$), and $\nu_{kk'}\in[0,1]$, the fraction of physicists with more expertise than average in a research area $k'$ among those who have more expertise than average in $k$ (see Eq. \eqref{eq:social_capital_nu}, Section \ref{sec:model}). $\delta_{kk'}$ decreases by 1.5 unit of standard deviation on average as $\nu_{kk'}$ goes from 0 (nobody holds expertise in both research areas $k$ and $k'$) to 1 (everyone with expertise in $k$ has expertise in $k'$). Social capital plays a more important role in shifts between cognitively distant research areas, in line with a previous finding by \citeauthor{Tripodi2020} \citealt{Tripodi2020}. These general patterns (the association between the concentration of intellectual capital in one area and the commitment to this area, and the increasing effect of social capital with cognitive dissimilarity) are insensitive to the temporal segmentation in place (see Appendix \ref{appendix:model_parameters} for similar Figures based on a different segmentation).

\section{Discussion}

This paper addressed the conflict between specialization and adaptability in science. To this end, an unsupervised Bayesian approach was developed, based on the idea that transformations in the scientific landscape prompt scientists to efficiently repurpose their prior knowledge. The model simultaneously measures transfers of attention across research areas and the effect of various variables on the evolution of scientists' research portfolios, in particular intellectual and social capital.



The model was applied to a cohort of $N=2\,094$ high-energy physicists between the years 2000 and 2019. At the macroscopic level, it reveals the decline of neutrinos physics due to migrations towards the electroweak sector (explored at the \gls{lhc}) and more importantly towards dark matter. Similarly, many physicists have shifted resources from the electroweak sector towards dark matter. Moreover, string theory \& supergravity has started to disintegrate into black holes and AdS/CFT research. The cohort has therefore responded to new experimental opportunities as well as to theoretical developments in quantum gravity. 

Then, leveraging the connection between our model and \glsreset{ot}\gls{ot}, we showed that the reallocation of research efforts among scientists is shaped by learning costs. Indeed, under changing circumstances, scientific institutions must address an \gls{ot} problem by efficiently reallocating research efforts in a way that balances learning costs and the imperative to adapt to new circumstances. This has the effect of enhancing the utility of the scientific capital disseminated amongst scientists as the perceived payoff of certain research areas change. \gls{ot} structurally explains path dependency, as individuals experiencing pressures to adapt seize the nearest opportunity available to them. \gls{ot} is also methodologically useful: Inverse \gls{ot} allows one to derive cost functions from observed behavior, and thus offers a potential way to better connect empirical data with evolutionary agent-based models of scientists' behavior that postulate latent utility functions \citep{Wu2023}. Moreover, it has the potential of informing policy by identifying potential bottlenecks if research efforts were to be reallocated in certain ways (as one can use \gls{ot} to estimate the ``cost'' of various counterfactual scenarios). Overall, the \gls{ot} approach illustrates that the adaptability of epistemic communities is constrained by how knowledge is distributed among individuals (specialization).

The longitudinal comparative analysis of physicists' trajectories revealed that the diversity of intellectual and social capital is positively associated with change: diversity promotes adaptability under new circumstances, and therefore diversifying research portfolios is a reasonable strategy when the future is uncertain. However, enforcing diversity can lead to suboptimal allocations of research efforts under stable circumstances \citep{Schimmelpfennig2021}. Differences among research areas are found: physicists expert in particle colliders have remained particularly conservative, possibly because they have secured long-term research opportunities (thanks to very large investments in particle accelerators like the \gls{lhc}). There is also evidence that physicists specialized in dark matter have been consolidating their specialization, presumably because the increasing popularity of the topic encouraged them to double down their investment in this research area. Higher concentrations of intellectual capital in certain research areas generate stronger commitment towards these areas; therefore, specialized scientists are more at risk of being trapped in a sunk cost fallacy as their expertise becomes unsuitable for new circumstances. However, specialized scientists can offset the risks associated with specialization by diversifying their social network. This raises the possibility of free-riding, as scientists are encouraged to focus on what seems most promising at the time and let their peers take the risk of exploring alternatives until their value is established \citep{Kummerfeld2016}. Additionally, social capital plays an increasingly important role as scientists expand their research agenda further beyond their specialization, as observed by \citealt{Tripodi2020}, suggesting that collaborations are crucial in overcoming cognitive barriers between research areas. Unlike diversity, ``power'' is associated with more stable research interests: presumably, cooperation can safeguard individuals from adaptive pressures, and most importantly minimizes the cost of remaining invested in certain areas. 


We have described renewal strategies of research portfolios according to a typology of incremental change developed in the context of historical institutionalism, which has been shown to account for how organizations like DESY and SLAC have transitioned from \gls{hep} to multi-purpose photon science. Adaptation strategies include the ``conversion'' of prior knowledge to new purposes, the ``layering'' of additional research interests via the acquisition of new knowledge, or the ``displacement'' of former research commitments resulting in a loss of knowledge. The connection with institutional change stems from the fact that institutions can face a challenge similar to that experienced by specialized scientists confronted with transformations in their scientific landscape, in that they too must sometimes adapt and redirect accumulated capital in directions that may not have been foreseen at the time of their ``design''. The \gls{lhc} itself has evolved through similar processes of gradual repurposing of prior infrastructure, including the accelerator's tunnels \citep{Smith2015}: adaptation prompts individuals (and collectives) to efficiently repurpose capital previously accumulated in different forms (cultural, social, economic, etc.). Following \citealt{Galesic2023}, we conclude that a better understanding of collective adaptation benefits from the pooling of diverse insights: while studies of institutional change have documented strategies of gradual adaptation that progressively leverage and repurpose accumulated capital when change is difficult, the literature on cultural evolvability stresses the critical roles of diversity and social learning. In return, this work provides an empirical contribution to the literature that treats science as a cultural evolutionary system \citep{Wu2023}.

Throughout the paper, various methodological limitations were raised. First, the requirement on the quantity of authors' publications makes the cohort atypical. A second issue, already noted by \citet{Gieryn1978}, lies in the arbitrariness in the choice of temporal and cognitive scopes for measuring change. In this paper, the choice in temporal segmentation was driven by the time-scale associated with the transformations in the landscape of experimental opportunities. For shorter time-scales, the very notion of research portfolio -- as operationalized in this paper --  may break down and lose any predictive force. As per the cognitive scope, previous works (e.g. \citealt{Jia2017,Aleta2019,Tripodi2020}) have typically relied on the hierarchical \gls{pacs} classification of physics literature, such that cognitive scales were imposed. For this work, the literature was divided into 15 topics that captured the features of change in the epistemic landscape discussed above, i.e. the rise of new probes of the cosmos, but other scales could have been considered. 
Moreover, although the unsupervised topic model approach is arguably a better proxy of cognitive change, it introduces noise which could in part explain the low predictive power of the models of change in scientists' research agendas. Additionally, the topic model was trained on the entire time range covered by the analysis. Changes in the relationships between topics and in their own vocabulary distributions are not considered, even though they constitute another interesting linguistic dimension of adaptive patterns that would deserve further investigation. Morever, the cost of shifting from one research area to another, is itself time-varying quantity in reality. Finally, one must be cautious before drawing strong causal conclusions from these findings. While the causal pathway ``power $\to$ collaborations $\to$ stabilization of research interests'' seems to be a reasonable interpretation of the results, the relationship between diversity and change is less clear; they could both be confounded by a shared latent trait observed among certain researchers (e.g. the ``explorers'' in \citealt{Chakresh2023}). Moreover, the sample size $(N=2\,094)$ is insufficient to explore sophisticated interactions or potential moderators in the comparative analysis. 


\printglossaries

\section*{Declarations}

\paragraph{Availability of data and material}{The data and code used for this publication is available at \url{https://github.com/lucasgautheron/specialization-adaptation/}. The repository will be archived on Zenodo upon publication.}

\paragraph{Competing Interests}{The author declares no competing interest.}

\paragraph{Funding}{This work has received support from the G-Research Grant for Early-Career Researchers.}

\paragraph{Authors' contributions}{The sole author of the paper completed all of the work.}

\paragraph{Acknowledgements}{I thank Thomas Heinze, Radin Dardashti, and Elisa Omodei for useful discussions; Marc Santolini for rich feedback; Georges Ricci for raising my interest in Optimal Transport; Marco Schirone for providing useful references; the LATTICE lab in Montrouge for the opportunity to discuss this work; and Elizabeth Zanghi for corrections.}

\printbibliography

\newpage

\appendix

\section{Supplementary material}

\subsection{\label{appendix:landscape}Transformations in high-energy physics}

\begin{figure}[H]
    \centering
    \includegraphics[width=\textwidth]{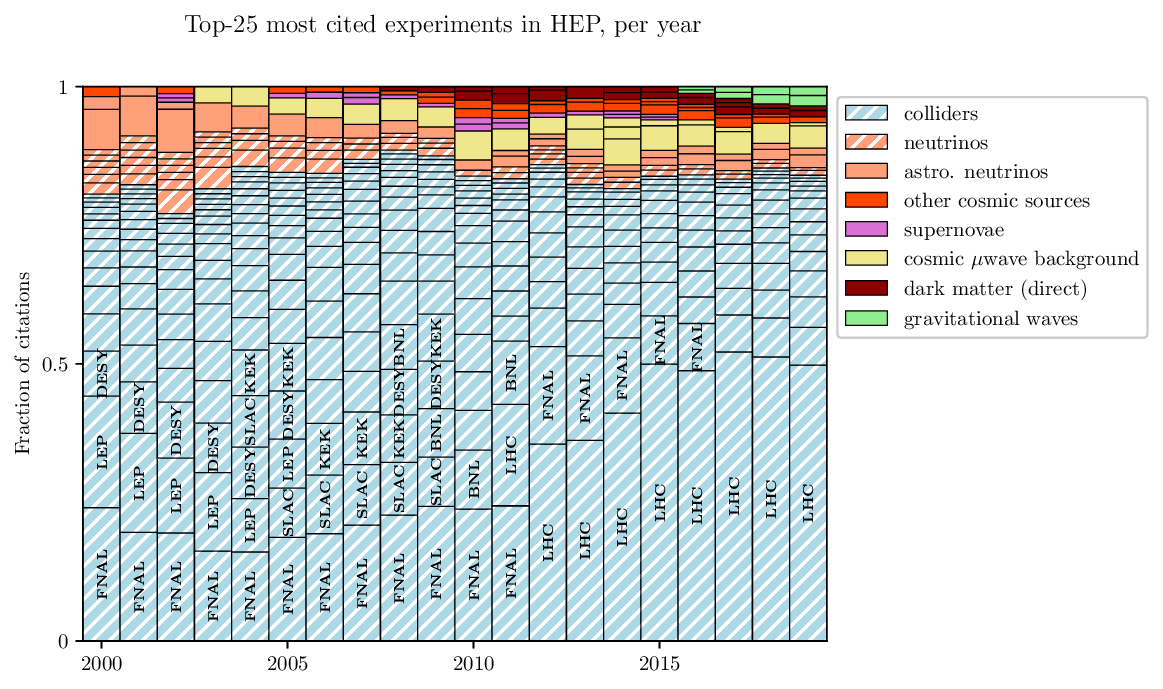}
    \caption{\textbf{Share of citations of each of the most cited experiments in high-energy physics literature, between 2000 and 2019}. Hatched rectangles correspond to experiments observing particles produced in colliders or nuclear reactors; other rectangles correspond to observations of phenomena or particles of astrophysical origin. We use citation and experiment data from the Inspire HEP database \citep{InspireAPI}. The classification of these experiments is our own. }
    \label{fig:experiments}
\end{figure}

\subsection{\label{appendix:sample_characteristics}Cohort characteristics}

\begin{figure}[H]
    \centering
    \includegraphics[width=0.8\textwidth]{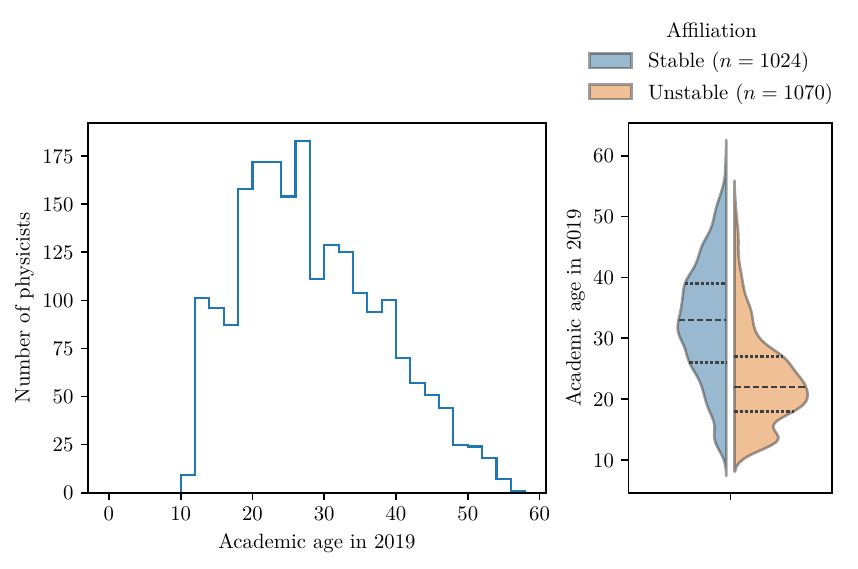}
    \caption{\textbf{Distribution of academic age (in 2019) and affiliation stability within the considered cohort}.  Academic age (left plot) is evaluated as the time passed since the first publication. There is a cut-off in the distribution academic age due to the requirement on publication counts between 2000 and 2009. The median academic age is 27 years in 2019. 49\% of the cohort has had at least one permanent affiliation spanning the period under consideration (right plot). As expected, permanent affiliations are associated with higher academic ages (bars indicate min/max/quartiles). }
    \label{fig:sample_characteristics}
\end{figure}

\subsection{\label{appendix:topics}Topics}

\subsubsection{\label{appendix:word2vec}Word embeddings dimension}

The word2vec skip-gram models is trained using different values for $L$ the dimension of the embeddings' space. $L=50$ (the choice made in the present paper) generally lies somewhere between under-fitting and over-fitting. The latter is a concern due to the small sample size (the model is trained on abstracts rather than full-texts).

\begin{figure}[H]
    \centering
    \includegraphics[width=0.8\linewidth]{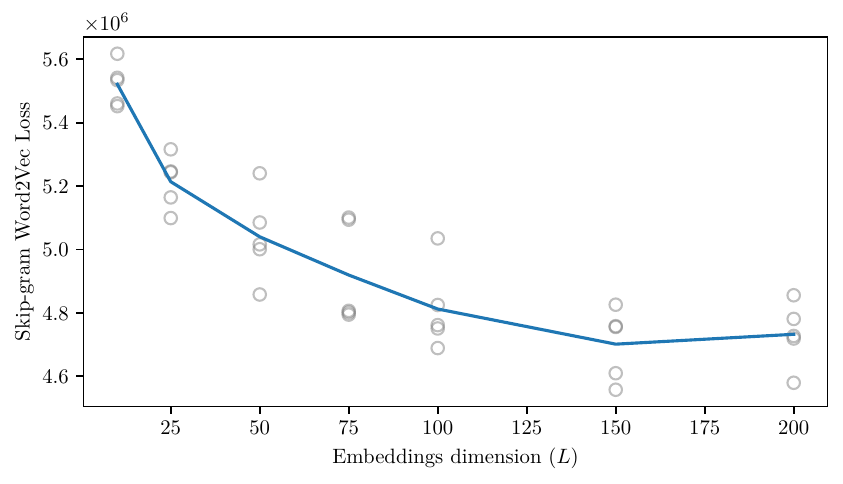}
    \caption{Loss of the word2vec model as a function of the embeddings dimension ($L$). }
    \label{fig:word2vec}
\end{figure} 

\subsubsection{\label{appendix:keywords}Keyword classification}

In the Embedding Topic Model used in the present paper -- as in topic models in general--, documents are mixtures of several topics; moreover, keywords may belong to several topics. This is a desirable feature: certain concepts serve different purposes depending on the context, and some concepts do not clearly belong to any research area. More importantly, this is crucial to the present approach, given that we must be sensitive to situations in which scientists have repurposed certain concepts to new research goals; i.e., instances where the same resources are applied in a new context (a new research area).
For that, each keyword $i$ from a document $d$ is assigned a research area $z_{di} := \argmax_{k=1,\dots,K} P(z_{di}=k|w_{di},\bm{\theta}_{d})$, which is the research area most probably associated with the keyword $i$ given the topic distribution of the document $\bm{\theta}_d$ -- i.e. the context. In fact:

\begin{equation}
    z_{di} := \argmax_{k=1,\dots,K} P(z_{di}=k|w_{di},\bm{\theta}_{d}) =  \argmax_{k=1,\dots,K} \dfrac{P(w_{di}|z_{di}=k)P(z_{di}=k|\bm{\theta}_d)}{P(w_{di}|\bm{\theta}_d)}
\end{equation}

In the process, we discard ambiguous keywords for which $H(z_{di})\geq \log{2}$, where $H$ denotes the entropy of the distribution $P(z_{di}=k|w_{di})$. Either such keywords do not carry any scientific content, or the context is insufficient to disambiguate among the possible research areas to which they might belong.

The average effective amount of topics in documents according to the topic model -- measured as $\exp H(\bm{\theta}_{d})$ where $H$ is the Shannon entropy -- is 7.5, which is unrealistically high. The filtering and classification procedure reduces the average effective amount of topics to 3.1 per document, which is much more informative. This procedure is especially important for short texts such as abstracts, which poorly constrain the latent topic distribution $\bm{\theta}_{d}$.

\subsubsection{List of topics}

\fontsize{6}{7}\selectfont\input{Table1}\normalsize

\subsubsection{\label{appendix:citation_validation}Topic validation using the citation network}

In order to validate the consistency and scientific dimension of the topics that were recovered, we verify that papers from a given topic tend to cite more papers from the same topic. Let $N_{k,k'}$ be the amount of citations of articles that belong to topic $k'$ originating from articles that belong to $k$, and $N=\sum_{k,k'}N_{k,k'}$ the total number of citations. From this matrix, a normalized pointwise mutual correlation $\mathrm{npmi}(k,k')$ is calculated:

\begin{equation}
    \mathrm{npmi}(k,k') = \log {\dfrac{N_{k,k'}/N}{(\sum_{i} N_{k,i}/N)(\sum_{i} N_{i,k'}/N)}}
\end{equation}

\begin{figure}[H]
    \centering
    \includegraphics[width=0.8\textwidth]{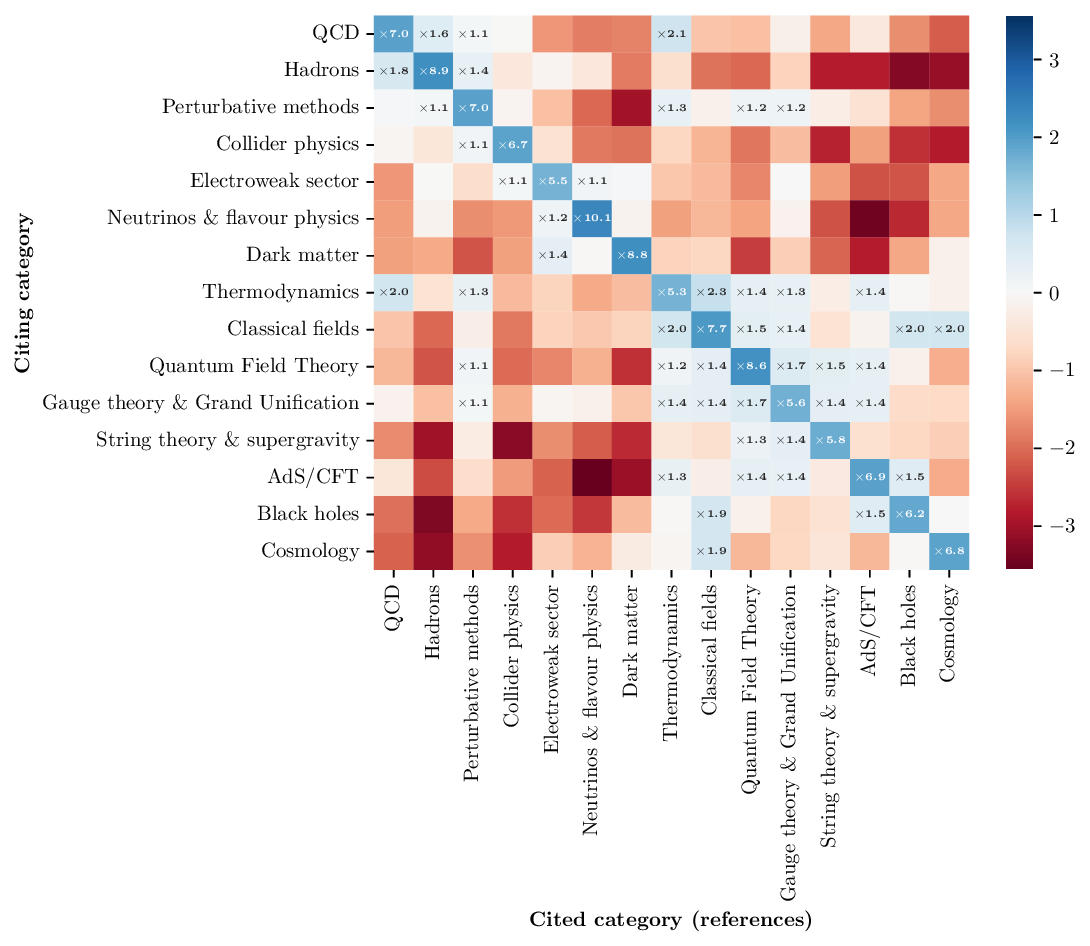}
    \caption{\textbf{Normalized pointwise mutual information $\mathrm{npmi}(k,k')$ of the citation matrix $N_{k,k'}$}. Positive values (in blue) indicate that the y-axis research area cites the x-axis research area more often than expected by chance; negative values indicate that citations occur less than expected by chance. When shown, individual values indicate how many times citations occur compared to chance alone. For instance, papers about Hadrons cite papers about Perturbative methods 1.4 times what would be expected if citations were uniform across research areas.}
    \label{fig:topic_citation_matrix}
\end{figure}

$\mathrm{npmi}(k,k')$ is shown in Figure \ref{fig:topic_citation_matrix}. It measures how frequent citations from $k$ to $k'$ are, relative to what would be expected if citations were uniformly distributed. The diagonal values are positive, indicating that the topics we retrieved tend to refer to themselves significantly more than expected by chance alone, providing further evidence of their scientific content and coherence.

\subsubsection{\label{appendix:topic_comparison}A comparison of three topic models}

\begin{figure}[H]
    \centering
    \includegraphics[width=0.9\textwidth]{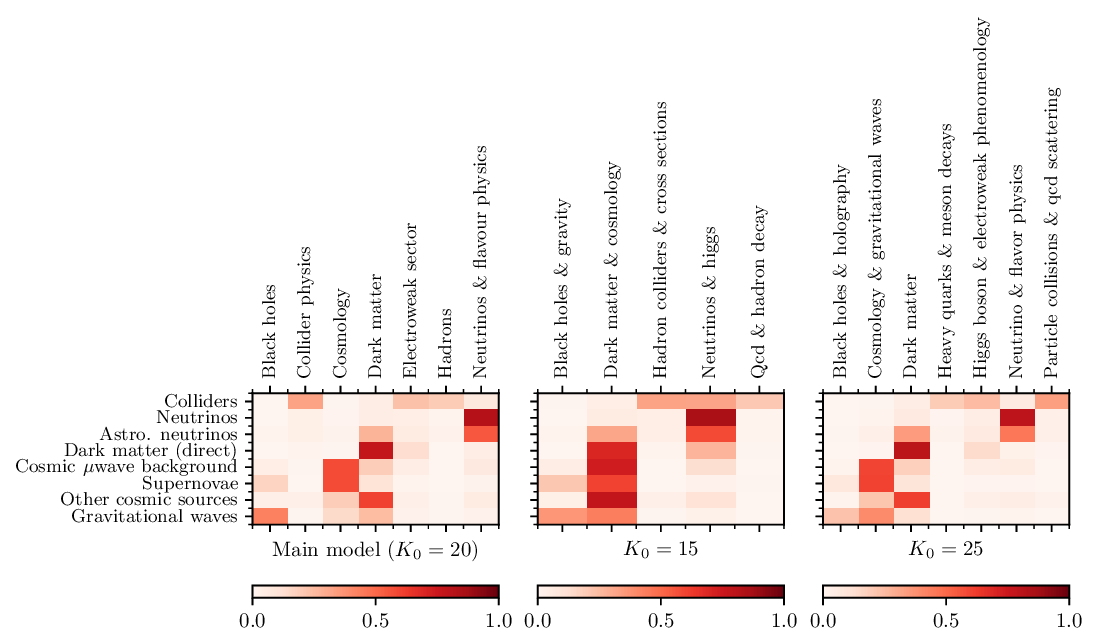}
    \caption{\label{fig:topic_experiments}Origin of papers citing each type of experiment, according to three topic models: the model considered throughout the present paper (to the left), a model with fewer topics ($K_0=15$), and a model with more topics ($K_0=25$). A value of one (dark red) indicates that 100\% of the papers citing a certain type of experiment (cf. rows) originate from a given topic (cf. columns).}
\end{figure}

Figure \ref{fig:topic_experiments} compares the ability of three topic models to measure the transformations in high-energy physics research resulting from changes in the landscape of experimental opportunities. 
In the coarse-grained model ($K_0=15$, in the middle), many types of experiments are lumped together into a single topic. It is therefore ill-suited for assessing the impact of the transformations in the landscape of experimental opportunities. The model used throughout the paper is well able to distinguish neutrino research for dark matter research, which have both undergone significant transformations according to Figure \ref{fig:experiments}. It is also better able to separate black hole phenomenology and cosmology, compared to the fine-grained model ($K_0=25$). 

\subsubsection{\label{appendix:pacs_validation}Topic validation using the PACS classification}

\begin{figure}[H]
    \centering
    \includegraphics[height=0.72\paperheight
]{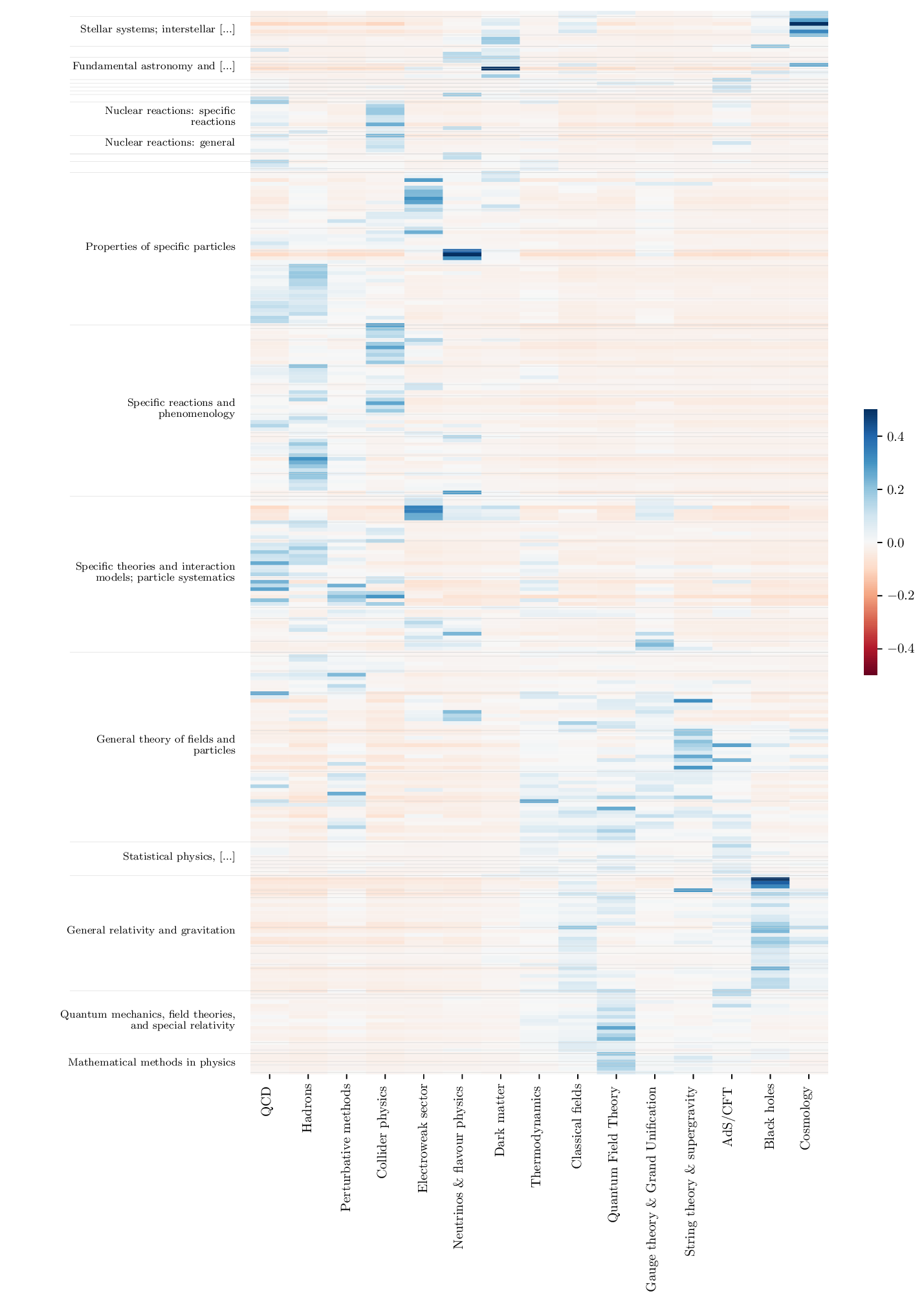}
    \caption{Correlation between \gls{pacs} categories present in $\geq 100$ publications in the corpus (rows), and the research areas recovered by the topic model (columns). Each colored cell indicates the correlation between a leaf category of the hierarchical \gls{pacs} classification and one of the topics from the topic model.}
\end{figure}

Blue cells show that the research areas recovered from the topic model correlate with the PACS classification (which further confirms their scientific dimension), but also that they can give a different picture. For instance, each of the topics  ``dark matter'', ``black holes'' and ``cosmology'' span over several higher-level categories of the PACS classification (e.g. ``fundamental astronomy \dots'' and ``specific theories and interaction models \dots'' for dark matter).

\subsection{\label{appendix:capital_validation}Measures of capital}

\begin{figure}[h]
    \centering
    \includegraphics[width=0.6\textwidth]{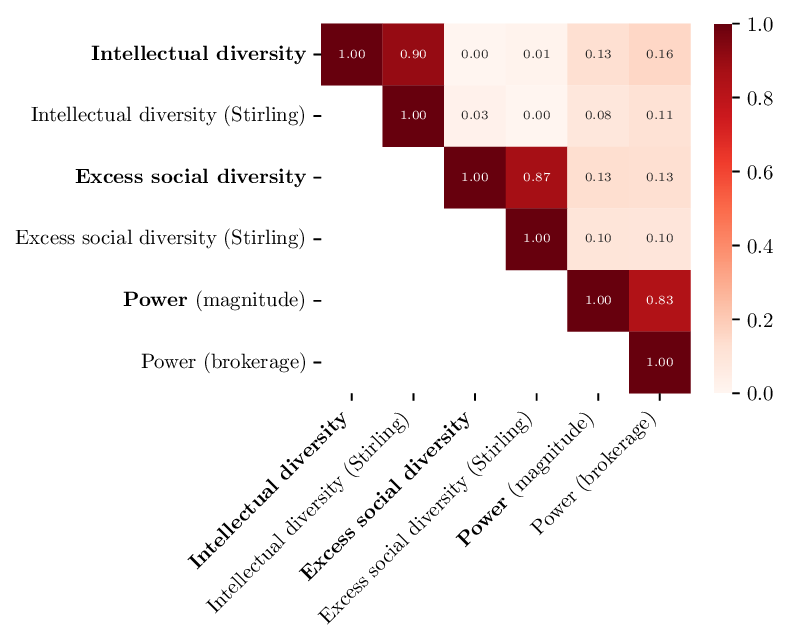}
    \caption{\textbf{Correlation between different measures of capital}. Measures considered in priority in this analysis are shown in bold. Alternative measures are shown for comparison purposes. By construction, excess social diversity is orthogonal to intellectual diversity ($R=0$). }
    \label{fig:capital_measures}
\end{figure}

Figure \ref{fig:capital_measures} shows the Pearson correlation between different measures of the diversity of intellectual and social capital and of power, as evaluated among the cohort of high-energy physicists. For comparison purposes, an alternative measure of diversity based on Stirling's index \citep{Stirling2007}, with prior applications to studies of interdisciplinarity \citep{Porter2007,Leahey2016} is evaluated\footnote{The Stirling-based diversity measured is evaluated as: \begin{equation}
    D_{\text{Stirling}} = 1-\sum_{k,k'}d_{kk'}I_{ak}I_{ak'}
\end{equation}

Where $d_{kk'}$ is the fraction of scientists who have more expertise than average in both $k$ and $k'$ among those that have expertise in one or the other (i.e., a similarity matrix). This  follows from previous approaches for measuring research interdisciplinarity \citealt{Porter2007,Leahey2016}.}. A measure of brokerage is also considered\footnote{We evaluated brokerage as the amount of pairs of scientists that have collaborated with a given physicist while having no common collaborator except for this physicist. This effectively measures the extent to which this physicist connects otherwise disconnected scientists.}. 

As shown in Figure \ref{fig:capital_measures}, the entropic measure of diversity considered in this paper correlates strongly with the Stirling measure. The magnitude of social capital (which is similar to degree centrality) correlates weakly with excess social diversity, thus emphasizing that power and diversity are partially orthogonal aspects of social capital. The magnitude of social capital is strongly correlated with brokerage; indeed, strongly connected scientists, with higher degree centrality, are also those scientists who initiate collaborations between otherwise disconnected scientists, as measured by brokerage.

\subsection{\label{appendix:model-performance}Model performance over multiple corpora, temporal segmentations, and topic granularities}

The predictive power of the model can be assessed by evaluating the total variation distance between the true distribution $\bm{y_a}$ and the predicted distribution $\bm{y_a^{\text{pred}}}$. This performance metric is calculated via 10-fold cross-validation. It is compared  to a baseline model that predicts no change in the research agenda ($\bm{y_a^{\text{baseline}}}=\bm{x_a}$). The results are shown in Table \ref{table:performance}. For the cohort of high-energy physicists, the model performs only marginally better than the baseline, given that individuals have remained quite conservative on average, most of the fluctuations being difficult to predict. Table \ref{table:performance} also considers a cohort of scientists from the ACL anthology corpus  of computational linguistics research \citep{acl_anthology_corpus}, by running the same pipeline (the measurement of research portfolios during two consecutive time periods using our topic model approach and the training of the trajectory model). Although the data are of significantly lesser quality, research portfolios have undergone much more significant transformations in this dataset (see Appendix \ref{appendix:hep_vs_acl}, Figure \ref{fig:sankey_acl}). Consequently, our model performs much better than the baseline for this cohort.

\input{Table2}

\subsection{\label{appendix:optimal_transport}Learning costs and optimal transport}

The MCMC algorithm from \citealt{pmlr-v162-chiu22b} is run on 1\,000\,000 iterations of the ``MetroMC'' algorithm, using what the authors call a ``P1'' prior (that is, a prior such that $\sum_{kk'}C_{kk'}=C_0=\mathrm{cst}$\footnote{We chose the minimum value of $C_0$ for which the system admitted a solution.}). More precisely, we assume that:

\begin{align}
    P(c_{kk'}|p_{kk'}) &= \dfrac{1}{Z}\dfrac{1}{\displaystyle\prod_{kk'} c_{kk'}^{1/2}} \exp{\left(-\alpha D_{KL}(c_{kk'}||p_{kk'})\right)} \text{ with } c_{kk'} = C_{kk'}/C_0\label{eq:entropic}\\
    \text{ and } p_{kk'} &= \mathrm{softmax}(\beta (1-\nu_{kk'}))\\
\end{align}

\begin{figure}[H]
    \centering
    \includegraphics[width=0.8\textwidth]{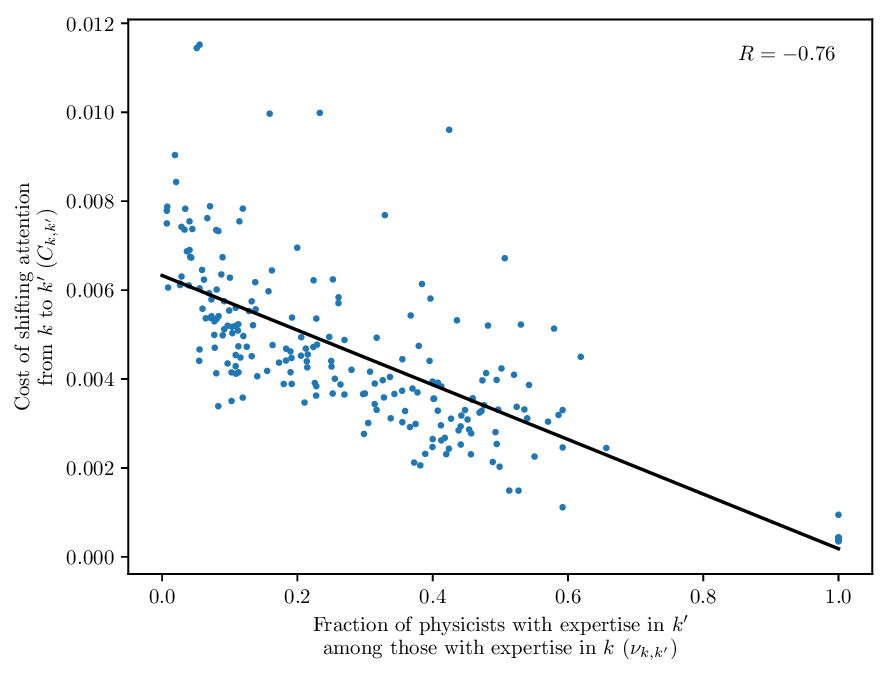}
    \caption{Cost of shifting a unit of cohort's research attention from $k$ to $k'$ as a function of the fraction of physicists with expertise in $k'$ among those with expertise in $k$ ($\nu_{kk'}$).}
    \label{fig:cost_knowledge}
\end{figure}

\eqref{eq:entropic} is sometimes referred to as the entropic prior \citep{skilling1991bayesian,MacKay1995}. The mean posterior values of $C_{k,k'}$ are shown in Figure \ref{fig:cost_knowledge}, as a function of the knowledge gap from $k$ to $k'$. The knowledge gap  $1-\nu_{kk'}$ is the fraction of experts in $k$ who do not hold significant expertise in $k'$ ($\nu$ is shown in Figure \ref{fig:nu}).  A significant correlation is found ($R=-0.76$). This is true also for the replication dataset of computational linguistics research (Appendix \ref{appendix:hep_vs_acl}), for which $R=-0.63$.

\subsection{Effect of capital on strategies of change}

\subsubsection{\label{appendix:change-model}Model for the magnitude of change}

\begin{figure}[h]
    \centering     \includegraphics[width=0.8\textwidth]{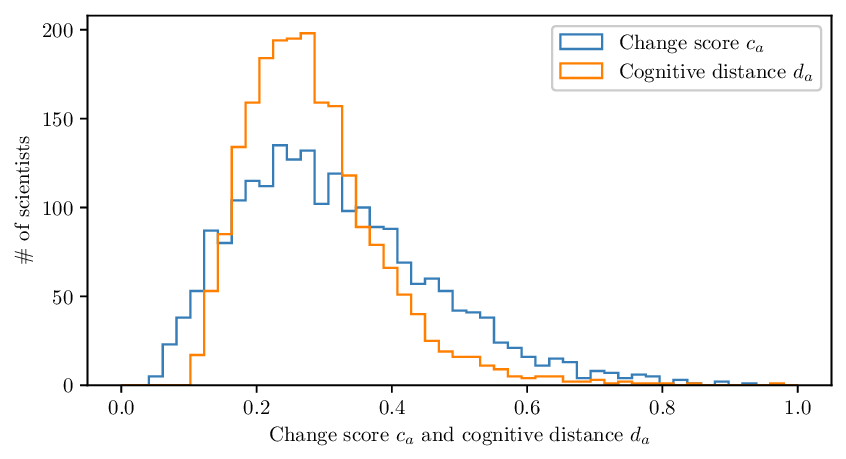}
    \caption{\textbf{Distribution of change and cognitive distances in the cohort}. Higher values correspond to more drastic changes in a scientists' research agenda.}    
    \label{fig:change_scores}
\end{figure}

The model for $c_a$ is:

\begin{align*}
    z(c_a) &\sim \mathcal{N}(\mu_a,\sigma)\\
    \mu_a &= \beta^{\text{int-div}} z(D(\bm{I_a}))+\beta^{\text{soc-div}}z(D^{\ast}(\bm{S_a})) + \beta^{\text{power}} z(P(\bm{S_a})) + \beta^{\text{stability}} p_a + \beta^{\text{age}} z(a_a) + \beta^{\text{prod}} z(\pi_a) + \mu^{\text{area}}_{k_a} + \mu\\
    k_a &= \argmax_k x_{ak}\\
    \beta,\mu &\sim \mathcal{N}(0, 1)\\
    |\mu^{\text{area}}_k| &\sim \mathrm{Exponential}(\tau)\\
    \tau,\sigma &\sim \mathrm{Exponential}(1)
\end{align*}

Where $z(\cdot)$ denotes standardized variables.

\subsubsection{\label{appendix:enter-exit-model}Model for the probability of having  entered/exited a research area}

The model for the probability $p_a$ of having entered a new field:

\begin{align*}
    \mathrm{logit}(p_a) &= \beta^{\text{int-div}} z(D(\bm{I_a}))+\beta^{\text{soc-div}}z(D^{\ast}(\bm{S_a})) + \beta^{\text{power}} z(P(\bm{S_a})) + \beta^{\text{stability}} p_a  + \beta^{\text{age}} z(a_a) + \beta^{\text{prod}} z(\pi_a) + \mu^{\text{area}}_{k_a} + \mu\\
    k_a &= \argmax_k x_{ak}\\
    \beta,\mu &\sim \mathcal{N}(0, 1)\\
    |\mu^{\text{area}}_k| &\sim \mathrm{Exponential}(\tau)\\
    \tau,\sigma &\sim \mathrm{Exponential}(1)
\end{align*}

The same model structure is used for the probability of having exited a research area.

\subsubsection{\label{appendix:robustness}Effect of capital and robustness checks}

\input{Table3}
\input{Table4}


\subsection{\label{section:robustness}Additional robustness checks}

The robustness of the results of the comparative analysis is assessed by varying different parameters:

\begin{itemize}
    \item The papers included in each authors' portfolio (any paper versus first-authored and last-authored papers only).
    \item The amount of topics in the topic model ($K_0$).
    \item The amount of dimensions for the word embeddings ($L$).
    \item The temporal segmentation for the early and late research portfolios.
\end{itemize}

\input{full_summary}

\subsection{\label{appendix:model_parameters}Trajectory model parameters evaluated on a different time period}

To further assess the robustness of the findings, the effect of intellectual and social capital on individual trajectories is measured on a different temporal segmentation (2000-2004 to 2005-2009). We make similar findings: the concentration of intellectual capital in one area promotes either commitment to this research area (or transfers in related areas). Social capital, on the other hand, matters increasingly as cognitive distance increases.

\begin{figure}[H]
\begin{subfigure}{.5\textwidth}
    \includegraphics[width=1\textwidth]{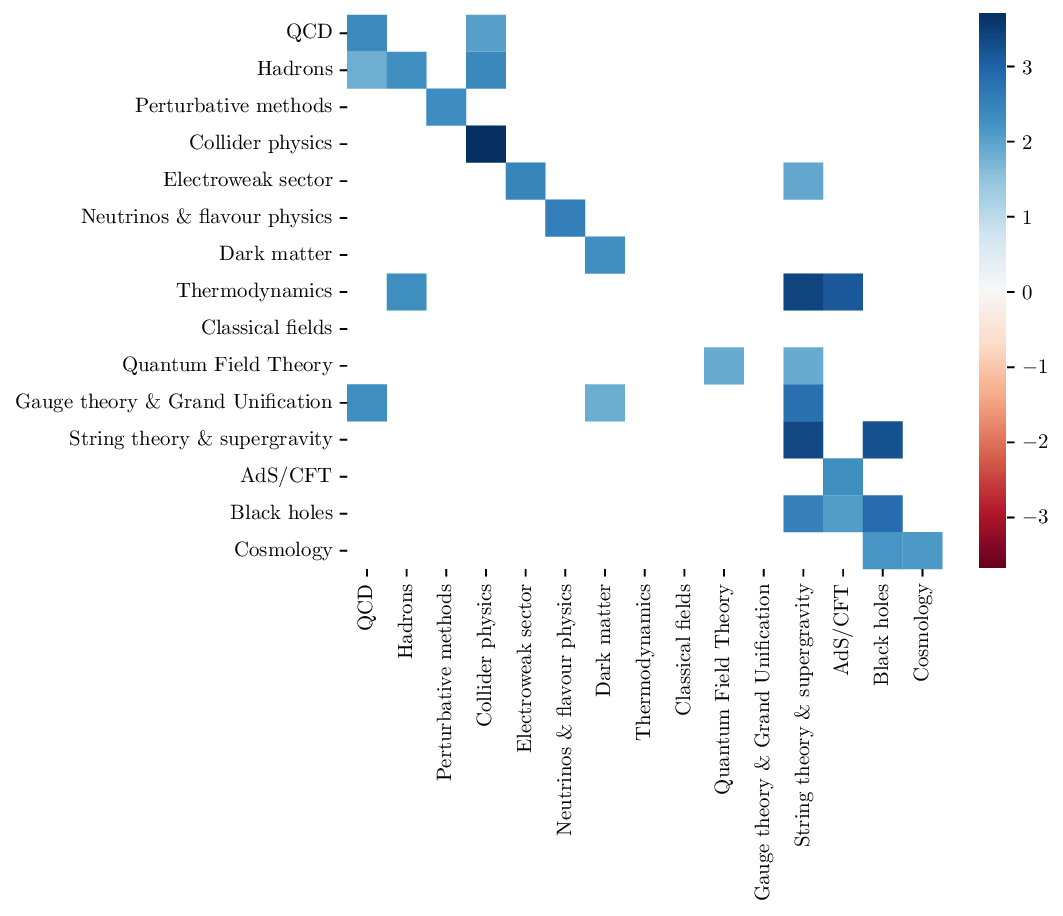}
    \caption{Effect of intellectual capital ($\gamma_{kk'}$).}
    \label{fig:intellectual-capital-effect_2000-2009}
\end{subfigure}%
\begin{subfigure}{.5\textwidth}
    \includegraphics[width=1\textwidth]{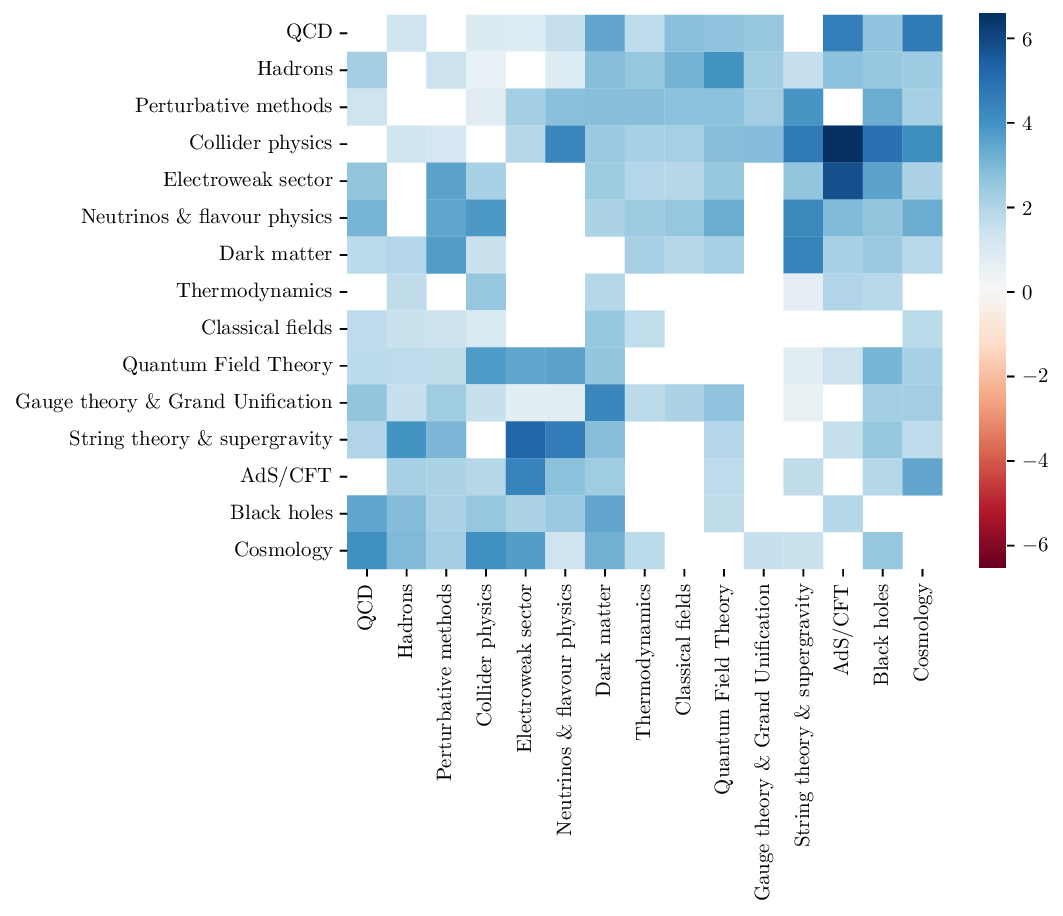}
    \caption{Effect of social capital ($\delta_{kk'}$).}
    \label{fig:social-capital-effect_2000-2009}
\end{subfigure}
\caption{\textbf{Effect of intellectual and social capital on transfers across research areas}. Rows represent research areas of origin and columns represent target areas. Effect sizes are expressed in log-odds ($\log {[\theta_{akk'}/(1-\theta_{akk'})]}$, where $\theta_{akk'}$ is the fraction of attention to $k$ redirected to $k'$) per unit of intellectual or social capital in the target research area $k'$. Effects that are not significant (at the 95\% credible level) are displayed in white for purposes of clarity.}
\end{figure}

\subsection{\label{appendix:hep_vs_acl}Replication corpus}

For purposes of testing and replication, certain analyses have been reproduced on the ACL anthology corpus of Computational Linguistics research .

\begin{figure}[h]
     \centering
     \includegraphics[width=\textwidth]{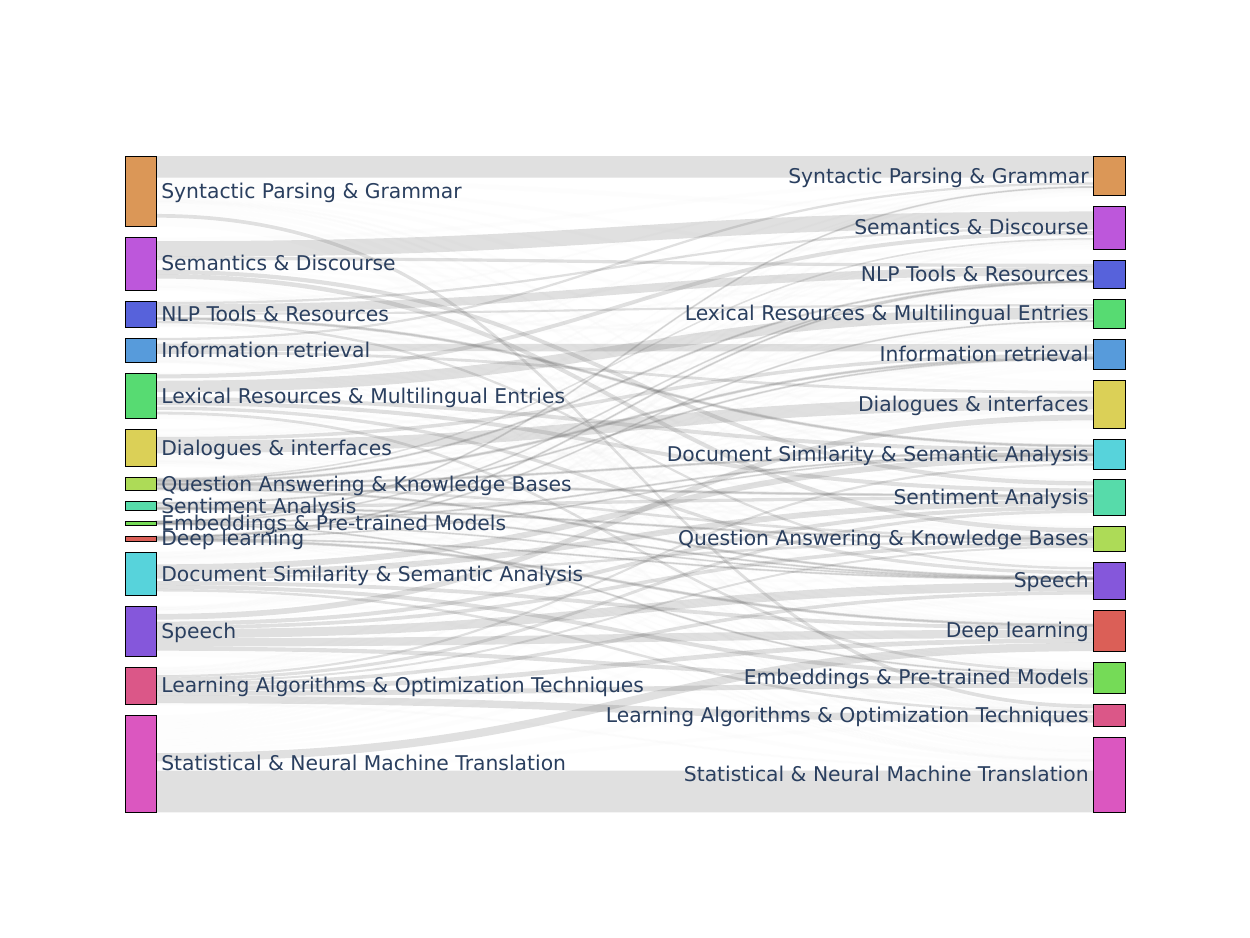}
     \caption{\textbf{Aggregate transfers of attention across research areas in the ACL anthology corpus of Computational Linguistics, between 2002-2011 (to the left) and 2012-2022 (to the right)}. Widths of flows are proportional to $\sum_a X_{ak}\theta_{akk'}$. Transfers less frequent than expected by chance alone are transparent. }
     \label{fig:sankey_acl}
 \end{figure}

The transfers of attention are shown in Figure \ref{fig:sankey_acl}. Compared to the high-energy physics corpus, it features significant disruptions (e.g., the emergence of new topics, such as ``deep learning'', ``sentiment analysis'' and ``embeddings \& pre-trained models'').

\end{document}

%% file: acronyms_en.tex
\newacronym{lhc}{LHC}{Large Hadron Collider}

\newacronym{desy}{DESY}{\textit{Deutsches Elektronen-Synchrotron}}
\newacronym{slac}{SLAC}{Stanford Linear Accelerator Center}
\newacronym{cern}{CERN}{Conseil Européen de Recherche Nucléaire}

\newacronym{aps}{APS}{American Physical Society}
\newacronym{aip}{AIP}{American Institute of Physics}
\newacronym{pacs}{PACS}{Physics and Astronomy Classification Scheme\textregistered}

\newacronym{hep}{HEP}{High-Energy Physics}

\newacronym{ot}{OT}{Optimal Transport}

%% file: Table1.tex
\begin{longtable}{p{0.15\textwidth}|b{0.425\textwidth}|b{0.425\textwidth}}
\caption{Research areas, their top-words, and their correlation with a standard classification (PACS).}
\label{table:research_areas}\\
\toprule
                    Research area &                                                                                                                                                                                                                                                Top words &                                                                                                                                                                                                                        Most correlated PACS categories \\
\midrule
\endfirsthead
\caption[]{Research areas, their top-words, and their correlation with a standard classification (PACS).} \\
\toprule
                    Research area &                                                                                                                                                                                                                                                Top words &                                                                                                                                                                                                                        Most correlated PACS categories \\
\midrule
\endhead
\midrule
\multicolumn{3}{r}{{Continued on next page}} \\
\midrule
\endfoot

\bottomrule
\endlastfoot
                          AdS/CFT &              boundary, holographic, flow, bulk, critical, conformal, critical\_point, boundary\_theory, cfts, point, bootstrap, central, conformal\_anomaly, strip, fixed, free, entanglement\_entropy, conformal\_field\_theory, criticality, condition &                                               \shortstack[l]{Gauge/string duality (0.27)\\ Conformal field theory, algebraic [...] (0.23)\\ Theory of quantized fields (0.15)\\ Theories and models of [...] (0.14)\\ Critical point phenomena (0.14)}\\ \hline
                      Black holes &                                            hole, black\_hole, gravity, black, horizon, geometry, gravitational, spacetimes, spacetime, curvature, thermodynamics, einstein, schwarzschild, metric, ad, hair, relativity, space\_time, observer, graviton &                                                    \shortstack[l]{Quantum aspects of black holes, [...] (0.51)\\ Classical black holes (0.41)\\ Physics of black holes (0.32)\\ Exact solutions (0.24)\\ Higher-dimensional black holes, [...] (0.23)}\\ \hline
                 Classical fields &                                                    scalar, general, first, scalar\_field, massless, real, explicit, dynamical, second, exact, special, linear, full, symmetric, static, electromagnetic, classical, nonlinear, approximate, non\_trivial &                                  \shortstack[l]{Modified theories of gravity (0.19)\\ Lorentz and Poincaré invariance (0.16)\\ Nonlinear or nonlocal theories and [...] (0.11)\\ Exact solutions (0.10)\\ Higher-dimensional gravity and [...] (0.10)}\\ \hline
                 Collider physics &                      distribution, collision, production, cross\_sections, section, parton, hadron, cross, cross\_section, process, hadronic\_collision, scattering, correction, fragmentation, partons, kinematics, transverse, impact, event, partonic &                 \shortstack[l]{Perturbative calculations (0.29)\\ Polarization in interactions and [...] (0.28)\\ Inclusive production with [...] (0.27)\\ Total and inclusive cross sections [...] (0.26)\\ Relativistic heavy-ion collisions (0.25)}\\ \hline
                        Cosmology &                           constant, cosmological, inflation, cosmic, perturbation, vacuum, universe, inflationary, cosmology, fluctuation, inhomogeneity, tension, lambda, planck, inflaton, cosmological\_perturbation, era, epoch, density, background &                                                                 \shortstack[l]{Particle-theory and field-theory [...] (0.59)\\ Cosmology (0.32)\\ Observational cosmology (including [...] (0.28)\\ Dark energy (0.25)\\ Background radiations (0.21)}\\ \hline
                      Dark matter &                                                matter, dark, dark\_matter, detection, dm, signal, abundance, observation, relic, direct, constraint, candidate, wimp, asymmetric, prospect, dark\_matter\_particle, center, detectable, cold, contribute &                                                                                                     \shortstack[l]{Dark matter (0.74)\\ γ-ray (0.22)\\ Cosmic rays (0.19)\\ γ-ray sources; γ-ray bursts (0.17)\\ Elementary particle processes (0.17)}\\ \hline
               Electroweak sector &                                                 standard, higgs, boson, particle, standard\_model, physic, lhc, sm, top, tev, collider, mssm, electroweak, minimal, phenomenology, extension, extra, supersymmetric\_model, superpartners, new\_particle &                             \shortstack[l]{Extensions of electroweak Higgs sector (0.34)\\ Supersymmetric models (0.33)\\ Non-standard-model Higgs bosons (0.30)\\ Supersymmetric partners of known [...] (0.28)\\ Standard-model Higgs bosons (0.27)}\\ \hline
Gauge theory \& Grand Unification &                                                                              dimension, coupling, scale, structure, operator, fermion, value, matrix, number, su, charge, sector, spin, group, topological, anomalous, breaking, anomaly, global, flavor & \shortstack[l]{Unified theories and models of [...] (0.22)\\ Unification of couplings; mass relations (0.17)\\ Quark and lepton masses and mixing (0.13)\\ Unified field theories and models (0.12)\\ Field theories in dimensions other [...] (0.12)}\\ \hline
                          Hadrons &                                                      decay, data, channel, bound, resonance, gamma, meson, width, experimental\_data, collaboration, kaon, prediction, experimental, measurement, admixture, narrow, process, hadronic\_decay, s0, ratio &                                                        \shortstack[l]{Decays of bottom mesons (0.30)\\ Decays of J/ψ, Υ, and other quarkonia (0.24)\\ Meson-meson interactions (0.21)\\ Decays of bottom mesons (0.20)\\ Bottom mesons (|B|>0) (0.19)}\\ \hline
     Neutrinos \& flavour physics &                                neutrino, violation, oscillation, flavor, cp, angle, mixing, experiment, lepton, flavour, hierarchy, majorana, cp\_violation, beta, leptogenesis, asymmetry, neutrino\_mass, neutrino\_oscillation, smallness, generation &                                                    \shortstack[l]{Neutrino mass and mixing (0.74)\\ Non-standard-model neutrinos, [...] (0.41)\\ Ordinary neutrinos (0.30)\\ Neutrino interactions (0.28)\\ Quark and lepton masses and mixing (0.23)}\\ \hline
             Perturbative methods &                                  amplitude, qcd, loop, diagram, sum, contribution, perturbative, expansion, vertex, rule, light\_cone, perturbative\_qcd, propagator, approach, correlator, one\_loop, evaluation, nonperturbative, kernel, diagrammatic &                                             \shortstack[l]{General properties of perturbation [...] (0.25)\\ Other nonperturbative calculations (0.24)\\ Sum rules (0.22)\\ Perturbative calculations (0.21)\\ General properties of QCD [...] (0.16)}\\ \hline
                              QCD &                                               quark, chiral, magnetic, baryon, relativistic, moment, qcd, light\_quark, strong, heavy, heavy\_quark, lattice, magnetic\_field, electric, deconfinement, chromodynamics, current, diquarks, plasma, color &                                                                            \shortstack[l]{Lattice QCD calculations (0.27)\\ Chiral symmetries (0.26)\\ Chiral Lagrangians (0.25)\\ Quark-gluon plasma (0.23)\\ General properties of QCD [...] (0.20)}\\ \hline
             Quantum Field Theory & quantum, group, quantum\_field, representation, quantisation, mechanic, quantum\_field\_theory, transformation, hamiltonians, algebra, finite\_dimensional, quantization, commutator, algebraic, arbitrary, operator, qft, invariant, analog, associated &                                                                                      \shortstack[l]{Algebraic methods (0.26)\\ Noncommutative field theory (0.25)\\ Quantum mechanics (0.22)\\ Noncommutative geometry (0.19)\\ Quantum groups (0.18)}\\ \hline
    String theory \& supergravity &                                   string, supersymmetric, superstring, six\_dimensional, modulus, super, instantons, supergravity, dyons, n2, mathcaln, superpotentials, heterotic, sigma\_models, n1, n4, gauged, space, deformation, compactifications &                                                                                                     \shortstack[l]{Supersymmetry (0.31)\\ Strings and branes (0.29)\\ Supergravity (0.29)\\ Compactification and four- [...] (0.25)\\ D branes (0.20)}\\ \hline
                   Thermodynamics &               potential, effective, interaction, limit, temperature, action, finite, local, freedom, approximation, level, weak, chemical, force, effective\_field\_theory, lagrangian, finite\_temperature, effective\_field, degree, effective\_theory &                                                                         \shortstack[l]{Finite-temperature field theory (0.26)\\ Chiral symmetries (0.09)\\ Nuclear forces (0.08)\\ Quark-gluon plasma (0.08)\\ General properties of QCD [...] (0.08)}\\ \hline
                  Uninterpretable &                                                 approach, method, analysis, recent, calculation, numerical, formalism, study, prediction, sigma, previous, work, theoretical, systematic, comparison, uncertainty, agreement, good, investigation, paper &                                                \shortstack[l]{Lattice QCD calculations (0.07)\\ Baryon resonances (S=C=B=0) (0.05)\\ Other nonperturbative calculations (0.05)\\ Few-body systems (0.05)\\ Lagrangian and Hamiltonian approach (0.05)}\\ \hline
                  Uninterpretable &                                                                        solution, equation, phase, space, time, system, transition, region, condition, constraint, dynamic, class, background, configuration, wave, range, motion, set, star, instability &                                       \shortstack[l]{Exact solutions (0.14)\\ Nonlinear or nonlocal theories and [...] (0.11)\\ Extended classical solutions; [...] (0.10)\\ Relativistic wave equations (0.10)\\ Modified theories of gravity (0.09)}\\ \hline
                  Uninterpretable &                                                                 form, correction, momentum, tensor, mode, relation, higher, factor, vector, invariant, formula, angular, part, theorem, spectrum, power, dimensional, invariance, expression, derivative &                                                                 \shortstack[l]{Electromagnetic form factors (0.17)\\ Protons and neutrons (0.10)\\ Lorentz and Poincaré invariance (0.08)\\ Gauge field theories (0.06)\\ Dispersion relations (0.06)}\\ \hline
                  Uninterpretable &                                                     spectrum, low, problem, low\_energy, important, high, property, high\_energy, small, physical, soft, fundamental, behavior, analytic, behaviour, spectral, dispersion, essential, phenomenon, regime &                                                            \shortstack[l]{General properties of QCD [...] (0.07)\\ Regge formalism (0.07)\\ Wave propagation, transmission and [...] (0.05)\\ Elastic scattering (0.05)\\ Lattice gauge theory (0.05)}\\ \hline
                  Uninterpretable &                                                          different, possible, particular, present, various, mechanism, type, example, massive, several, scenario, simple, single, similar, consistent, addition, hand, different\_type, interesting, way &                                                          \shortstack[l]{Particle-theory and field-theory [...] (0.07)\\ Modified theories of gravity (0.07)\\ Field theories in dimensions other [...] (0.05)\\ Cosmology (0.05)\\ Dark energy (0.05)}\\ \hline
\end{longtable}

%% file: Table2.tex
\begin{table}[H]
\caption{Performance of the actual model versus that of the baseline model for various corpora, temporal segmentations, topic model parameters, and authorship criteria.}
\label{table:performance}
\begin{tabular}{llrrrlll}
\toprule
Corpus & Authorship & $K$ & $L$ & \makecell{Cohort \\ size} & \makecell{Temporal \\ segmentation} & \makecell{Model \\ $\mu(d_{{\mathrm{{TV}}}}(\bm{{y_{{a}}}}, \bm{{y_{{a}}}}^{{\text{{pred}}}}))$} & \makecell{Baseline \\ $\mu(d_{{\mathrm{{TV}}}}(\bm{{y_{{a}}}}, \bm{{x_{{a}}}}))$} \\
\midrule
\gls{hep} & Any & 20 & 50 & 2108 & \makecell{2000--2009 \\ 2015--2019} & \textbf{0.306} & 0.316 \\
\gls{hep} & 1st/last & 20 & 50 & 2108 & \makecell{2000--2009 \\ 2015--2019} & \textbf{0.306} & 0.316 \\
\gls{hep} & Any & 20 & 50 & 1836 & \makecell{2000--2004 \\ 2005--2009} & \textbf{0.262} & 0.262 \\
\gls{hep} & Any & 20 & 50 & 2530 & \makecell{2005--2009 \\ 2010--2014} & \textbf{0.261} & 0.265 \\
\gls{hep} & Any & 20 & 50 & 3816 & \makecell{2010--2014 \\ 2015--2019} & 0.246 & \textbf{0.244} \\
\gls{hep} & Any & 15 & 50 & 2375 & \makecell{2000--2009 \\ 2015--2019} & \textbf{0.293} & 0.297 \\
\gls{hep} & Any & 25 & 50 & 2109 & \makecell{2000--2009 \\ 2010--2019} & \textbf{0.315} & 0.328 \\
\gls{hep} & Any & 15 & 50 & 2069 & \makecell{2000--2009 \\ 2015--2019} & \textbf{0.290} & 0.295 \\
\gls{hep} & Any & 20 & 150 & 2169 & \makecell{2000--2009 \\ 2015--2019} & \textbf{0.309} & 0.318 \\
ACL Anthology & Any & 20 & 50 & 578 & \makecell{2002--2011 \\ 2012--2022} & \textbf{0.337} & 0.466 \\
\bottomrule
\end{tabular}
\end{table}

%% file: Table3.tex
\begin{table}[H]
\caption{Effect of each variable on (a) the change score and (b) the cognitive distance for each model. The reference model uses entropy as the diversity measure $D$ and the magnitude of intellectual capital as a measure of power $P$. Values indicate the mean posterior effect size and the 95\% credible interval. Significant effects are shown in bold.}
\label{table:summary_change_disruption}
\renewcommand{\arraystretch}{2}\fontsize{6}{7}\selectfont\begin{tabular}{lllllll}
\toprule
Dep. variable & \multicolumn{3}{c}{Change score ($c_a$)} & \multicolumn{3}{c}{Cognitive distance ($d_a$)} \\
Model & Reference & $D=\text{Stirling}$ & $P=\text{Brokerage}$ & Reference & $D=\text{Stirling}$ & $P=\text{Brokerage}$ \\
Predictor &  &  &  &  &  &  \\
\midrule
\textbf{Intellectual capital (diversity)} & $\bm{+0.28}\substack{+0.044 \\ -0.044}$ & $\bm{+0.28}\substack{+0.042 \\ -0.043}$ & $\bm{+0.27}\substack{+0.044 \\ -0.043}$ & $\bm{+0.33}\substack{+0.043 \\ -0.042}$ & $\bm{+0.34}\substack{+0.042 \\ -0.042}$ & $\bm{+0.32}\substack{+0.043 \\ -0.043}$ \\
\textbf{Social capital (diversity)} & $\bm{+0.09}\substack{+0.04 \\ -0.04}$ & $\bm{+0.07}\substack{+0.04 \\ -0.04}$ & $\bm{+0.08}\substack{+0.04 \\ -0.04}$ & $\bm{+0.11}\substack{+0.04 \\ -0.041}$ & $\bm{+0.09}\substack{+0.04 \\ -0.04}$ & $\bm{+0.1}\substack{+0.04 \\ -0.04}$ \\
\textbf{Social capital (power)} & $\bm{-0.09}\substack{+0.06 \\ -0.06}$ & $\bm{-0.07}\substack{+0.06 \\ -0.06}$ & $-0.02\substack{+0.05 \\ -0.05}$ & $\bm{-0.14}\substack{+0.061 \\ -0.061}$ & $\bm{-0.12}\substack{+0.06 \\ -0.06}$ & $-0.05\substack{+0.05 \\ -0.05}$ \\
\textbf{Stable affiliation} & $-0.01\substack{+0.09 \\ -0.09}$ & $-0.009\substack{+0.09 \\ -0.09}$ & $-0.0008\substack{+0.09 \\ -0.09}$ & $-0.007\substack{+0.09 \\ -0.09}$ & $+0.0009\substack{+0.09 \\ -0.09}$ & $+0.01\substack{+0.09 \\ -0.09}$ \\
\textbf{Academic age} & $\bm{-0.1}\substack{+0.05 \\ -0.05}$ & $\bm{-0.1}\substack{+0.05 \\ -0.05}$ & $\bm{-0.1}\substack{+0.047 \\ -0.047}$ & $\bm{-0.07}\substack{+0.05 \\ -0.05}$ & $\bm{-0.07}\substack{+0.047 \\ -0.047}$ & $\bm{-0.08}\substack{+0.05 \\ -0.05}$ \\
\textbf{Productivity (co-authored)} & $\bm{-0.12}\substack{+0.058 \\ -0.059}$ & $\bm{-0.12}\substack{+0.058 \\ -0.058}$ & $\bm{-0.17}\substack{+0.052 \\ -0.053}$ & $\bm{-0.1}\substack{+0.06 \\ -0.06}$ & $\bm{-0.1}\substack{+0.058 \\ -0.056}$ & $\bm{-0.17}\substack{+0.053 \\ -0.052}$ \\
\textbf{Productivity (solo-authored)} & $\bm{-0.05}\substack{+0.041 \\ -0.04}$ & $\bm{-0.05}\substack{+0.04 \\ -0.04}$ & $\bm{-0.06}\substack{+0.04 \\ -0.04}$ & $-0.04\substack{+0.04 \\ -0.04}$ & $-0.03\substack{+0.04 \\ -0.04}$ & $\bm{-0.04}\substack{+0.04 \\ -0.04}$ \\
\hline Hadrons & $-0.009\substack{+0.2 \\ -0.2}$ & $-0.11\substack{+0.18 \\ -0.2}$ & $-0.008\substack{+0.2 \\ -0.2}$ & $+0.03\substack{+0.1 \\ -0.1}$ & $-0.05\substack{+0.1 \\ -0.2}$ & $+0.04\substack{+0.1 \\ -0.1}$ \\
String theory \& supergravity & $\bm{+0.28}\substack{+0.15 \\ -0.15}$ & $\bm{+0.34}\substack{+0.18 \\ -0.18}$ & $\bm{+0.25}\substack{+0.15 \\ -0.15}$ & $+0.11\substack{+0.13 \\ -0.11}$ & $\bm{+0.21}\substack{+0.15 \\ -0.15}$ & $+0.07\substack{+0.1 \\ -0.1}$ \\
Perturbative methods & $+0.12\substack{+0.22 \\ -0.18}$ & $+0.06\substack{+0.2 \\ -0.2}$ & $+0.13\substack{+0.22 \\ -0.18}$ & $+0.15\substack{+0.22 \\ -0.17}$ & $+0.1\substack{+0.21 \\ -0.17}$ & $+0.16\substack{+0.23 \\ -0.18}$ \\
Classical fields & $-0.25\substack{+0.36 \\ -0.59}$ & $-0.21\substack{+0.38 \\ -0.55}$ & $-0.23\substack{+0.35 \\ -0.57}$ & $-0.19\substack{+0.27 \\ -0.58}$ & $-0.17\substack{+0.3 \\ -0.51}$ & $-0.16\substack{+0.26 \\ -0.55}$ \\
Collider physics & $\bm{-0.19}\substack{+0.16 \\ -0.17}$ & $\bm{-0.34}\substack{+0.19 \\ -0.19}$ & $\bm{-0.2}\substack{+0.17 \\ -0.17}$ & $-0.02\substack{+0.1 \\ -0.1}$ & $\bm{-0.16}\substack{+0.15 \\ -0.16}$ & $-0.03\substack{+0.1 \\ -0.1}$ \\
Neutrinos \& flavour physics & $\bm{+0.21}\substack{+0.18 \\ -0.17}$ & $+0.17\substack{+0.2 \\ -0.18}$ & $\bm{+0.18}\substack{+0.18 \\ -0.17}$ & $+0.11\substack{+0.16 \\ -0.13}$ & $+0.1\substack{+0.2 \\ -0.1}$ & $+0.09\substack{+0.2 \\ -0.1}$ \\
Black holes & $+0.06\substack{+0.2 \\ -0.2}$ & $+0.15\substack{+0.22 \\ -0.19}$ & $+0.05\substack{+0.2 \\ -0.2}$ & $-0.03\substack{+0.1 \\ -0.2}$ & $+0.06\substack{+0.2 \\ -0.1}$ & $-0.04\substack{+0.1 \\ -0.2}$ \\
Gauge theory \& Grand Unification & $-0.07\substack{+0.3 \\ -0.4}$ & $-0.05\substack{+0.3 \\ -0.4}$ & $-0.06\substack{+0.3 \\ -0.3}$ & $-0.02\substack{+0.2 \\ -0.3}$ & $+0.003\substack{+0.3 \\ -0.3}$ & $-0.008\substack{+0.2 \\ -0.3}$ \\
Dark matter & $\bm{-0.27}\substack{+0.24 \\ -0.25}$ & $\bm{-0.32}\substack{+0.26 \\ -0.27}$ & $\bm{-0.28}\substack{+0.24 \\ -0.25}$ & $-0.13\substack{+0.17 \\ -0.23}$ & $-0.18\substack{+0.2 \\ -0.24}$ & $-0.13\substack{+0.17 \\ -0.23}$ \\
Thermodynamics & $+0.14\substack{+0.36 \\ -0.26}$ & $+0.27\substack{+0.41 \\ -0.32}$ & $+0.16\substack{+0.37 \\ -0.27}$ & $+0.1\substack{+0.34 \\ -0.2}$ & $+0.25\substack{+0.41 \\ -0.29}$ & $+0.12\substack{+0.36 \\ -0.21}$ \\
Cosmology & $-0.02\substack{+0.16 \\ -0.17}$ & $+0.01\substack{+0.2 \\ -0.2}$ & $-0.03\substack{+0.2 \\ -0.2}$ & $-0.06\substack{+0.1 \\ -0.2}$ & $-0.02\substack{+0.1 \\ -0.2}$ & $-0.06\substack{+0.13 \\ -0.17}$ \\
Electroweak sector & $-0.14\substack{+0.15 \\ -0.16}$ & $\bm{-0.21}\substack{+0.18 \\ -0.18}$ & $\bm{-0.17}\substack{+0.16 \\ -0.16}$ & $-0.04\substack{+0.1 \\ -0.1}$ & $-0.1\substack{+0.13 \\ -0.15}$ & $-0.07\substack{+0.1 \\ -0.1}$ \\
QCD & $-0.009\substack{+0.2 \\ -0.2}$ & $-0.07\substack{+0.2 \\ -0.2}$ & $+0.001\substack{+0.2 \\ -0.2}$ & $+0.02\substack{+0.1 \\ -0.1}$ & $-0.03\substack{+0.14 \\ -0.16}$ & $+0.04\substack{+0.1 \\ -0.1}$ \\
Quantum Field Theory & $+0.04\substack{+0.2 \\ -0.2}$ & $+0.09\substack{+0.2 \\ -0.2}$ & $+0.04\substack{+0.21 \\ -0.19}$ & $-0.09\substack{+0.2 \\ -0.2}$ & $-0.04\substack{+0.2 \\ -0.2}$ & $-0.08\substack{+0.2 \\ -0.2}$ \\
AdS/CFT & $+0.003\substack{+0.2 \\ -0.2}$ & $+0.07\substack{+0.3 \\ -0.2}$ & $+0.005\substack{+0.2 \\ -0.2}$ & $+0.02\substack{+0.2 \\ -0.2}$ & $+0.11\substack{+0.27 \\ -0.2}$ & $+0.03\substack{+0.2 \\ -0.2}$ \\
\bottomrule
\end{tabular}\normalsize\renewcommand{\arraystretch}{1}
\end{table}

%% file: Table4.tex
\begin{table}[H]
\caption{Effect of each variable on (a) the probability of having entered a new research area and (b) the probability of having exited a research area, for each model. The reference model uses entropy as the diversity measure $D$ and the magnitude of intellectual capital as a measure of power $P$. Values indicate the mean posterior effect size and the 95\% credible interval. Significant effects are shown in bold.}
\label{table:summary_entered_exited}
\renewcommand{\arraystretch}{2}\fontsize{6}{7}\selectfont\begin{tabular}{lllllll}
\toprule
Dep. variable & \multicolumn{3}{c}{Entered a new research area} & \multicolumn{3}{c}{Exited a research area} \\
Model & Reference & $D=\text{Stirling}$ & $P=\text{Brokerage}$ & Reference & $D=\text{Stirling}$ & $P=\text{Brokerage}$ \\
Predictor &  &  &  &  &  &  \\
\midrule
\textbf{Intellectual capital (diversity)} & $\bm{+0.2}\substack{+0.11 \\ -0.11}$ & $\bm{+0.17}\substack{+0.1 \\ -0.1}$ & $\bm{+0.19}\substack{+0.11 \\ -0.11}$ & $\bm{+1}\substack{+0.14 \\ -0.14}$ & $\bm{+0.85}\substack{+0.12 \\ -0.12}$ & $\bm{+1}\substack{+0.14 \\ -0.14}$ \\
\textbf{Social capital (diversity)} & $\bm{+0.22}\substack{+0.1 \\ -0.1}$ & $\bm{+0.18}\substack{+0.1 \\ -0.1}$ & $\bm{+0.22}\substack{+0.099 \\ -0.1}$ & $+0.04\substack{+0.1 \\ -0.1}$ & $+0.04\substack{+0.1 \\ -0.1}$ & $+0.04\substack{+0.1 \\ -0.1}$ \\
\textbf{Social capital (power)} & $+0.006\substack{+0.1 \\ -0.1}$ & $+0.03\substack{+0.15 \\ -0.15}$ & $+0.04\substack{+0.1 \\ -0.1}$ & $-0.03\substack{+0.2 \\ -0.2}$ & $+0.02\substack{+0.2 \\ -0.2}$ & $+0.03\substack{+0.1 \\ -0.1}$ \\
\textbf{Stable affiliation} & $-0.19\substack{+0.22 \\ -0.22}$ & $-0.18\substack{+0.22 \\ -0.22}$ & $-0.19\substack{+0.22 \\ -0.22}$ & $+0.04\substack{+0.2 \\ -0.2}$ & $+0.06\substack{+0.2 \\ -0.2}$ & $+0.04\substack{+0.24 \\ -0.24}$ \\
\textbf{Academic age} & $+0.04\substack{+0.12 \\ -0.11}$ & $+0.04\substack{+0.1 \\ -0.1}$ & $+0.04\substack{+0.1 \\ -0.1}$ & $\bm{-0.21}\substack{+0.12 \\ -0.12}$ & $\bm{-0.21}\substack{+0.12 \\ -0.12}$ & $\bm{-0.22}\substack{+0.12 \\ -0.12}$ \\
\textbf{Productivity (co-authored)} & $-0.07\substack{+0.1 \\ -0.1}$ & $-0.08\substack{+0.1 \\ -0.1}$ & $-0.09\substack{+0.1 \\ -0.1}$ & $\bm{-0.28}\substack{+0.15 \\ -0.14}$ & $\bm{-0.28}\substack{+0.15 \\ -0.15}$ & $\bm{-0.31}\substack{+0.13 \\ -0.13}$ \\
\textbf{Productivity (solo-authored)} & $-0.05\substack{+0.1 \\ -0.09}$ & $-0.05\substack{+0.1 \\ -0.09}$ & $-0.06\substack{+0.1 \\ -0.1}$ & $-0.02\substack{+0.1 \\ -0.1}$ & $-0.007\substack{+0.1 \\ -0.1}$ & $-0.03\substack{+0.1 \\ -0.1}$ \\
\hline Hadrons & $-0.14\substack{+0.29 \\ -0.36}$ & $-0.24\substack{+0.33 \\ -0.39}$ & $-0.14\substack{+0.29 \\ -0.35}$ & $+0.03\substack{+0.3 \\ -0.2}$ & $-0.09\substack{+0.3 \\ -0.4}$ & $+0.04\substack{+0.3 \\ -0.2}$ \\
String theory \& supergravity & $\bm{+0.32}\substack{+0.32 \\ -0.3}$ & $\bm{+0.4}\substack{+0.35 \\ -0.33}$ & $\bm{+0.32}\substack{+0.32 \\ -0.29}$ & $\bm{+0.3}\substack{+0.34 \\ -0.3}$ & $\bm{+0.65}\substack{+0.39 \\ -0.38}$ & $+0.28\substack{+0.34 \\ -0.29}$ \\
Perturbative methods & $+0.11\substack{+0.45 \\ -0.35}$ & $+0.09\substack{+0.5 \\ -0.4}$ & $+0.11\substack{+0.45 \\ -0.35}$ & $-0.03\substack{+0.29 \\ -0.34}$ & $-0.13\substack{+0.4 \\ -0.48}$ & $-0.03\substack{+0.3 \\ -0.3}$ \\
Classical fields & $+0.22\substack{+1.1 \\ -0.6}$ & $+0.34\substack{+1.4 \\ -0.7}$ & $+0.22\substack{+1.1 \\ -0.6}$ & $-0.07\substack{+0.4 \\ -0.6}$ & $-0.07\substack{+0.7 \\ -0.8}$ & $-0.07\substack{+0.4 \\ -0.6}$ \\
Collider physics & $\bm{-0.43}\substack{+0.33 \\ -0.34}$ & $\bm{-0.61}\substack{+0.33 \\ -0.34}$ & $\bm{-0.42}\substack{+0.33 \\ -0.34}$ & $-0.01\substack{+0.2 \\ -0.2}$ & $-0.28\substack{+0.32 \\ -0.37}$ & $-0.02\substack{+0.2 \\ -0.2}$ \\
Neutrinos \& flavour physics & $+0.08\substack{+0.3 \\ -0.3}$ & $+0.04\substack{+0.3 \\ -0.3}$ & $+0.07\substack{+0.3 \\ -0.3}$ & $-0.21\substack{+0.25 \\ -0.35}$ & $-0.31\substack{+0.35 \\ -0.41}$ & $-0.21\substack{+0.26 \\ -0.36}$ \\
Black holes & $-0.0006\substack{+0.3 \\ -0.3}$ & $+0.06\substack{+0.4 \\ -0.3}$ & $-0.003\substack{+0.3 \\ -0.3}$ & $+0.08\substack{+0.4 \\ -0.3}$ & $+0.43\substack{+0.53 \\ -0.45}$ & $+0.08\substack{+0.4 \\ -0.3}$ \\
Gauge theory \& Grand Unification & $-0.04\substack{+0.6 \\ -0.6}$ & $-0.04\substack{+0.6 \\ -0.7}$ & $-0.03\substack{+0.6 \\ -0.6}$ & $-0.08\substack{+0.4 \\ -0.6}$ & $-0.11\substack{+0.64 \\ -0.79}$ & $-0.08\substack{+0.4 \\ -0.6}$ \\
Dark matter & $\bm{-0.62}\substack{+0.55 \\ -0.56}$ & $\bm{-0.68}\substack{+0.55 \\ -0.56}$ & $\bm{-0.63}\substack{+0.55 \\ -0.56}$ & $-0.05\substack{+0.3 \\ -0.4}$ & $-0.11\substack{+0.42 \\ -0.5}$ & $-0.05\substack{+0.3 \\ -0.4}$ \\
Thermodynamics & $-0.03\substack{+0.5 \\ -0.6}$ & $+0.01\substack{+0.62 \\ -0.58}$ & $-0.02\substack{+0.5 \\ -0.6}$ & $-0.05\substack{+0.4 \\ -0.6}$ & $+0.03\substack{+0.7 \\ -0.6}$ & $-0.05\substack{+0.41 \\ -0.53}$ \\
Cosmology & $-0.07\substack{+0.3 \\ -0.4}$ & $-0.03\substack{+0.3 \\ -0.4}$ & $-0.07\substack{+0.3 \\ -0.4}$ & $+0.09\substack{+0.4 \\ -0.3}$ & $+0.36\substack{+0.56 \\ -0.43}$ & $+0.09\substack{+0.4 \\ -0.3}$ \\
Electroweak sector & $+0.05\substack{+0.3 \\ -0.3}$ & $+0.009\substack{+0.3 \\ -0.3}$ & $+0.05\substack{+0.3 \\ -0.3}$ & $-0.003\substack{+0.2 \\ -0.2}$ & $-0.04\substack{+0.3 \\ -0.3}$ & $-0.009\substack{+0.2 \\ -0.2}$ \\
QCD & $+0.008\substack{+0.3 \\ -0.3}$ & $-0.03\substack{+0.3 \\ -0.4}$ & $+0.01\substack{+0.3 \\ -0.3}$ & $+0.04\substack{+0.32 \\ -0.26}$ & $+0.04\substack{+0.4 \\ -0.4}$ & $+0.04\substack{+0.3 \\ -0.3}$ \\
Quantum Field Theory & $+0.15\substack{+0.52 \\ -0.38}$ & $+0.22\substack{+0.57 \\ -0.41}$ & $+0.15\substack{+0.52 \\ -0.38}$ & $+0.05\substack{+0.4 \\ -0.3}$ & $+0.26\substack{+0.64 \\ -0.44}$ & $+0.05\substack{+0.4 \\ -0.3}$ \\
AdS/CFT & $+0.14\substack{+0.6 \\ -0.42}$ & $+0.22\substack{+0.67 \\ -0.47}$ & $+0.14\substack{+0.59 \\ -0.42}$ & $-0.06\substack{+0.4 \\ -0.5}$ & $+0.07\substack{+0.6 \\ -0.5}$ & $-0.06\substack{+0.3 \\ -0.5}$ \\
\bottomrule
\end{tabular}\normalsize\renewcommand{\arraystretch}{1}
\end{table}

%% file: full_summary.tex
\begin{table}[H]
\caption{Summary of the effect of each predictor on the change score ($c_a$) across topic models and temporal segmentations. Values indicate the mean posterior effect size and the 95\% credible interval. Significant effects are shown in bold.}
\label{table:full_summary_change}
\renewcommand{\arraystretch}{2}\fontsize{6}{7}\selectfont\begin{tabular}{lrrllllllll}
\toprule
\makecell{Author-\\ ship} & $K_0$ & $L$ & \makecell{Temporal \\ segmentation} & \makecell{Intell. capital \\ (diversity)} & \makecell{Soc. capital \\ (diversity)} & \makecell{Soc. capital \\ (power)} & \makecell{Stable \\ affiliation} & \makecell{Academic \\ age} & \makecell{Prod. \\ (co-auth.)} & \makecell{Prod. \\ (solo-auth.)} \\
\midrule
Any & 20 & 50 & \makecell{2000-2009 \\ 2015-2019} & $\bm{+0.28}\substack{+0.044 \\ -0.044}$ & $\bm{+0.09}\substack{+0.04 \\ -0.04}$ & $\bm{-0.09}\substack{+0.06 \\ -0.06}$ & $-0.01\substack{+0.09 \\ -0.09}$ & $\bm{-0.1}\substack{+0.05 \\ -0.05}$ & $\bm{-0.12}\substack{+0.058 \\ -0.059}$ & $\bm{-0.05}\substack{+0.041 \\ -0.04}$ \\
1st/last & 20 & 50 & \makecell{2000-2009 \\ 2015-2019} & $\bm{+0.25}\substack{+0.055 \\ -0.056}$ & $\bm{+0.09}\substack{+0.05 \\ -0.05}$ & $-0.01\substack{+0.08 \\ -0.08}$ & $+0.04\substack{+0.1 \\ -0.1}$ & $\bm{-0.12}\substack{+0.059 \\ -0.06}$ & $\bm{-0.16}\substack{+0.075 \\ -0.074}$ & $-0.05\substack{+0.05 \\ -0.05}$ \\
Any & 20 & 50 & \makecell{2000-2004 \\ 2005-2009} & $\bm{+0.37}\substack{+0.045 \\ -0.045}$ & $\bm{+0.12}\substack{+0.043 \\ -0.043}$ & $-0.02\substack{+0.06 \\ -0.06}$ & $-0.02\substack{+0.1 \\ -0.1}$ & $\bm{-0.08}\substack{+0.05 \\ -0.05}$ & $\bm{-0.21}\substack{+0.059 \\ -0.059}$ & $\bm{-0.06}\substack{+0.04 \\ -0.04}$ \\
Any & 20 & 50 & \makecell{2010-2014 \\ 2015-2019} & $\bm{+0.36}\substack{+0.039 \\ -0.039}$ & $\bm{+0.08}\substack{+0.04 \\ -0.04}$ & $\bm{-0.06}\substack{+0.051 \\ -0.052}$ & $\bm{-0.11}\substack{+0.087 \\ -0.086}$ & $\bm{-0.06}\substack{+0.04 \\ -0.04}$ & $\bm{-0.21}\substack{+0.051 \\ -0.051}$ & $\bm{-0.04}\substack{+0.04 \\ -0.04}$ \\
Any & 20 & 50 & \makecell{2000-2009 \\ 2010-2019} & $\bm{+0.37}\substack{+0.033 \\ -0.033}$ & $\bm{+0.06}\substack{+0.03 \\ -0.03}$ & $\bm{-0.06}\substack{+0.05 \\ -0.05}$ & $-0.02\substack{+0.07 \\ -0.07}$ & $\bm{-0.04}\substack{+0.036 \\ -0.035}$ & $\bm{-0.22}\substack{+0.046 \\ -0.046}$ & $\bm{-0.04}\substack{+0.03 \\ -0.03}$ \\
Any & 20 & 50 & \makecell{2000-2009 \\ 2015-2019} & $\bm{+0.32}\substack{+0.044 \\ -0.044}$ & $\bm{+0.09}\substack{+0.04 \\ -0.04}$ & $\bm{-0.09}\substack{+0.06 \\ -0.06}$ & $+0.02\substack{+0.09 \\ -0.09}$ & $\bm{-0.12}\substack{+0.047 \\ -0.046}$ & $\bm{-0.15}\substack{+0.057 \\ -0.057}$ & $\bm{-0.05}\substack{+0.04 \\ -0.04}$ \\
Any & 15 & 50 & \makecell{2000-2009 \\ 2015-2019} & $\bm{+0.33}\substack{+0.04 \\ -0.04}$ & $\bm{+0.09}\substack{+0.04 \\ -0.04}$ & $\bm{-0.12}\substack{+0.057 \\ -0.057}$ & $-0.04\substack{+0.09 \\ -0.09}$ & $\bm{-0.12}\substack{+0.05 \\ -0.051}$ & $\bm{-0.1}\substack{+0.054 \\ -0.054}$ & $\bm{-0.06}\substack{+0.04 \\ -0.04}$ \\
Any & 25 & 50 & \makecell{2000-2009 \\ 2015-2019} & $\bm{+0.35}\substack{+0.042 \\ -0.042}$ & $\bm{+0.13}\substack{+0.04 \\ -0.04}$ & $\bm{-0.15}\substack{+0.058 \\ -0.058}$ & $-0.002\substack{+0.09 \\ -0.09}$ & $\bm{-0.15}\substack{+0.052 \\ -0.052}$ & $\bm{-0.12}\substack{+0.056 \\ -0.056}$ & $\bm{-0.05}\substack{+0.04 \\ -0.04}$ \\
Any & 15 & 150 & \makecell{2000-2009 \\ 2015-2019} & $\bm{+0.35}\substack{+0.044 \\ -0.044}$ & $\bm{+0.07}\substack{+0.04 \\ -0.04}$ & $\bm{-0.08}\substack{+0.06 \\ -0.06}$ & $-0.02\substack{+0.089 \\ -0.089}$ & $\bm{-0.08}\substack{+0.05 \\ -0.05}$ & $\bm{-0.1}\substack{+0.056 \\ -0.057}$ & $-0.02\substack{+0.04 \\ -0.04}$ \\
Any & 20 & 150 & \makecell{2000-2009 \\ 2015-2019} & $\bm{+0.34}\substack{+0.044 \\ -0.043}$ & $\bm{+0.08}\substack{+0.04 \\ -0.04}$ & $\bm{-0.09}\substack{+0.06 \\ -0.06}$ & $+0.008\substack{+0.09 \\ -0.09}$ & $\bm{-0.1}\substack{+0.046 \\ -0.046}$ & $\bm{-0.12}\substack{+0.056 \\ -0.056}$ & $\bm{-0.06}\substack{+0.04 \\ -0.04}$ \\
Any & 25 & 150 & \makecell{2000-2009 \\ 2015-2019} & $\bm{+0.34}\substack{+0.044 \\ -0.044}$ & $\bm{+0.06}\substack{+0.04 \\ -0.04}$ & $\bm{-0.09}\substack{+0.06 \\ -0.06}$ & $-0.009\substack{+0.09 \\ -0.09}$ & $\bm{-0.11}\substack{+0.047 \\ -0.047}$ & $\bm{-0.13}\substack{+0.057 \\ -0.057}$ & $-0.02\substack{+0.04 \\ -0.04}$ \\
\bottomrule
\end{tabular}\normalsize\renewcommand{\arraystretch}{1}
\end{table}

\begin{table}[H]
\caption{Summary of the effect of each predictor on the cognitive distance ($d_a$) across topic models and temporal segmentations. Values indicate the mean posterior effect size and the 95\% credible interval. Significant effects are shown in bold.}
\label{table:full_summary_disruption}
\renewcommand{\arraystretch}{2}\fontsize{6}{7}\selectfont\begin{tabular}{lrrllllllll}
\toprule
\makecell{Author-\\ ship} & $K_0$ & $L$ & \makecell{Temporal \\ segmentation} & \makecell{Intell. capital \\ (diversity)} & \makecell{Soc. capital \\ (diversity)} & \makecell{Soc. capital \\ (power)} & \makecell{Stable \\ affiliation} & \makecell{Academic \\ age} & \makecell{Prod. \\ (co-auth.)} & \makecell{Prod. \\ (solo-auth.)} \\
\midrule
Any & 20 & 50 & \makecell{2000-2009 \\ 2015-2019} & $\bm{+0.33}\substack{+0.043 \\ -0.042}$ & $\bm{+0.11}\substack{+0.04 \\ -0.041}$ & $\bm{-0.14}\substack{+0.061 \\ -0.061}$ & $-0.007\substack{+0.09 \\ -0.09}$ & $\bm{-0.07}\substack{+0.05 \\ -0.05}$ & $\bm{-0.1}\substack{+0.06 \\ -0.06}$ & $-0.04\substack{+0.04 \\ -0.04}$ \\
1st/last & 20 & 50 & \makecell{2000-2009 \\ 2015-2019} & $\bm{+0.3}\substack{+0.056 \\ -0.055}$ & $\bm{+0.1}\substack{+0.053 \\ -0.052}$ & $-0.06\substack{+0.08 \\ -0.08}$ & $-0.01\substack{+0.1 \\ -0.1}$ & $\bm{-0.07}\substack{+0.06 \\ -0.06}$ & $\bm{-0.12}\substack{+0.076 \\ -0.075}$ & $-0.04\substack{+0.05 \\ -0.05}$ \\
Any & 20 & 50 & \makecell{2000-2004 \\ 2005-2009} & $\bm{+0.37}\substack{+0.044 \\ -0.044}$ & $\bm{+0.13}\substack{+0.043 \\ -0.044}$ & $-0.05\substack{+0.06 \\ -0.06}$ & $-0.04\substack{+0.1 \\ -0.1}$ & $\bm{-0.07}\substack{+0.05 \\ -0.05}$ & $\bm{-0.19}\substack{+0.06 \\ -0.06}$ & $\bm{-0.07}\substack{+0.04 \\ -0.04}$ \\
Any & 20 & 50 & \makecell{2010-2014 \\ 2015-2019} & $\bm{+0.37}\substack{+0.037 \\ -0.037}$ & $\bm{+0.09}\substack{+0.04 \\ -0.04}$ & $\bm{-0.07}\substack{+0.05 \\ -0.05}$ & $\bm{-0.12}\substack{+0.086 \\ -0.086}$ & $-0.03\substack{+0.04 \\ -0.04}$ & $\bm{-0.21}\substack{+0.05 \\ -0.051}$ & $-0.04\substack{+0.04 \\ -0.04}$ \\
Any & 20 & 50 & \makecell{2000-2009 \\ 2010-2019} & $\bm{+0.4}\substack{+0.031 \\ -0.031}$ & $\bm{+0.06}\substack{+0.031 \\ -0.031}$ & $\bm{-0.09}\substack{+0.05 \\ -0.05}$ & $+0.01\substack{+0.07 \\ -0.07}$ & $\bm{-0.05}\substack{+0.035 \\ -0.035}$ & $\bm{-0.19}\substack{+0.046 \\ -0.046}$ & $\bm{-0.03}\substack{+0.03 \\ -0.03}$ \\
Any & 20 & 50 & \makecell{2000-2009 \\ 2015-2019} & $\bm{+0.36}\substack{+0.043 \\ -0.043}$ & $\bm{+0.11}\substack{+0.04 \\ -0.04}$ & $\bm{-0.14}\substack{+0.059 \\ -0.06}$ & $+0.003\substack{+0.09 \\ -0.09}$ & $\bm{-0.09}\substack{+0.05 \\ -0.05}$ & $\bm{-0.14}\substack{+0.056 \\ -0.056}$ & $\bm{-0.05}\substack{+0.04 \\ -0.04}$ \\
Any & 15 & 50 & \makecell{2000-2009 \\ 2015-2019} & $\bm{+0.28}\substack{+0.041 \\ -0.041}$ & $\bm{+0.1}\substack{+0.04 \\ -0.04}$ & $\bm{-0.18}\substack{+0.057 \\ -0.058}$ & $-0.04\substack{+0.09 \\ -0.09}$ & $\bm{-0.09}\substack{+0.05 \\ -0.05}$ & $-0.05\substack{+0.05 \\ -0.06}$ & $-0.03\substack{+0.04 \\ -0.04}$ \\
Any & 25 & 50 & \makecell{2000-2009 \\ 2015-2019} & $\bm{+0.25}\substack{+0.043 \\ -0.043}$ & $\bm{+0.12}\substack{+0.041 \\ -0.04}$ & $\bm{-0.18}\substack{+0.059 \\ -0.058}$ & $+0.06\substack{+0.09 \\ -0.09}$ & $\bm{-0.12}\substack{+0.053 \\ -0.053}$ & $\bm{-0.06}\substack{+0.06 \\ -0.06}$ & $-0.03\substack{+0.04 \\ -0.04}$ \\
Any & 15 & 150 & \makecell{2000-2009 \\ 2015-2019} & $\bm{+0.28}\substack{+0.048 \\ -0.047}$ & $\bm{+0.07}\substack{+0.04 \\ -0.04}$ & $\bm{-0.09}\substack{+0.06 \\ -0.06}$ & $+0.03\substack{+0.09 \\ -0.1}$ & $\bm{-0.07}\substack{+0.05 \\ -0.05}$ & $\bm{-0.08}\substack{+0.06 \\ -0.06}$ & $-0.02\substack{+0.04 \\ -0.04}$ \\
Any & 20 & 150 & \makecell{2000-2009 \\ 2015-2019} & $\bm{+0.27}\substack{+0.046 \\ -0.045}$ & $\bm{+0.04}\substack{+0.04 \\ -0.04}$ & $\bm{-0.1}\substack{+0.062 \\ -0.061}$ & $+0.06\substack{+0.09 \\ -0.09}$ & $\bm{-0.1}\substack{+0.05 \\ -0.05}$ & $\bm{-0.09}\substack{+0.06 \\ -0.06}$ & $\bm{-0.04}\substack{+0.04 \\ -0.04}$ \\
Any & 25 & 150 & \makecell{2000-2009 \\ 2015-2019} & $\bm{+0.26}\substack{+0.047 \\ -0.047}$ & $\bm{+0.07}\substack{+0.043 \\ -0.043}$ & $\bm{-0.15}\substack{+0.063 \\ -0.064}$ & $+0.02\substack{+0.09 \\ -0.09}$ & $\bm{-0.08}\substack{+0.05 \\ -0.05}$ & $\bm{-0.08}\substack{+0.06 \\ -0.06}$ & $-0.003\substack{+0.04 \\ -0.04}$ \\
\bottomrule
\end{tabular}\normalsize\renewcommand{\arraystretch}{1}
\end{table}

\begin{table}[H]
\caption{Summary of the effect of each predictor on the probability of having entered a research area across topic models and temporal segmentations. Values indicate the mean posterior effect size and the 95\% credible interval. Significant effects are shown in bold.}
\label{table:full_summary_entered}
\renewcommand{\arraystretch}{2}\fontsize{6}{7}\selectfont\begin{tabular}{lrrllllllll}
\toprule
\makecell{Author-\\ ship} & $K_0$ & $L$ & \makecell{Temporal \\ segmentation} & \makecell{Intell. capital \\ (diversity)} & \makecell{Soc. capital \\ (diversity)} & \makecell{Soc. capital \\ (power)} & \makecell{Stable \\ affiliation} & \makecell{Academic \\ age} & \makecell{Prod. \\ (co-auth.)} & \makecell{Prod. \\ (solo-auth.)} \\
\midrule
Any & 20 & 50 & \makecell{2000-2009 \\ 2015-2019} & $\bm{+0.2}\substack{+0.11 \\ -0.11}$ & $\bm{+0.22}\substack{+0.1 \\ -0.1}$ & $+0.006\substack{+0.1 \\ -0.1}$ & $-0.19\substack{+0.22 \\ -0.22}$ & $+0.04\substack{+0.12 \\ -0.11}$ & $-0.07\substack{+0.1 \\ -0.1}$ & $-0.05\substack{+0.1 \\ -0.09}$ \\
1st/last & 20 & 50 & \makecell{2000-2009 \\ 2015-2019} & $\bm{+0.25}\substack{+0.14 \\ -0.13}$ & $\bm{+0.19}\substack{+0.13 \\ -0.13}$ & $\bm{+0.21}\substack{+0.19 \\ -0.19}$ & $+0.12\substack{+0.29 \\ -0.29}$ & $-0.07\substack{+0.1 \\ -0.1}$ & $\bm{-0.33}\substack{+0.18 \\ -0.18}$ & $-0.0006\substack{+0.1 \\ -0.1}$ \\
Any & 20 & 50 & \makecell{2000-2004 \\ 2005-2009} & $\bm{+0.16}\substack{+0.11 \\ -0.11}$ & $\bm{+0.2}\substack{+0.11 \\ -0.11}$ & $+0.01\substack{+0.15 \\ -0.15}$ & $-0.007\substack{+0.2 \\ -0.2}$ & $-0.06\substack{+0.1 \\ -0.1}$ & $\bm{-0.14}\substack{+0.14 \\ -0.14}$ & $-0.05\substack{+0.1 \\ -0.1}$ \\
Any & 20 & 50 & \makecell{2010-2014 \\ 2015-2019} & $\bm{+0.24}\substack{+0.095 \\ -0.095}$ & $\bm{+0.18}\substack{+0.091 \\ -0.091}$ & $-0.06\substack{+0.1 \\ -0.1}$ & $+0.05\substack{+0.2 \\ -0.21}$ & $+0.07\substack{+0.1 \\ -0.1}$ & $\bm{-0.18}\substack{+0.12 \\ -0.12}$ & $-0.02\substack{+0.09 \\ -0.09}$ \\
Any & 20 & 50 & \makecell{2000-2009 \\ 2010-2019} & $\bm{+0.27}\substack{+0.083 \\ -0.083}$ & $\bm{+0.27}\substack{+0.076 \\ -0.076}$ & $-0.01\substack{+0.1 \\ -0.1}$ & $-0.08\substack{+0.2 \\ -0.2}$ & $+0.04\substack{+0.09 \\ -0.09}$ & $\bm{-0.16}\substack{+0.11 \\ -0.11}$ & $-0.004\substack{+0.07 \\ -0.07}$ \\
Any & 20 & 50 & \makecell{2000-2009 \\ 2015-2019} & $\bm{+0.24}\substack{+0.11 \\ -0.11}$ & $\bm{+0.2}\substack{+0.1 \\ -0.1}$ & $+0.03\substack{+0.15 \\ -0.14}$ & $-0.07\substack{+0.2 \\ -0.2}$ & $-0.01\substack{+0.1 \\ -0.1}$ & $\bm{-0.19}\substack{+0.13 \\ -0.14}$ & $-0.06\substack{+0.1 \\ -0.09}$ \\
Any & 15 & 50 & \makecell{2000-2009 \\ 2015-2019} & $\bm{+0.13}\substack{+0.11 \\ -0.1}$ & $\bm{+0.13}\substack{+0.096 \\ -0.095}$ & $-0.005\substack{+0.1 \\ -0.1}$ & $-0.03\substack{+0.2 \\ -0.2}$ & $-0.04\substack{+0.1 \\ -0.1}$ & $\bm{-0.16}\substack{+0.13 \\ -0.13}$ & $+0.002\substack{+0.09 \\ -0.09}$ \\
Any & 25 & 50 & \makecell{2000-2009 \\ 2015-2019} & $\bm{+0.18}\substack{+0.12 \\ -0.12}$ & $\bm{+0.17}\substack{+0.11 \\ -0.11}$ & $-0.009\substack{+0.2 \\ -0.2}$ & $-0.07\substack{+0.2 \\ -0.2}$ & $-0.04\substack{+0.1 \\ -0.1}$ & $\bm{-0.15}\substack{+0.15 \\ -0.15}$ & $-0.02\substack{+0.1 \\ -0.1}$ \\
Any & 15 & 150 & \makecell{2000-2009 \\ 2015-2019} & $\bm{+0.12}\substack{+0.1 \\ -0.1}$ & $\bm{+0.19}\substack{+0.095 \\ -0.095}$ & $-0.08\substack{+0.1 \\ -0.1}$ & $-0.02\substack{+0.2 \\ -0.2}$ & $-0.02\substack{+0.1 \\ -0.1}$ & $-0.04\substack{+0.1 \\ -0.1}$ & $-0.06\substack{+0.09 \\ -0.09}$ \\
Any & 20 & 150 & \makecell{2000-2009 \\ 2015-2019} & $\bm{+0.23}\substack{+0.11 \\ -0.11}$ & $+0.07\substack{+0.1 \\ -0.1}$ & $+0.12\substack{+0.15 \\ -0.15}$ & $-0.06\substack{+0.2 \\ -0.2}$ & $-0.08\substack{+0.1 \\ -0.1}$ & $\bm{-0.21}\substack{+0.14 \\ -0.14}$ & $-0.05\substack{+0.095 \\ -0.093}$ \\
Any & 25 & 150 & \makecell{2000-2009 \\ 2015-2019} & $\bm{+0.31}\substack{+0.13 \\ -0.13}$ & $+0.01\substack{+0.1 \\ -0.1}$ & $-0.09\substack{+0.2 \\ -0.2}$ & $-0.14\substack{+0.25 \\ -0.25}$ & $+0.06\substack{+0.1 \\ -0.1}$ & $+0.001\substack{+0.2 \\ -0.2}$ & $-0.02\substack{+0.1 \\ -0.1}$ \\
\bottomrule
\end{tabular}\normalsize\renewcommand{\arraystretch}{1}
\end{table}

\begin{table}[H]
\caption{Summary of the effect of each predictor on the probability of having exited a research area across topic models and temporal segmentations. Values indicate the mean posterior effect size and the 95\% credible interval. Significant effects are shown in bold.}
\label{table:full_summary_exited}
\renewcommand{\arraystretch}{2}\fontsize{6}{7}\selectfont\begin{tabular}{lrrllllllll}
\toprule
\makecell{Author-\\ ship} & $K_0$ & $L$ & \makecell{Temporal \\ segmentation} & \makecell{Intell. capital \\ (diversity)} & \makecell{Soc. capital \\ (diversity)} & \makecell{Soc. capital \\ (power)} & \makecell{Stable \\ affiliation} & \makecell{Academic \\ age} & \makecell{Prod. \\ (co-auth.)} & \makecell{Prod. \\ (solo-auth.)} \\
\midrule
Any & 20 & 50 & \makecell{2000-2009 \\ 2015-2019} & $\bm{+1}\substack{+0.14 \\ -0.14}$ & $+0.04\substack{+0.1 \\ -0.1}$ & $-0.03\substack{+0.2 \\ -0.2}$ & $+0.04\substack{+0.2 \\ -0.2}$ & $\bm{-0.21}\substack{+0.12 \\ -0.12}$ & $\bm{-0.28}\substack{+0.15 \\ -0.14}$ & $-0.02\substack{+0.1 \\ -0.1}$ \\
1st/last & 20 & 50 & \makecell{2000-2009 \\ 2015-2019} & $\bm{+1.1}\substack{+0.18 \\ -0.18}$ & $+0.01\substack{+0.1 \\ -0.1}$ & $-0.12\substack{+0.2 \\ -0.2}$ & $-0.05\substack{+0.3 \\ -0.3}$ & $-0.15\substack{+0.16 \\ -0.16}$ & $-0.12\substack{+0.19 \\ -0.19}$ & $-0.01\substack{+0.1 \\ -0.1}$ \\
Any & 20 & 50 & \makecell{2000-2004 \\ 2005-2009} & $\bm{+1}\substack{+0.15 \\ -0.14}$ & $+0.04\substack{+0.1 \\ -0.1}$ & $\bm{-0.18}\substack{+0.16 \\ -0.16}$ & $-0.07\substack{+0.3 \\ -0.3}$ & $\bm{-0.18}\substack{+0.13 \\ -0.13}$ & $\bm{-0.25}\substack{+0.15 \\ -0.15}$ & $+0.04\substack{+0.1 \\ -0.1}$ \\
Any & 20 & 50 & \makecell{2010-2014 \\ 2015-2019} & $\bm{+0.89}\substack{+0.12 \\ -0.11}$ & $+0.04\substack{+0.09 \\ -0.1}$ & $+0.1\substack{+0.1 \\ -0.1}$ & $\bm{-0.22}\substack{+0.22 \\ -0.22}$ & $-0.04\substack{+0.1 \\ -0.1}$ & $\bm{-0.34}\substack{+0.13 \\ -0.13}$ & $\bm{-0.09}\substack{+0.09 \\ -0.09}$ \\
Any & 20 & 50 & \makecell{2000-2009 \\ 2010-2019} & $\bm{+0.88}\substack{+0.093 \\ -0.09}$ & $\bm{+0.16}\substack{+0.077 \\ -0.076}$ & $-0.01\substack{+0.1 \\ -0.1}$ & $+0.07\substack{+0.2 \\ -0.2}$ & $-0.06\substack{+0.09 \\ -0.08}$ & $\bm{-0.32}\substack{+0.11 \\ -0.11}$ & $\bm{-0.09}\substack{+0.07 \\ -0.07}$ \\
Any & 20 & 50 & \makecell{2000-2009 \\ 2015-2019} & $\bm{+0.91}\substack{+0.13 \\ -0.13}$ & $-0.05\substack{+0.1 \\ -0.1}$ & $-0.07\substack{+0.2 \\ -0.2}$ & $+0.001\substack{+0.2 \\ -0.2}$ & $\bm{-0.16}\substack{+0.12 \\ -0.12}$ & $\bm{-0.28}\substack{+0.14 \\ -0.14}$ & $-0.002\substack{+0.1 \\ -0.1}$ \\
Any & 15 & 50 & \makecell{2000-2009 \\ 2015-2019} & $\bm{+1.1}\substack{+0.14 \\ -0.14}$ & $-0.003\substack{+0.1 \\ -0.1}$ & $-0.15\substack{+0.15 \\ -0.15}$ & $-0.19\substack{+0.23 \\ -0.23}$ & $-0.05\substack{+0.1 \\ -0.1}$ & $\bm{-0.21}\substack{+0.14 \\ -0.14}$ & $-0.07\substack{+0.1 \\ -0.1}$ \\
Any & 25 & 50 & \makecell{2000-2009 \\ 2015-2019} & $\bm{+1.1}\substack{+0.16 \\ -0.16}$ & $-0.07\substack{+0.1 \\ -0.1}$ & $\bm{+0.17}\substack{+0.17 \\ -0.17}$ & $-0.06\substack{+0.3 \\ -0.3}$ & $\bm{-0.21}\substack{+0.14 \\ -0.14}$ & $\bm{-0.39}\substack{+0.15 \\ -0.16}$ & $+0.09\substack{+0.1 \\ -0.1}$ \\
Any & 15 & 150 & \makecell{2000-2009 \\ 2015-2019} & $\bm{+1.3}\substack{+0.15 \\ -0.14}$ & $+0.05\substack{+0.1 \\ -0.1}$ & $-0.11\substack{+0.15 \\ -0.15}$ & $+0.04\substack{+0.2 \\ -0.2}$ & $-0.06\substack{+0.1 \\ -0.1}$ & $\bm{-0.23}\substack{+0.15 \\ -0.15}$ & $+0.04\substack{+0.1 \\ -0.1}$ \\
Any & 20 & 150 & \makecell{2000-2009 \\ 2015-2019} & $\bm{+1.1}\substack{+0.14 \\ -0.14}$ & $+0.09\substack{+0.1 \\ -0.1}$ & $-0.09\substack{+0.2 \\ -0.2}$ & $+0.09\substack{+0.2 \\ -0.2}$ & $-0.11\substack{+0.12 \\ -0.12}$ & $\bm{-0.21}\substack{+0.15 \\ -0.15}$ & $-0.06\substack{+0.1 \\ -0.1}$ \\
Any & 25 & 150 & \makecell{2000-2009 \\ 2015-2019} & $\bm{+1.1}\substack{+0.17 \\ -0.17}$ & $+0.1\substack{+0.1 \\ -0.1}$ & $+0.16\substack{+0.18 \\ -0.18}$ & $+0.24\substack{+0.27 \\ -0.27}$ & $\bm{-0.14}\substack{+0.14 \\ -0.14}$ & $\bm{-0.44}\substack{+0.16 \\ -0.16}$ & $-0.05\substack{+0.1 \\ -0.1}$ \\
\bottomrule
\end{tabular}\normalsize\renewcommand{\arraystretch}{1}
\end{table}